\documentclass[prx,aps,twocolumn,superscriptaddress,floatfix,showpacs]{revtex4-1}

\usepackage{graphicx}
\usepackage{amssymb}
\usepackage{amsmath}
\usepackage{changes}
\usepackage{hyperref}
\usepackage{comment}

\usepackage{notes2bib}
\bibnotesetup{
	note-name = ,
	use-sort-key = false
}

\usepackage{filecontents}

\newcommand{\timerev}[1]{} %Show or not the non-TR QPC


\begin{document}

\title{Spin-resolved spectroscopy of helical Andreev bound states} %dc-Shapiro steps without ac-current bias by Majorana fusion rule}
\author{Alessio Calzona}
\affiliation{Institute for Theoretical Physics and Astrophysics, University of W\"urzburg, D-97074 W\"urzburg, Germany}
\author{Bj\"orn Trauzettel}
\affiliation{Institute for Theoretical Physics and Astrophysics, University of W\"urzburg, D-97074 W\"urzburg, Germany}
\affiliation{W\"urzburg-Dresden Cluster of Excellence ct.qmat, Germany}
\date{\today}

\begin{abstract}
	We propose a versatile setup that allows performing spin-resolved spectroscopy of helical Andreev bound states, proving their existence and therefore the topological nature of the Josephson junction that hosts them. The latter is realized on the helical edge of a 2D topological insulator, proximitized with two superconducting electrodes. The spectroscopic analysis is enabled by a quantum point contact, which couples the junction with another helical edge acting as a spin-sensitive probe. By means of straightforward transport measurements, it is possible to detect the particular spin structure of the helical Andreev bound state and to shine light on the mechanism responsible for their existence, i.e. the presence of protected Andreev reflection within the Josephson junction. We discuss the robustness of this helical Andreev spectrometer with respect to processes that weaken spin to charge conversion.
\end{abstract}
\maketitle

\textit{Introduction. 	\textemdash }
	Two-dimensional topological insulators (2DTIs) have been receiving a lot of attention since their first discovery \cite{Konig07,Bernevig06,Kane05,Kane05bis} and represent a prominent example of topological systems. They feature a 2D insulating bulk with topologically protected 1D helical edge states, which propagate in opposite directions and have opposite spin orientation \cite{Qi2011}. Because of this remarkable feature, known as spin-momentum locking, 2DTIs are a precious resource in view of their functionalities in spintronics \cite{He2019,Zhou2014,Linder2015,Breunig2018}. Moreover, helical edges proximitized by superconductors are predicted to realize topological superconductivity \cite{Fu2008,alicea12,Sato2017}, a fundamentally interesting phenomenon with multiple applications, ranging from topological quantum computation \cite{Lutchyn10,kitaev01,nayak08,Aasen16,Oreg10} to low-temperature thermal devices \cite{Sothmann2017,Scharf2020,Blasi2021,Keidel2020,Gresta2021}. To date, 2DTIs have been proposed and realized in a variety of materials \cite{Reis17,Shi2019,Wu2018,Tang2017,Li2018,Liu2020,Knez11,Liu08,Pribiag2015}. One of the most mature platforms is represented by HgTe quantum wells where robust ballistic transport on the edges has been experimentally observed \cite{Konig07, Roth2009,Shamim2021,Bruene2012} and superconducting contacts have been successfully fabricated \cite{Hart2014}. The latter advancement has allowed for the realization of 2DTI-based Josephson junctions (JJs) and to the subsequent observation of fractional Josephson effect \cite{Wiedenmann2016,Bocquillon2016,Bocquillon2018} and superconducting edge transport \cite{Hart2014,Bocquillon2016}. 
	
	A distinctive feature of a JJ defined in a 2DTI is the presence of helical Andreev bound states (hABSs) \cite{Tkachov2013,Beenakker2013,Bocquillon2016,Sothmann2016,Tanaka2009}. They consist of two particular forms of mid-gap states with a specific spin structure and a protected crossing at zero energy. Due to spin-momentum locking and perfect Andreev reflection (AR), each hABS is either a perfect superposition of polarized electrons with spin up and holes with spin down or vice-versa. %These two classes of hABSs are related by time-reversal (TR) and therefore feature distinct energy-phase relations. 
	Their existence represents a hallmark of 2DTI-based JJs and eventually leads to a rich phenomenology, including the theoretically predicted appearance of a zero-energy Majorana Kramers pair \cite{Li2016,Pikulin2016}, as well as the experimentally observed $4\pi$ periodic current-phase relation \cite{Laroche2019,Wiedenmann2016,Bocquillon2016,Bocquillon2018}, and the even-odd effect in the diffraction pattern \cite{Tkachov2015,Baxevanis2015,Pribiag2015}. Unfortunately, experimental signatures of these phenomena can be obscured by non-topological effects \cite{Dartiailh2021,deVries2018,deVries2019}. It is therefore an open question how to directly probe the existence of hABSs and thus the topological origin of the JJ \cite{Blasi2020,Galambos2020,Vigliotti2022}. %The great experimental challenges associated with the creation of a JJ and the subsequent spin-sensitive inspection of its mid-gap states, however, have so far prevented direct observation of hABSs in experiments. 
	
	We propose a feasible setup to perform spectroscopy of Andreev bound states \cite{Danon2020,Nichele2017,Schindele2014} in a spin-sensitive way, by taking advantage of a recent experimental breakthrough in HgTe-based 2DTIs, the realization of a quantum point contact (QPC) that  couples the two helical edges of a 2DTI \cite{Strunz2019}. On one of the edges, say the upper one, we define a JJ by adding two superconducting contacts. On the lower edge, we exploit helicity to detect the spin orientation of electrons by purely electrical means. The resulting setup allows for a spin-resolved DC tunneling spectroscopy  of hABSs within a single integrated device. We demonstrate the robustness of our proposal by investigating the influence of spin-flipping tunneling events in the QPC as well as the effect of spurious electronic backscattering within the JJ.

	%, which result in the non-trivial injection of electron pairs on the lower edge \cite{Pikulin2016,Li2016}. Moreover, we argue that our setup is able to detect the presence of unwanted conventional electronic backscattering within the JJ, thus allowing for an in-depth characterization of the latter.   
	
	 %Importantly, our proposed scheme is robust with respect to the moderate presence of spin-flip tunnelings at the QPC, which is associated with the intriguing injection of non-conventional Cooper pairs into the lower helical edge \cite{Pikulin2016,Li2016,Fleckenstein2021}, and allows for direct detection of PAR at the superconducting interfaces. %Such a complete characterization of the JJ is achieved by means of purely electrical DC measurements. Finally, we stress that the observation of hABSs with our setup also represents a strong evidence of helical transport on the non-proximitized lower edge of the 2DTI, which complements the experimental signature of helicity obtained by piercing the 2DTI with a magnetic field \cite{Bruene2012}. 
 
\begin{figure}
	\centering
	\includegraphics[width=0.95\linewidth]{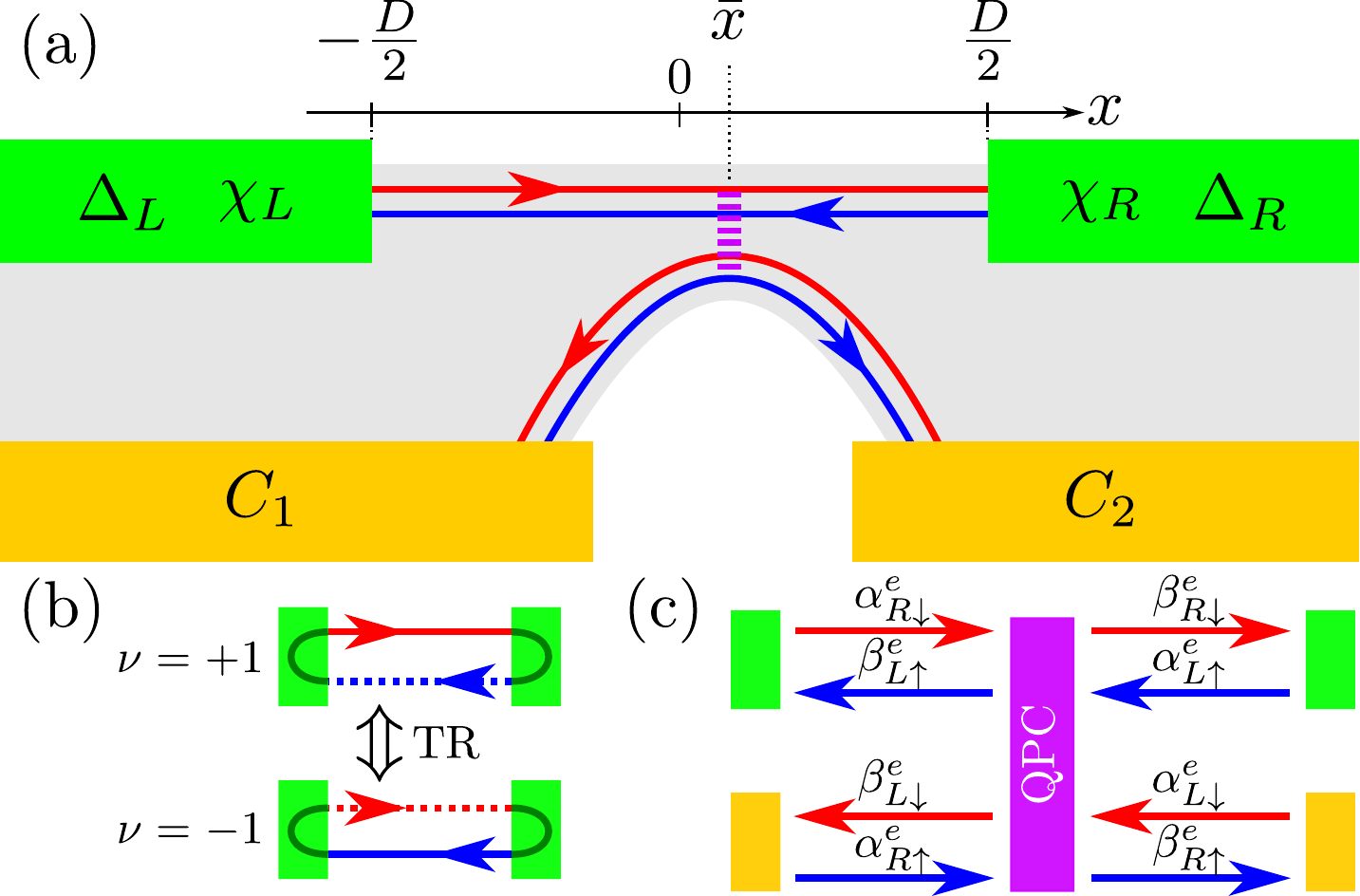}
	\caption{(a) Schematic of the setup. Green (yellow) rectangles represent superconducting (normal) leads. The 2DTI (gray) features helical edge states, shown with blue (spin-up) and red (spin-down) line/arrows. (b) The two classes of hABSs: Solid (dotted) lines represent electrons (holes). (c) Labeling of the electronic amplitudes, incoming ($\alpha$) or outgoing ($\beta$) with respect to the QPC.}
	\label{fig:setup}
\end{figure}

\textit{Setup. \textemdash }
The setup is sketched in Fig.~\ref{fig:setup}(a). It consists of a 2DTI (gray region) featuring helical edge states (blue and red lines). Two standard s-wave superconducting electrodes (depicted in green) induce superconducting correlations on two parts of the upper helical edge of the 2DTI via the proximity effect \cite{Tkachov2015} and thus define a JJ that extends from $x=-D/2$ to $x=D/2$. The tunneling of Cooper pairs between each superconducting electrode and the helical channels underneath locally gaps the latter by inducing an effective pairing potential $\Delta_r \, e^{i\chi_r}$, with $r=R/L$ indicating the left/right superconductor \bibnote{We assume the superconducting regions to be much larger than the superconducting coherence lengths $\xi_r = v/(\pi\Delta_r)$ ($v$ is the Fermi velocity) so that the upper helical edge is completed decoupled from other gapless parts of the 2DTI on the outer side of each superconducting electrode.}. The phase difference  $\chi = \chi_R-\chi_L$, can be controlled by connecting the two electrodes with a superconducting bridge [not shown in the schematic] %, see the Supplemental Material (SM) \cite{SuppMat}] 
and tuning the magnetic flux which threads the resulting asymmetric SQUID \cite{Delagrange2015,Li2018_4pi,Murani2017}. As for the bottom edge, it is contacted with two metallic leads (depicted in yellow and labeled $C_1$ and $C_2$) which allow performing transport measurements. %They are decoupled from the superconductors thanks to the presence of two additional leads, labeled $C_3$ and $C_4$.
We assume the JJ to be long enough so that it can accommodate a QPC, located at $x=\bar x$, that tunnel couples the lower edge with the upper one (see the purple lines).

The upper and lower helical edges are described by the free Hamiltonian with linear dispersion
\begin{equation}
\label{eq:h0}
	H_0 = v \sum_{\sigma}\sum_r \int dx\;  \psi^\dagger_{r\sigma} (x) (-ir\partial_x) \psi_{r\sigma}(x).
\end{equation}
Throughout the paper, we set $\hbar = 1$ unless stated differently. Without loss of generality, we assume channels with positive helicity ($r\sigma > 0$) to be on the lower edge and vice-versa. The $x$ coordinate describes the position along each edge. The operator $\psi_{r\sigma}^\dagger$ creates an electron with spin $\sigma =\, \uparrow,\downarrow\, \equiv \pm1$ on right-moving ($r=R\equiv 1$) or left-moving ($r=L\equiv -1$) channels. 

\textit{Helical Andreev bound states. \textemdash } Within our JJ, particles/holes impinging against a superconducting region necessarily undergo perfect AR \cite{Crepin2014}, i.e. they are completely reflected back as holes/particles with opposite spin, as schematically depicted in Fig.~\ref{fig:setup}(b). That is indeed the only allowed single-particle process %when a single helical edge is contacted with a topological superconductor,
in absence of TR-breaking scatterers and for a large enough bulk gap of the 2DTI. %which completely decouples the single helical edge from other gapless regions in the device. % Our goal is to investigate how PAR affect the transport properties of the system and, in turn, to clarify how the latter can be used to verify and confirm the topological nature of the Josephson junction on the upper edge and its helical nature. 
The amplitude $\alpha_{r\sigma}^c$ of a reflected electron/hole ($c=e/h\equiv \pm 1$) with spin $\sigma$ and propagation direction $r=R/L$, is thus related to the one of the incoming hole/electron simply by a phase shift \cite{SuppMat}
\begin{equation}
\label{eq:ar}
	\alpha_{r\sigma}^c = -i \sigma c \; e^{-i c\, \chi_{\tilde r}} e^{i\Phi_{\tilde r}(E)} \beta_{\tilde r\tilde \sigma}^{\tilde c}
\end{equation}
where $E$ is the energy and the tilde notation indicates the opposite values, i.e. $\tilde e = h$, $\tilde R = L$, $\tilde \uparrow = \downarrow$ and vice-versa. The amplitude of the hole/particle impinging on the $\tilde r$ superconductor is denoted by $\beta_{\tilde r\tilde \sigma}^{\tilde c}$ and we introduce the function $\Phi_r(E) = \pi/2 + EDv^{-1} -\arccos (E/\Delta_r)$. A schematic overview of the scattering amplitude labeling is provided in Fig.~\ref{fig:setup}(c) \bibnote{The labeling for the hole amplitudes is completely analogous.}. 

Due to the helical nature of the weak link, and the consequent presence of perfect AR, the mid-gap states hosted by the JJ acquire a particular structure and are dubbed hABSs. To better highlight their properties, we  temporarily ignore the presence of the QPC. In this case, given Eq.~\eqref{eq:ar}, it is straightforward to show that the JJ hosts a bound state whenever the compatibly condition
$-\exp[i (\nu \chi + \Phi_R(E) + \Phi_L(E))] = 1$ is satisfied \cite{SuppMat, Crepin2014, Beenakker1991}. The quantity $\nu=rc=\pm1$ distinguishes between two decoupled classes of hABSs, related by TR, featuring the energy-phase relations
	\begin{equation}
	\label{eq:Epr}
		\chi= \pm (\pi-\Phi_L(E)-\Phi_R(E))\quad  \text{(mod $2\pi$)}. 
	\end{equation} 
Importantly, each class has a specific spin structure: hABSs with $\nu=+1$ ($\nu=-1$) consist only of right-moving spin-down electrons (holes) and left-moving spin-up holes (electrons), illustrated in Fig.~\ref{fig:setup}(b). 

\textit{Spin-resolved Andreev spectroscopy. \textemdash} We now demonstrate how the intriguing structure of hABSs can be directly probed by means of the QPC sketched in Fig.~\ref{fig:setup}(a). We model it by the point-like tunneling Hamiltonian \cite{Ferraro2014,Calzona2019,Stroem2009,Inhofer2013,Dolcini2010}
\begin{equation}
\label{eq:h_qpc}
H_{\rm QPC} = 2v \lambda_p \sum_{\sigma} \psi_{R\sigma}^\dagger \psi_{L\sigma} + 2v\lambda_f \sum_r r\,  \psi_{r\uparrow}^\dagger \psi_{r\downarrow} + h.c., 
\end{equation}
where the fermionic operators are evaluated at the position $\bar x$ of the QPC. This Hamiltonian consists of both spin-preserving (amplitude $\lambda_p$) and spin-flipping (amplitude $\lambda_f$) tunneling terms. The latter can exists only if spin-axial symmetry is broken, a condition that cannot be excluded at the QPC where the pinching of the edges can induce a local modification of the spin-orbit coupling \cite{Dolcini2010,Ferraro2014}. Therefore, we also consider the effects of a small-but-finite $\lambda_f$. For the QPC Hamiltonian to be TR invariant, which is the case considered here, the amplitudes $\lambda_p$ and $\lambda_f$ must be real.

\begin{figure}
	\centering
	\includegraphics[width=\linewidth]{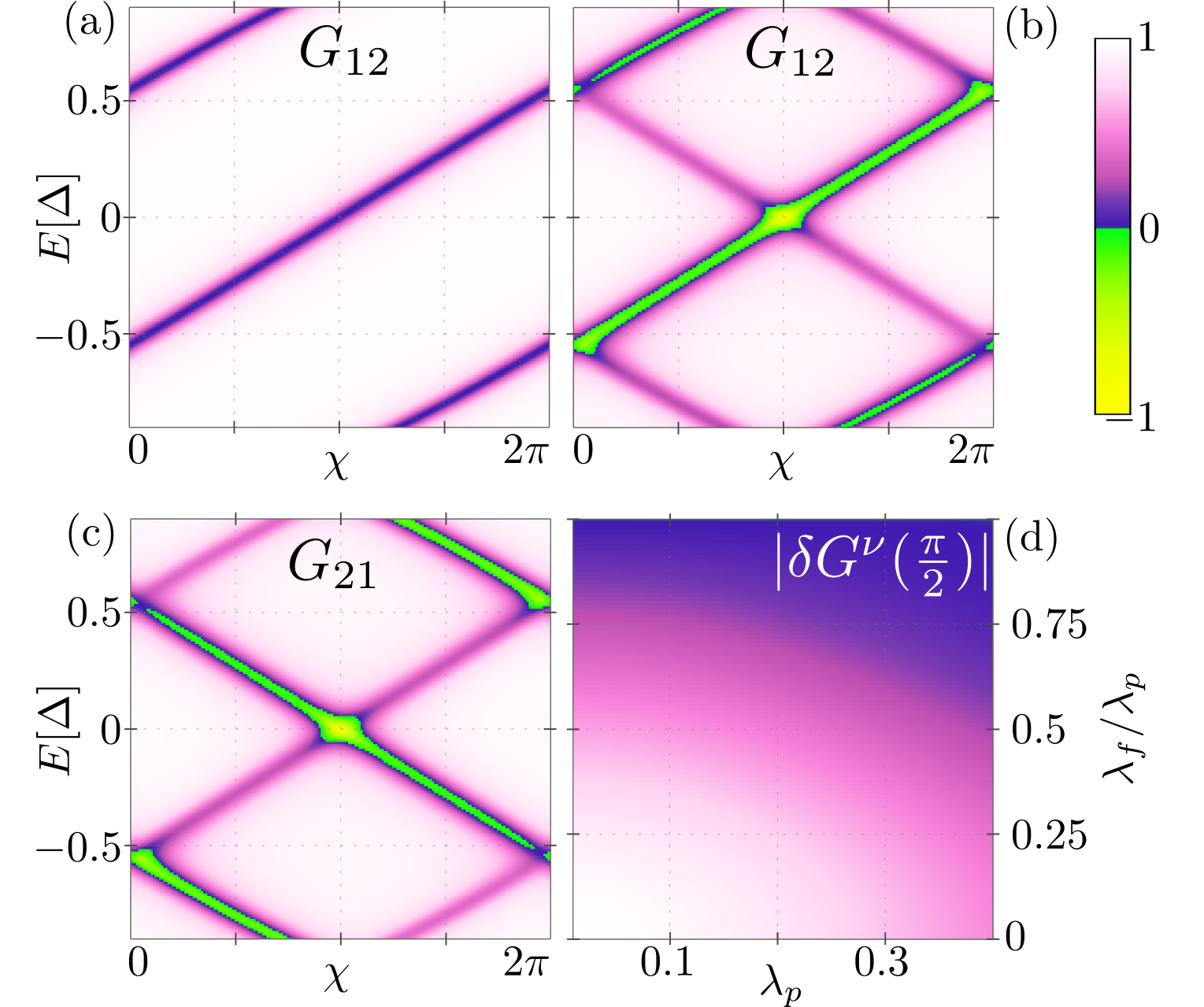}
	\caption{(a-c) Non-local differential conductances $G_{12/21}(E,\chi)$ for the parameters: $\lambda_p=0.25$, $\bar x = 1\xi_L$, $\Delta_L=\Delta$, $\Delta_R=1.2 \Delta$, {$D = 6\xi_L$, with the coherence length $\xi_L=v/(\pi\Delta_L)$}.  In (a) we plot $G_{12}$ for $\lambda_f = 0$. In (b) and (c) we plot $G_{12}$ and $G_{21}$, respectively, with $\lambda_f=0.15$. In (d) we plot $|\delta G^\nu|(\pi/2)|$ {[see Eq.~\eqref{eq:deltaG}]} as a function of $\lambda_p$ and $\lambda_f/\lambda_p$. {All plots share the same color bar, in units of $e^2/h$.}}
	\label{fig:plot}
\end{figure}

We derive the transport properties on the lower edge using the scattering matrix formalism. The combined effect of the QPC and AR within the JJ results in the relation
% The QPC scattering matrix, derived from the Hamiltonian $H_0+H_{\rm QPC}$, relates the amplitudes $\alpha^{e/h}_{r\sigma}$ of electrons/holes impinging on the QPC, on $r$-moving channels with spin $\sigma$, with the amplitudes $\beta^{e/h}_{r\sigma}$ of the outgoing electrons \cite{SuppMat}.  %which reads \cite{SuppMat,Ferraro2014,Calzona2019}
%\begin{equation} 
%\label{eq:sQPC}
%\begin{pmatrix}
%\beta_{L\uparrow}^e\\
%\beta_{L\downarrow}^e\\
%\beta_{R\uparrow}^e\\
%\beta_{R\downarrow}^e
%\end{pmatrix} = \frac{1}{ \zeta}
%\begin{pmatrix}
%\Lambda_{\rm pf} & \Lambda_{\rm ff} & \Lambda_{\rm pb}& 0\\
%\Lambda_{\rm ff}&\Lambda_{\rm pf} &0&\Lambda_{\rm pb}\\
%-\Lambda_{\rm pb}^*&0&\Lambda_{\rm pf} &-\Lambda_{\rm ff}\\
%0&-\Lambda_{\rm pb}^*&-\Lambda_{\rm ff}&\Lambda_{\rm pf} \\
%\end{pmatrix}
%\begin{pmatrix}
%\alpha_{L\uparrow}^e\\
%\alpha_{L\downarrow}^e\\
%\alpha_{R\uparrow}^e\\
%\alpha_{R\downarrow}^e
%\end{pmatrix},
%\end{equation}
%with $\zeta=1+\lambda_{f}^2+\lambda_{p}^2$. The three parameters $\Lambda_{\rm pf} = (1-\lambda_{f}^2-\lambda_{p}^2)$, $\Lambda_{\rm ff} = 2 i \lambda_f$, and $\Lambda_{\rm pb}= -{2 i \lambda_p e^{2iE\bar x/v}}$ are associated with spin-preserving forward scattering, spin-flipping forward scattering, and spin-preserving backscattering processes respectively \cite{Inhofer2013}. 
%Its combination with the perfect AR constraints in Eq.~\eqref{eq:ar} allows us to relate outgoing and incoming amplitudes on the lower edge via the scattering matrix \cite{SuppMat} (we suppress the redundant spin index)
\begin{equation}
\label{eq:SG}
\begin{pmatrix}
\beta_R^e\\
\beta_L^e\\
\beta_R^h\\
\beta_L^h
\end{pmatrix}
 = \begin{pmatrix}
 t^{ee}_{\rightarrow} &   r^{ee}_{\hookrightarrow}&  c^{eh}_{\rightarrow} & a^{eh}_{\hookrightarrow} \\
 r^{ee}_{\hookleftarrow}	& t^{ee}_{\leftarrow}& a^{eh}_{\hookleftarrow}& c^{eh}_{\leftarrow}\\
 c^{he}_{\rightarrow}&  a^{he}_{\hookrightarrow}& t^{hh}_{\rightarrow}& r^{hh}_{\hookrightarrow}\\
 a^{he}_{\hookleftarrow}& c^{he}_{\leftarrow}& r^{hh}_{\hookleftarrow}& t^{hh}_{\leftarrow}	
\end{pmatrix}
\begin{pmatrix}
\alpha_R^e\\
\alpha_L^e\\
\alpha_R^h\\
\alpha_L^h
\end{pmatrix}
\end{equation}
between outgoing and incoming amplitudes on the lower edge (we suppress the redundant spin index) \cite{SuppMat,Ferraro2014,Calzona2019}. It features four kinds of energy-dependent coefficients, which refer to four distinct physical processes: transmission of electrons/holes (t), backscattering of electrons/holes (r), Andreev reflection (a), and {Andreev transmission} (c). Those coefficients, whose analytical expressions are presented in the Supplemental Material \cite{SuppMat}, are directly related to transport properties of the lower edge. At first, we focus on the non-local differential conductances
\begin{equation}
\label{eq:Gs}
\begin{split}
	G_{12}(V_1, \chi) &= \frac{d I_2}{d V_1} = \frac{e^2}{h} (|t^{ee}_\rightarrow|^2 - |c_\rightarrow^{he}|^2), \\
	G_{21}(V_2, \chi) &= \frac{d I_1}{d V_2} = \frac{e^2}{h} (|t^{ee}_\leftarrow|^2 - |c_\leftarrow^{he}|^2),
	\end{split}
\end{equation}
describing the transmission between contacts $C_1$ and $C_2$ \bibnote{Note that our setup is effectively a multi-terminal device, featuring two metallic and two superconducting electrodes. The latter ones can act as sinks/sources of pairs of electrons via Andreev reflection.}. If the two edges are decoupled (i.e. $\lambda_f=\lambda_p=0$), $G_{12/21}$ are both quantized at $e^2/h$. By contrast, when tunneling across the QPC is allowed, we expect deviations from the quantized values whenever the hABSs formed on the upper edge couple to the lower one. This is clearly shown in Fig.~\ref{fig:plot}, where we plot $G_{12/21}$ for a variety of system parameters. 

If only spin-preserving tunneling is allowed (i.e. $\lambda_f=0$), the QPC operates as a perfect spin-resolved Andreev spectrometer: $G_{12}$ ($G_{21}$) is indeed only sensitive to electronic states on the upper edge with spin-up (spin-down). This allows us to perfectly distinguish between the two classes of hABSs, as $G_{12}$ ($G_{21}$) exclusively reveals the energy-phase relation of the $\nu=-1$ ($\nu=+1$) class. In particular, we predict $G_{12}(E,\chi) = G_{21}(E,-\chi)$ with \bibnote{Note that Andreev transmission processes necessarily involve both spin-preserving and spin-flipping tunnelings. Hence, for $\lambda_f=0$, we have $c^{he}_\leftrightarrow=0$ and $0\leq G_{12/21} \leq e^2/h$.}
\begin{equation}
\label{eq:Gideal}
G_{12}(E,\chi) = \left[1+\frac{8\lambda_p^4}{(\lambda_p^4-1)^2(1+\cos(\Phi_L+\Phi_R-\chi))}\right]^{-1}.
\end{equation}
This result is displayed in Fig.~\ref{fig:plot}(a) {where, to highlight the robustness of the setup against geometrical asymmetries, we consider a finite $\bar x\neq 0$ and {$\Delta_R\neq\Delta_L$ \cite{Chang1994} \bibnote{Since we consider $\Delta_R \sim \Delta_L$ and finite junction lengths in order to accomodate the QPC (e.g. $D=6\xi_L$ in Figs. 2 and 3), we expect to find ABS for all the values of the phase difference $\chi$ \cite{Chang1994}.}.} In this regime, our device can therefore provide a compelling proof of the existence of hABSs and directly probe their particular spin structure. 

Importantly, the proposed setup is robust with respect to the two main effects that lead to deviations from the ideal scenario discussed above: (i) Limited spin-resolving power of the detector due to a finite spin-flipping amplitude $\lambda_f$ at the QPC; (ii) Presence of spurious reflection mechanisms within the JJ. In the following, we carefully analyze both of them. 

\textit{Robustness against spin-flip events. \textemdash} Our device retains a strong spin-sensitivity even in presence of finite $\lambda_f$. This is shown, for example, in Figs.~\ref{fig:plot}(b,c), where we plot $G_{12}$ [panel (b)] and $G_{21}$ [panel (c)] for $\lambda_p=0.25$ and a relatively large $\lambda_f=0.15$. {Each conductance is sensitive to both classes of hABSs, since a given spin channel on the lower edge is now coupled to both spin orientations on the upper edge. However, reductions from the quantized value are particularly pronounced only for one of them.} {The qualitative distinction between the two classes is facilitated by the presence of regions featuring negative values of the conductance, resulting from the onset of the Andreev transmission processes discussed in the next paragraph. To achieve analytical progress, we consider the (reasonable) assumption $\Delta_R=\Delta_L=\Delta$} that allows us to derive a compact expression for $\delta G^\nu (\chi) = G_{12}(E_\nu,\chi)-G_{21}(E_\nu,\chi)$, 
	\begin{equation}
	\label{eq:deltaG}
	\delta G^\nu (\chi)= \frac{e^2}{h} \frac{\nu (\lambda_p^2-\lambda_f^2)(\Lambda^2-1)^2 \sin(\chi)^2}{4\Lambda^3 \cos(\chi)^2+\Lambda(\Lambda^2+1)^2 \sin(\chi)^2},
	\end{equation}  
with $\Lambda=\lambda_p^2+\lambda_f^2$ and $E_\nu$ satisfying the energy-phase relations $\Phi(E_\nu) = (\pi-\nu\chi)/2$ of the hABSs. As expected, the difference vanishes for $\chi=0,\pi$ (when the hABSs are degenerate) and for $\lambda_p=\lambda_f$ (when the tunneling at the QPC becomes spin-independent). By contrast, the maximum absolute value of $\delta G^\nu (\chi)$ is reached for $\chi=\pi/2, 3\pi/2$. In Fig.~\ref{fig:plot}(d), we plot $|\delta G^\nu (\pi/2)|$ as a function of $\lambda_p$ and $\lambda_f/\lambda_p$ and show that it remains significantly larger than zero even when $\lambda_f$ represents a sizable fraction of $\lambda_p$. % , and increases as $\lambda_p$ (and thus the hABSs line-width) is reduced. 
Therefore, even for finite $\lambda_f$, the asymmetry between $G_{12}$ and $G_{21}$ remains a good indicator of the existence of two classes of Andreev bound states, with complementary spin-structure and energy-phase relation. %We also stress that the very existence of such asymmetry can be regarded as strong evidence for helicity on the lower 2DTI edge.
{Only in the specific case $\lambda_p=\lambda_f$, tunneling at the QPC becomes completely spin-independent. Then, it is not possible anymore to distinguish between the two hABS classes.}

Figs.~\ref{fig:plot}(b,c) show another intriguing feature: When both tunneling amplitudes are finite, it is possible to observe negative values of $G_{12/21}$ (green-yellow regions). Indeed, for $\lambda_p,\lambda_f \neq 0$, Andreev transmission processes $c^{he}_\leftrightarrow$ are allowed at the QPC and they can dominate over the standard transmission. This is particularly evident close to the energy-phase relations of the hABSs, where the standard transmission is suppressed. Focusing on $E=0$ and $\chi=\pi$, we find the simple relation $0\geq (G_{12/21} \propto -\lambda_p^2\lambda_f^2) \geq -{e^2/h}$ \cite{SuppMat}, stemming from the presence of a Kramers pair of Majorana zero modes (obtained from a proper linear combination of the degenerate hABSs) \cite{Li2016,Pikulin2016}. In general, however, negative values of $G_{12/21}$ can be found at all energies below the gap, and even in absence of hABSs, see Figs.~\ref{fig:plot_M}(a,b) and Ref. \cite{Fleckenstein2021}. 

\begin{figure}
	\centering
	\includegraphics[width=\linewidth]{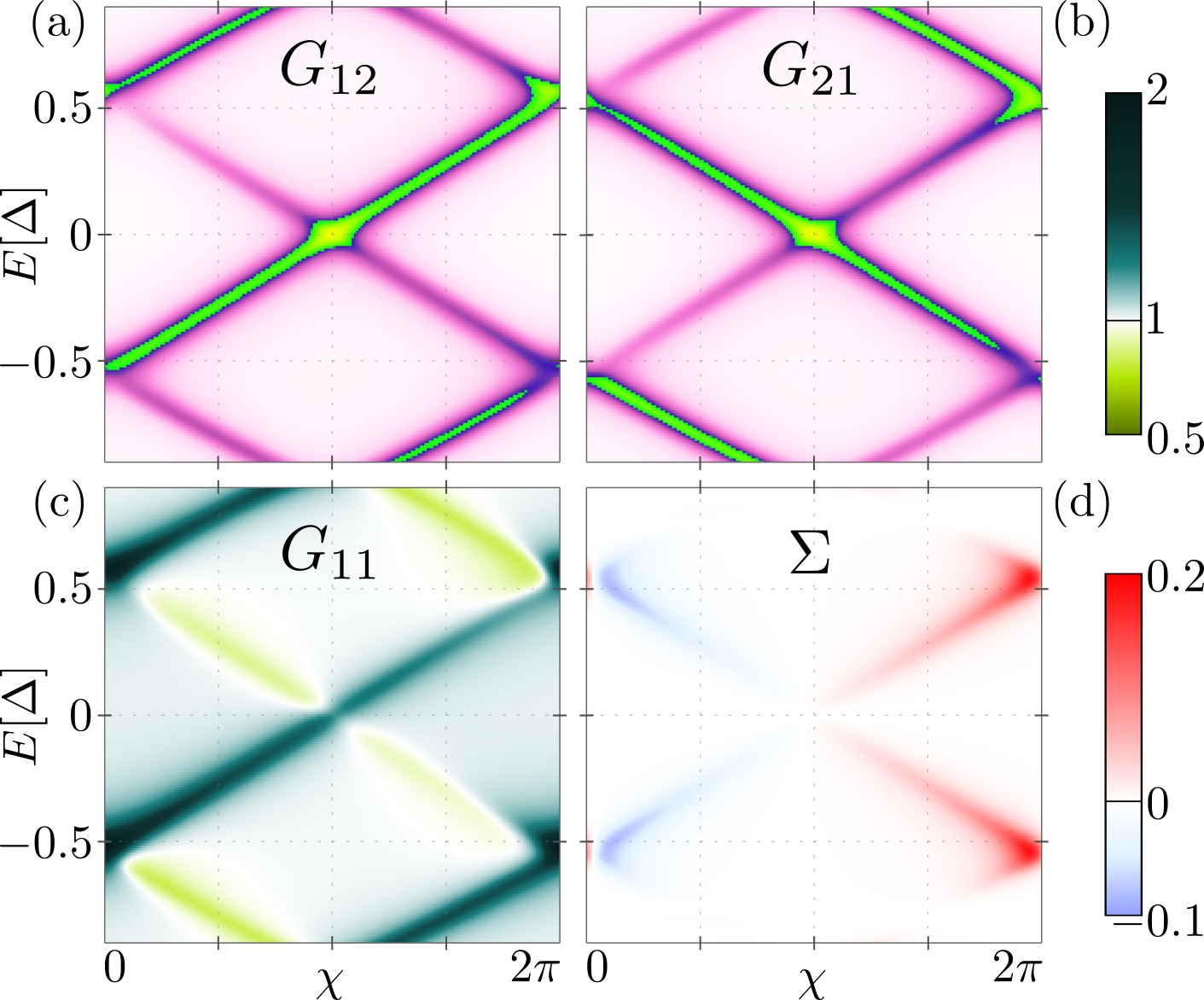}
	\caption{Differential conductances in presence of electronic backscattering on the upper edge, induced by a magnetic impurity with $m=0.05$ and located at $x_{\rm M}=2\xi_L$. The {other} parameters are the same as in Figs.\ref{fig:plot}(b,c). The conductances are in units of $e^2/h$. Plots (a,b) share the same color bar with Fig.~\ref{fig:plot}. }
	\label{fig:plot_M}
\end{figure}

\textit{Spurious backscattering terms. \textemdash} To study the effects of spurious backscattering mechanisms within the JJ, we begin by considering the example provided in Figs.~\ref{fig:plot_M}(a,b). There, we plot the conductances $G_{12/21}$ obtained for the same parameters used in Figs.~\ref{fig:plot}(b,c) but considering the additional presence of a weak magnetic impurity within the JJ. The latter, described by the Hamiltonian $H_M = 2vm\, \psi_{L\uparrow}^\dagger(x_M) \psi_{R\downarrow}(x_M) + h.c.$, induces electronic backscattering at $x=x_{\rm M}$ and therefore hybridizes the hABSs. For a small impurity strength $m=0.05\ll1$, %\textcolor{blue}{[Comment. From Eq. (E7) one can see that $m$ is directly related to the probability of a backscattering event induced by the impurity. In particular, for $m=0$ there is no scattering at all, while for $m=1$ the reflection amplitudes become $1$ and the transmission ones vanish. It's basically the same thing for $\lambda_p$ and $\lambda_f$ in Eq.~(B5-B8) However, here in the main text I just added $\ll 1$ in order not to break too much the flow.]}
 the resulting mid-gap states retain a significant spin-polarization, which emerges as a strong asymmetry between the two non-local conductances in Figs.~\ref{fig:plot_M}(a,b). While this underlines once more the high spin-sensitivity of our device, a direct comparison with Figs.~\ref{fig:plot}(b,c) shows that a qualitative analysis of $G_{12/21}$ alone might not be able to clearly distinguish between the two cases. {In this respect, we stress that a violation of
 	\begin{equation}
 		\label{eq:Gsymm}
 		G_{12/21}(E,\chi) = G_{12/21}(-E,-\chi)
 	\end{equation}
 represents a proof for the absence of hABSs. %, as Eq.~\eqref{eq:Gsymm} necessarily holds in presence of hABSs.
 However, the converse is not true  \cite{supp} %, since Eq.~\eqref{eq:Gsymm} can hold even in absence of hABSs \cite{supp}. 
 and this calls for a more in-depth analysis of the conductances.} 

Our device provides us with an additional tool that can efficiently detect the presence of spurious electronic backscattering within the JJ, allowing us to distinguish between the presence of hABSs and other mid-gap states. In particular, our setup naturally allows for the measurement of the local differential conductances
\begin{equation}
\label{eq:G11_22}
\begin{split}
G_{11}(V_1, \chi) &= \frac{d I_1}{d V_1} = \frac{e^2}{h} (1-|r^{ee}_{\hookleftarrow}|^2 + |a_{\hookleftarrow}^{he}|^2), \\
G_{22}(V_2, \chi) &= \frac{d I_2}{d V_2} = \frac{e^2}{h} (1-|r^{ee}_{\hookrightarrow}|^2 + |a_{\hookrightarrow}^{he}|^2),
\end{split}
\end{equation}
that can be combined to define 
\begin{equation}
\label{eq:Xi}
\begin{split}
\Sigma &= G_{12}+G_{11}-G_{21}-G_{22}\\
& {= 2\big(
|r^{ee}_{\hookrightarrow}|^2 + |c^{he}_{\leftarrow}|^2 - |r^{ee}_{\hookleftarrow}|^2 - |c^{he}_{\rightarrow}|^2 \big).}
\end{split}
\end{equation} 
{The four independent quantities $G_{12/21/11}$ and $\Sigma$ are plotted in Fig.~\ref{fig:plot_M}.} {The presence of hABSs implies $\Sigma=0$. Indeed, as each hABS consists only of electrons with a specific spin orientation, there is only one way for electrons to tunnel from (into) that hABS. In particular, the reflection of an electron emerging from contact $C_j$ necessarily involves one spin-flipping and one spin-preserving tunneling event, regardless of $j=1,2$. The same applies for the Andreev transmission and leads to $|r^{ee}_{\hookrightarrow}|^2=|r^{ee}_{\hookleftarrow}|^2$ and $|c^{he}_{\leftarrow}|^2=|c^{he}_{\rightarrow}|^2$ \cite{SuppMat}. By contrast, mid-gap states with non-helical features offer more channels for the electrons to tunnel between the edges. %Reflection and Andreev transmission amplitudes result from the interference between different paths and become strongly dependent on the contacts $C_j$ the electron originated from. 
This results in deviations from $\Sigma = 0$ as shown in Fig.~\ref{fig:plot_M}(d).}
%\textcolor{blue}{ hABS, the reflection of an incoming electron back to the lower edge necessarily involve a single spin-preserving and spin-flipping tunneling event at the QPC, regardless of the initial electron spin. The same applies for the Andreev transmission, leading to $|r^{ee}_{\hookrightarrow}|^2=|r^{ee}_{\hookleftarrow}|^2$ and $|c^{he}_{\leftarrow}|^2=|c^{he}_{\rightarrow}|^2$, and thus to $\Sigma=0$ \cite{SuppMat}. By contrast, mid-gap states with a more generic structure offer more channels for the electrons to tunnel between the edges \cite{SuppMat}. Reflection and Andreev transmission processes result from the interference between different paths and become strongly dependent on the initial spin of the incoming electron, leading to deviations from $\Sigma = 0$ as shown in Fig.~\ref{fig:plot_M}(d).}
%\textcolor{blue}{The structure of hABSs guarantees that the reflection and Andreev transmission of one incoming electron necessarily involve a spin-preserving and a spin-flipping tunneling event at the QPC, regardless of the electron spin. As a result, one has  $|r^{ee}_{\hookrightarrow}|^2=|r^{ee}_{\hookleftarrow}|^2$ and $|c^{he}_{\leftarrow}|^2=|c^{he}_{\rightarrow}|^2$, leading to $\Sigma=0$. By contrast, in presence of a more generic mid-gap state, the incoming electron can follow several different paths that lead to a reflection or Andreev transmission \cite{SuppMat}. In general, the interference between these multiple paths is strongly dependent on the electron spin, leading to deviations from $\Sigma = 0$ as shown in Fig.~\ref{fig:plot_M}(d).}
We thus claim that the consistent observation of a vanishing $\Sigma=0$ over a wide range of parameters represents a proof of the presence of hABSs.	

To properly demonstrate this, we consider the presence of generic backscattering processes between the two helical channels of the upper edge. To this end, we replace the AR-induced constraints on the scattering amplitudes in Eq.~\eqref{eq:ar} with the more general conditions 
\begin{equation}
\label{eq:ref_gen}
\begin{pmatrix}
\alpha_{r\sigma}^e\\
\alpha_{r{\sigma}}^h 
\end{pmatrix}
=\begin{pmatrix}
\cos(\theta_r) e^{i \xi^{ee}}_r & \sin(\theta_r) e^{i \xi^{eh}_r  }\\	
\sin(\theta_r) e^{i \xi^{he}}_r & \cos(\theta_r) e^{i \xi^{hh}_r  }\\
\end{pmatrix}
\begin{pmatrix}
\beta_{\tilde r\tilde\sigma}^e\\
\beta_{\tilde r\tilde{\sigma}}^h 
\end{pmatrix},
\end{equation}
describing generic complete reflection processes happening to the left ($r=L$) or to the right ($r=R$) of the QPC. The parameters $\theta_r$ (mod $\pi$) control the nature of these processes, which can continuously range from the pure electronic backscattering limit ($\theta_r=0$) to the pure AR scenario considered in Eq.~\eqref{eq:ar} and corresponding to $\theta_r = \pi/2$ (with $\xi^{eh/he}_r(E) =\mp (\chi_r + r \pi/2) + \Phi_r(E)$). The analytical expressions of the four differential conductances, determined by imposing Eq.~\eqref{eq:ref_gen}, show that the exclusive presence of AR (and thus of hABSs) implies a vanishing $\Sigma=0$ \cite{SuppMat}. Moreover, with the possible exception of fine-tuned points in the parameter space, we demonstrate that $\Sigma=0$ also guarantees $\theta_R=\theta_L=\pi/2$ \bibnote{Strictly speaking, the observation of $\Sigma=0$ implies either $\theta_R=\theta_L=0$ (i.e. pure electronic backscattering) or $\theta_R=\theta_L=\pi/2$ (i.e. pure AR). It is however straightforward to distinguish between these to limits. For example, in presence of pure electronic backscattering, no Andreev processes are present and $G_{12/21}$ cannot be negative. Moreover, in this case, the superconductors should not play any role and the conductances are thus expected not to depend on $\chi$.}, and thus the presence of pure AR \cite{SuppMat}. Therefore, the experimental observation of $\Sigma=0$ over a wide range of parameters, {e.g. over the $(E-\chi)$ diagram and for different samples featuring different JJ length, QPC transparency and position, represents a proof of the exclusive presence of AR within the JJ. This, in turn, assures that the mid-gap states probed by our spin-resolved Andreev spectrometer are indeed hABSs.}

\textit{Discussion. \textemdash} We argue that our proposal is within experimental reach, given the degree of maturity of HgTe-based 2DTI devices. The edge states exhibit ballistic mean free paths that exceed several micrometers \cite{Bocquillon2018,Konig07}. {We, therefore, expect that inelastic scattering, which would be detrimental both for the formation of hABS and for their detection, should not play a major role in a realization of our proposal based on HgTe.} Josephson junctions have been successfully realized, for example with Al contacts, featuring an estimated coherence length $\xi \simeq\,  600\, {\rm nm}$ and superconducting gap $\Delta \sim 40\, \mu {\rm eV}$ \cite{Bocquillon2016,Bocquillon2018}. As the physical width of a QPC is typically of the order of $100\, {\rm nm}$ \cite{Strunz2019}, the geometrical constraints of our proposal can be satisfied. Importantly, our results are robust with respect to geometrical asymmetries in the setup, as confirmed by Figs.~\ref{fig:plot} and \ref{fig:plot_M} that have been obtained for an asymmetric position of the QPC and asymmetric pairing potentials $\Delta_L\neq \Delta_R$.

 %\textcolor{red}{We do not consider effects of electron-electron interactions. They are predicted to be irrelevant at zero energy \cite{Pikulin2016} but they might affect the finite-energy behavior of our setup. In this case, the data provided by our device can represent a precious resource to shine more light on the still poorly understood topic of interaction effect in QPC defined on thick 2DTIs \cite{Strunz2019}}. 

%We proposed a versatile and feasible setup, which allows us to build a JJ and validate its topological nature by performing a spin-resolved spectroscopy of its hABSs. The latter, which has never been carried out to date, is allowed by a QPC that couples the helical edge states of a 2DTI. A qualitative distinction between the two classes of hABSs persists even for a moderate spin-flipping tunneling amplitude at the QPC, which is associated with the emergence of negative non-local conductances. 

\begin{acknowledgements}
	This work was supported by the W\"urzburg-Dresden Cluster of Excellence ct.qmat, EXC2147, project-id 390858490, and the DFG (SPP 1666 and SFB 1170). Additionally, we acknowledge support from the High Tech Agenda Bavaria. 
\end{acknowledgements}

\bibliography{Ref,notes}

\begin{thebibliography}{77}%
\makeatletter
\providecommand \@ifxundefined [1]{%
 \@ifx{#1\undefined}
}%
\providecommand \@ifnum [1]{%
 \ifnum #1\expandafter \@firstoftwo
 \else \expandafter \@secondoftwo
 \fi
}%
\providecommand \@ifx [1]{%
 \ifx #1\expandafter \@firstoftwo
 \else \expandafter \@secondoftwo
 \fi
}%
\providecommand \natexlab [1]{#1}%
\providecommand \enquote  [1]{``#1''}%
\providecommand \bibnamefont  [1]{#1}%
\providecommand \bibfnamefont [1]{#1}%
\providecommand \citenamefont [1]{#1}%
\providecommand \href@noop [0]{\@secondoftwo}%
\providecommand \href [0]{\begingroup \@sanitize@url \@href}%
\providecommand \@href[1]{\@@startlink{#1}\@@href}%
\providecommand \@@href[1]{\endgroup#1\@@endlink}%
\providecommand \@sanitize@url [0]{\catcode `\\12\catcode `\$12\catcode
  `\&12\catcode `\#12\catcode `\^12\catcode `\_12\catcode `\%12\relax}%
\providecommand \@@startlink[1]{}%
\providecommand \@@endlink[0]{}%
\providecommand \url  [0]{\begingroup\@sanitize@url \@url }%
\providecommand \@url [1]{\endgroup\@href {#1}{\urlprefix }}%
\providecommand \urlprefix  [0]{URL }%
\providecommand \Eprint [0]{\href }%
\providecommand \doibase [0]{http://dx.doi.org/}%
\providecommand \selectlanguage [0]{\@gobble}%
\providecommand \bibinfo  [0]{\@secondoftwo}%
\providecommand \bibfield  [0]{\@secondoftwo}%
\providecommand \translation [1]{[#1]}%
\providecommand \BibitemOpen [0]{}%
\providecommand \bibitemStop [0]{}%
\providecommand \bibitemNoStop [0]{.\EOS\space}%
\providecommand \EOS [0]{\spacefactor3000\relax}%
\providecommand \BibitemShut  [1]{\csname bibitem#1\endcsname}%
\let\auto@bib@innerbib\@empty
%</preamble>
\bibitem [{\citenamefont {K{\"o}nig}\ \emph {et~al.}(2007)\citenamefont
  {K{\"o}nig}, \citenamefont {Wiedmann}, \citenamefont {Br{\"u}ne},
  \citenamefont {Roth}, \citenamefont {Buhmann}, \citenamefont {Molenkamp},
  \citenamefont {Qi},\ and\ \citenamefont {Zhang}}]{Konig07}%
  \BibitemOpen
  \bibfield  {author} {\bibinfo {author} {\bibfnamefont {M.}~\bibnamefont
  {K{\"o}nig}}, \bibinfo {author} {\bibfnamefont {S.}~\bibnamefont {Wiedmann}},
  \bibinfo {author} {\bibfnamefont {C.}~\bibnamefont {Br{\"u}ne}}, \bibinfo
  {author} {\bibfnamefont {A.}~\bibnamefont {Roth}}, \bibinfo {author}
  {\bibfnamefont {H.}~\bibnamefont {Buhmann}}, \bibinfo {author} {\bibfnamefont
  {L.~W.}\ \bibnamefont {Molenkamp}}, \bibinfo {author} {\bibfnamefont {X.-L.}\
  \bibnamefont {Qi}}, \ and\ \bibinfo {author} {\bibfnamefont {S.-C.}\
  \bibnamefont {Zhang}},\ }\href {\doibase 10.1126/science.1148047} {\bibfield
  {journal} {\bibinfo  {journal} {Science}\ }\textbf {\bibinfo {volume}
  {318}},\ \bibinfo {pages} {766} (\bibinfo {year} {2007})}\BibitemShut
  {NoStop}%
\bibitem [{\citenamefont {Bernevig}\ \emph {et~al.}(2006)\citenamefont
  {Bernevig}, \citenamefont {Hughes},\ and\ \citenamefont
  {Zhang}}]{Bernevig06}%
  \BibitemOpen
  \bibfield  {author} {\bibinfo {author} {\bibfnamefont {B.~A.}\ \bibnamefont
  {Bernevig}}, \bibinfo {author} {\bibfnamefont {T.~L.}\ \bibnamefont
  {Hughes}}, \ and\ \bibinfo {author} {\bibfnamefont {S.-C.}\ \bibnamefont
  {Zhang}},\ }\href {\doibase 10.1126/science.1133734} {\bibfield  {journal}
  {\bibinfo  {journal} {Science}\ }\textbf {\bibinfo {volume} {314}},\ \bibinfo
  {pages} {1757} (\bibinfo {year} {2006})}\BibitemShut {NoStop}%
\bibitem [{\citenamefont {Kane}\ and\ \citenamefont
  {Mele}(2005{\natexlab{a}})}]{Kane05}%
  \BibitemOpen
  \bibfield  {author} {\bibinfo {author} {\bibfnamefont {C.~L.}\ \bibnamefont
  {Kane}}\ and\ \bibinfo {author} {\bibfnamefont {E.~J.}\ \bibnamefont
  {Mele}},\ }\href {\doibase 10.1103/PhysRevLett.95.226801} {\bibfield
  {journal} {\bibinfo  {journal} {Phys. Rev. Lett.}\ }\textbf {\bibinfo
  {volume} {95}},\ \bibinfo {pages} {226801} (\bibinfo {year}
  {2005}{\natexlab{a}})}\BibitemShut {NoStop}%
\bibitem [{\citenamefont {Kane}\ and\ \citenamefont
  {Mele}(2005{\natexlab{b}})}]{Kane05bis}%
  \BibitemOpen
  \bibfield  {author} {\bibinfo {author} {\bibfnamefont {C.~L.}\ \bibnamefont
  {Kane}}\ and\ \bibinfo {author} {\bibfnamefont {E.~J.}\ \bibnamefont
  {Mele}},\ }\href {\doibase 10.1103/PhysRevLett.95.146802} {\bibfield
  {journal} {\bibinfo  {journal} {Phys. Rev. Lett.}\ }\textbf {\bibinfo
  {volume} {95}},\ \bibinfo {pages} {146802} (\bibinfo {year}
  {2005}{\natexlab{b}})}\BibitemShut {NoStop}%
\bibitem [{\citenamefont {Qi}\ and\ \citenamefont {Zhang}(2011)}]{Qi2011}%
  \BibitemOpen
  \bibfield  {author} {\bibinfo {author} {\bibfnamefont {X.-L.}\ \bibnamefont
  {Qi}}\ and\ \bibinfo {author} {\bibfnamefont {S.-C.}\ \bibnamefont {Zhang}},\
  }\href {\doibase 10.1103/RevModPhys.83.1057} {\bibfield  {journal} {\bibinfo
  {journal} {Rev. Mod. Phys.}\ }\textbf {\bibinfo {volume} {83}},\ \bibinfo
  {pages} {1057} (\bibinfo {year} {2011})}\BibitemShut {NoStop}%
\bibitem [{\citenamefont {He}\ \emph {et~al.}(2019)\citenamefont {He},
  \citenamefont {Sun},\ and\ \citenamefont {He}}]{He2019}%
  \BibitemOpen
  \bibfield  {author} {\bibinfo {author} {\bibfnamefont {M.}~\bibnamefont
  {He}}, \bibinfo {author} {\bibfnamefont {H.}~\bibnamefont {Sun}}, \ and\
  \bibinfo {author} {\bibfnamefont {Q.~L.}\ \bibnamefont {He}},\ }\href
  {\doibase 10.1007/s11467-019-0893-4} {\bibfield  {journal} {\bibinfo
  {journal} {Frontiers of Physics}\ }\textbf {\bibinfo {volume} {14}} (\bibinfo
  {year} {2019}),\ 10.1007/s11467-019-0893-4}\BibitemShut {NoStop}%
\bibitem [{\citenamefont {Zhou}\ \emph {et~al.}(2014)\citenamefont {Zhou},
  \citenamefont {Ming}, \citenamefont {Liu}, \citenamefont {Wang},
  \citenamefont {Li},\ and\ \citenamefont {Liu}}]{Zhou2014}%
  \BibitemOpen
  \bibfield  {author} {\bibinfo {author} {\bibfnamefont {M.}~\bibnamefont
  {Zhou}}, \bibinfo {author} {\bibfnamefont {W.}~\bibnamefont {Ming}}, \bibinfo
  {author} {\bibfnamefont {Z.}~\bibnamefont {Liu}}, \bibinfo {author}
  {\bibfnamefont {Z.}~\bibnamefont {Wang}}, \bibinfo {author} {\bibfnamefont
  {P.}~\bibnamefont {Li}}, \ and\ \bibinfo {author} {\bibfnamefont
  {F.}~\bibnamefont {Liu}},\ }\href {\doibase 10.1073/pnas.1409701111}
  {\bibfield  {journal} {\bibinfo  {journal} {Proceedings of the National
  Academy of Sciences}\ }\textbf {\bibinfo {volume} {111}},\ \bibinfo {pages}
  {14378} (\bibinfo {year} {2014})}\BibitemShut {NoStop}%
\bibitem [{\citenamefont {Linder}\ and\ \citenamefont
  {Robinson}(2015)}]{Linder2015}%
  \BibitemOpen
  \bibfield  {author} {\bibinfo {author} {\bibfnamefont {J.}~\bibnamefont
  {Linder}}\ and\ \bibinfo {author} {\bibfnamefont {J.~W.~A.}\ \bibnamefont
  {Robinson}},\ }\href {\doibase 10.1038/nphys3242} {\bibfield  {journal}
  {\bibinfo  {journal} {Nature Physics}\ }\textbf {\bibinfo {volume} {11}},\
  \bibinfo {pages} {307} (\bibinfo {year} {2015})}\BibitemShut {NoStop}%
\bibitem [{\citenamefont {Breunig}\ \emph {et~al.}(2018)\citenamefont
  {Breunig}, \citenamefont {Burset},\ and\ \citenamefont
  {Trauzettel}}]{Breunig2018}%
  \BibitemOpen
  \bibfield  {author} {\bibinfo {author} {\bibfnamefont {D.}~\bibnamefont
  {Breunig}}, \bibinfo {author} {\bibfnamefont {P.}~\bibnamefont {Burset}}, \
  and\ \bibinfo {author} {\bibfnamefont {B.}~\bibnamefont {Trauzettel}},\
  }\href {\doibase 10.1103/PhysRevLett.120.037701} {\bibfield  {journal}
  {\bibinfo  {journal} {Phys. Rev. Lett.}\ }\textbf {\bibinfo {volume} {120}},\
  \bibinfo {pages} {037701} (\bibinfo {year} {2018})}\BibitemShut {NoStop}%
\bibitem [{\citenamefont {Fu}\ and\ \citenamefont {Kane}(2008)}]{Fu2008}%
  \BibitemOpen
  \bibfield  {author} {\bibinfo {author} {\bibfnamefont {L.}~\bibnamefont
  {Fu}}\ and\ \bibinfo {author} {\bibfnamefont {C.~L.}\ \bibnamefont {Kane}},\
  }\href {\doibase 10.1103/PhysRevLett.100.096407} {\bibfield  {journal}
  {\bibinfo  {journal} {Phys. Rev. Lett.}\ }\textbf {\bibinfo {volume} {100}},\
  \bibinfo {pages} {096407} (\bibinfo {year} {2008})}\BibitemShut {NoStop}%
\bibitem [{\citenamefont {Alicea}(2012)}]{alicea12}%
  \BibitemOpen
  \bibfield  {author} {\bibinfo {author} {\bibfnamefont {J.}~\bibnamefont
  {Alicea}},\ }\href {http://arxiv.org/abs/1202.1293} {\bibfield  {journal}
  {\bibinfo  {journal} {Rep. Prog. Phys.}\ }\textbf {\bibinfo {volume} {75}},\
  \bibinfo {pages} {076501} (\bibinfo {year} {2012})}\BibitemShut {NoStop}%
\bibitem [{\citenamefont {Sato}\ and\ \citenamefont {Ando}(2017)}]{Sato2017}%
  \BibitemOpen
  \bibfield  {author} {\bibinfo {author} {\bibfnamefont {M.}~\bibnamefont
  {Sato}}\ and\ \bibinfo {author} {\bibfnamefont {Y.}~\bibnamefont {Ando}},\
  }\href {\doibase 10.1088/1361-6633/aa6ac7} {\bibfield  {journal} {\bibinfo
  {journal} {Reports on Progress in Physics}\ }\textbf {\bibinfo {volume}
  {80}},\ \bibinfo {pages} {076501} (\bibinfo {year} {2017})}\BibitemShut
  {NoStop}%
\bibitem [{\citenamefont {Lutchyn}\ \emph {et~al.}(2010)\citenamefont
  {Lutchyn}, \citenamefont {Sau},\ and\ \citenamefont {Das~Sarma}}]{Lutchyn10}%
  \BibitemOpen
  \bibfield  {author} {\bibinfo {author} {\bibfnamefont {R.~M.}\ \bibnamefont
  {Lutchyn}}, \bibinfo {author} {\bibfnamefont {J.~D.}\ \bibnamefont {Sau}}, \
  and\ \bibinfo {author} {\bibfnamefont {S.}~\bibnamefont {Das~Sarma}},\ }\href
  {\doibase 10.1103/PhysRevLett.105.077001} {\bibfield  {journal} {\bibinfo
  {journal} {Phys. Rev. Lett.}\ }\textbf {\bibinfo {volume} {105}},\ \bibinfo
  {pages} {077001} (\bibinfo {year} {2010})}\BibitemShut {NoStop}%
\bibitem [{\citenamefont {Kitaev}(2001)}]{kitaev01}%
  \BibitemOpen
  \bibfield  {author} {\bibinfo {author} {\bibfnamefont {A.~Y.}\ \bibnamefont
  {Kitaev}},\ }\href {\doibase 10.1070/1063-7869/44/10S/S29} {\bibfield
  {journal} {\bibinfo  {journal} {Phys. Usp.}\ }\textbf {\bibinfo {volume}
  {44}},\ \bibinfo {pages} {131} (\bibinfo {year} {2001})}\BibitemShut
  {NoStop}%
\bibitem [{\citenamefont {Nayak}\ \emph {et~al.}(2008)\citenamefont {Nayak},
  \citenamefont {Simon}, \citenamefont {Stern}, \citenamefont {Freedman},\ and\
  \citenamefont {Das~Sarma}}]{nayak08}%
  \BibitemOpen
  \bibfield  {author} {\bibinfo {author} {\bibfnamefont {C.}~\bibnamefont
  {Nayak}}, \bibinfo {author} {\bibfnamefont {S.~H.}\ \bibnamefont {Simon}},
  \bibinfo {author} {\bibfnamefont {A.}~\bibnamefont {Stern}}, \bibinfo
  {author} {\bibfnamefont {M.}~\bibnamefont {Freedman}}, \ and\ \bibinfo
  {author} {\bibfnamefont {S.}~\bibnamefont {Das~Sarma}},\ }\href {\doibase
  10.1103/RevModPhys.80.1083} {\bibfield  {journal} {\bibinfo  {journal} {Rev.
  Mod. Phys.}\ }\textbf {\bibinfo {volume} {80}},\ \bibinfo {pages} {1083}
  (\bibinfo {year} {2008})}\BibitemShut {NoStop}%
\bibitem [{\citenamefont {Aasen}\ \emph {et~al.}(2016)\citenamefont {Aasen},
  \citenamefont {Hell}, \citenamefont {Mishmash}, \citenamefont {Higginbotham},
  \citenamefont {Danon}, \citenamefont {Leijnse}, \citenamefont {Jespersen},
  \citenamefont {Folk}, \citenamefont {Marcus}, \citenamefont {Flensberg},\
  and\ \citenamefont {Alicea}}]{Aasen16}%
  \BibitemOpen
  \bibfield  {author} {\bibinfo {author} {\bibfnamefont {D.}~\bibnamefont
  {Aasen}}, \bibinfo {author} {\bibfnamefont {M.}~\bibnamefont {Hell}},
  \bibinfo {author} {\bibfnamefont {R.~V.}\ \bibnamefont {Mishmash}}, \bibinfo
  {author} {\bibfnamefont {A.}~\bibnamefont {Higginbotham}}, \bibinfo {author}
  {\bibfnamefont {J.}~\bibnamefont {Danon}}, \bibinfo {author} {\bibfnamefont
  {M.}~\bibnamefont {Leijnse}}, \bibinfo {author} {\bibfnamefont {T.~S.}\
  \bibnamefont {Jespersen}}, \bibinfo {author} {\bibfnamefont {J.~A.}\
  \bibnamefont {Folk}}, \bibinfo {author} {\bibfnamefont {C.~M.}\ \bibnamefont
  {Marcus}}, \bibinfo {author} {\bibfnamefont {K.}~\bibnamefont {Flensberg}}, \
  and\ \bibinfo {author} {\bibfnamefont {J.}~\bibnamefont {Alicea}},\ }\href
  {\doibase 10.1103/PhysRevX.6.031016} {\bibfield  {journal} {\bibinfo
  {journal} {Phys. Rev. X}\ }\textbf {\bibinfo {volume} {6}},\ \bibinfo {pages}
  {031016} (\bibinfo {year} {2016})}\BibitemShut {NoStop}%
\bibitem [{\citenamefont {Oreg}\ \emph {et~al.}(2010)\citenamefont {Oreg},
  \citenamefont {Refael},\ and\ \citenamefont {von Oppen}}]{Oreg10}%
  \BibitemOpen
  \bibfield  {author} {\bibinfo {author} {\bibfnamefont {Y.}~\bibnamefont
  {Oreg}}, \bibinfo {author} {\bibfnamefont {G.}~\bibnamefont {Refael}}, \ and\
  \bibinfo {author} {\bibfnamefont {F.}~\bibnamefont {von Oppen}},\ }\href
  {\doibase 10.1103/PhysRevLett.105.177002} {\bibfield  {journal} {\bibinfo
  {journal} {Phys. Rev. Lett.}\ }\textbf {\bibinfo {volume} {105}},\ \bibinfo
  {pages} {177002} (\bibinfo {year} {2010})}\BibitemShut {NoStop}%
\bibitem [{\citenamefont {Sothmann}\ \emph {et~al.}()\citenamefont {Sothmann},
  \citenamefont {Giazotto},\ and\ \citenamefont {Hankiewicz}}]{Sothmann2017}%
  \BibitemOpen
  \bibfield  {author} {\bibinfo {author} {\bibfnamefont {B.}~\bibnamefont
  {Sothmann}}, \bibinfo {author} {\bibfnamefont {F.}~\bibnamefont {Giazotto}},
  \ and\ \bibinfo {author} {\bibfnamefont {E.~M.}\ \bibnamefont {Hankiewicz}},\
  }\href {\doibase 10.1088/1367-2630/aa60d4} {\bibfield  {journal} {\bibinfo
  {journal} {New J. Phys. 19, 023056 (2017)}\ }10.1088/1367-2630/aa60d4},\
  \Eprint {http://arxiv.org/abs/1610.06099v2} {1610.06099v2} \BibitemShut
  {NoStop}%
\bibitem [{\citenamefont {Scharf}\ \emph {et~al.}()\citenamefont {Scharf},
  \citenamefont {Braggio}, \citenamefont {Strambini}, \citenamefont
  {Giazotto},\ and\ \citenamefont {Hankiewicz}}]{Scharf2020}%
  \BibitemOpen
  \bibfield  {author} {\bibinfo {author} {\bibfnamefont {B.}~\bibnamefont
  {Scharf}}, \bibinfo {author} {\bibfnamefont {A.}~\bibnamefont {Braggio}},
  \bibinfo {author} {\bibfnamefont {E.}~\bibnamefont {Strambini}}, \bibinfo
  {author} {\bibfnamefont {F.}~\bibnamefont {Giazotto}}, \ and\ \bibinfo
  {author} {\bibfnamefont {E.~M.}\ \bibnamefont {Hankiewicz}},\ }\href
  {\doibase 10.1038/s42005-020-00463-6} {\bibfield  {journal} {\bibinfo
  {journal} {Communications Physics 3, 198 (2020)}\
  }10.1038/s42005-020-00463-6},\ \Eprint {http://arxiv.org/abs/2002.05492v2}
  {2002.05492v2} \BibitemShut {NoStop}%
\bibitem [{\citenamefont {Blasi}\ \emph {et~al.}(2021)\citenamefont {Blasi},
  \citenamefont {Taddei}, \citenamefont {Arrachea}, \citenamefont {Carrega},\
  and\ \citenamefont {Braggio}}]{Blasi2021}%
  \BibitemOpen
  \bibfield  {author} {\bibinfo {author} {\bibfnamefont {G.}~\bibnamefont
  {Blasi}}, \bibinfo {author} {\bibfnamefont {F.}~\bibnamefont {Taddei}},
  \bibinfo {author} {\bibfnamefont {L.}~\bibnamefont {Arrachea}}, \bibinfo
  {author} {\bibfnamefont {M.}~\bibnamefont {Carrega}}, \ and\ \bibinfo
  {author} {\bibfnamefont {A.}~\bibnamefont {Braggio}},\ }\href {\doibase
  10.1103/PhysRevB.103.235434} {\bibfield  {journal} {\bibinfo  {journal}
  {Phys. Rev. B}\ }\textbf {\bibinfo {volume} {103}},\ \bibinfo {pages}
  {235434} (\bibinfo {year} {2021})}\BibitemShut {NoStop}%
\bibitem [{\citenamefont {Keidel}\ \emph {et~al.}(2020)\citenamefont {Keidel},
  \citenamefont {Hwang}, \citenamefont {Trauzettel}, \citenamefont {Sothmann},\
  and\ \citenamefont {Burset}}]{Keidel2020}%
  \BibitemOpen
  \bibfield  {author} {\bibinfo {author} {\bibfnamefont {F.}~\bibnamefont
  {Keidel}}, \bibinfo {author} {\bibfnamefont {S.-Y.}\ \bibnamefont {Hwang}},
  \bibinfo {author} {\bibfnamefont {B.}~\bibnamefont {Trauzettel}}, \bibinfo
  {author} {\bibfnamefont {B.}~\bibnamefont {Sothmann}}, \ and\ \bibinfo
  {author} {\bibfnamefont {P.}~\bibnamefont {Burset}},\ }\href {\doibase
  10.1103/PhysRevResearch.2.022019} {\bibfield  {journal} {\bibinfo  {journal}
  {Phys. Rev. Research}\ }\textbf {\bibinfo {volume} {2}},\ \bibinfo {pages}
  {022019(R)} (\bibinfo {year} {2020})}\BibitemShut {NoStop}%
\bibitem [{\citenamefont {Gresta}\ \emph {et~al.}(2021)\citenamefont {Gresta},
  \citenamefont {Blasi}, \citenamefont {Taddei}, \citenamefont {Carrega},
  \citenamefont {Braggio},\ and\ \citenamefont {Arrachea}}]{Gresta2021}%
  \BibitemOpen
  \bibfield  {author} {\bibinfo {author} {\bibfnamefont {D.}~\bibnamefont
  {Gresta}}, \bibinfo {author} {\bibfnamefont {G.}~\bibnamefont {Blasi}},
  \bibinfo {author} {\bibfnamefont {F.}~\bibnamefont {Taddei}}, \bibinfo
  {author} {\bibfnamefont {M.}~\bibnamefont {Carrega}}, \bibinfo {author}
  {\bibfnamefont {A.}~\bibnamefont {Braggio}}, \ and\ \bibinfo {author}
  {\bibfnamefont {L.}~\bibnamefont {Arrachea}},\ }\href {\doibase
  10.1103/PhysRevB.103.075439} {\bibfield  {journal} {\bibinfo  {journal}
  {Phys. Rev. B}\ }\textbf {\bibinfo {volume} {103}},\ \bibinfo {pages}
  {075439} (\bibinfo {year} {2021})}\BibitemShut {NoStop}%
\bibitem [{\citenamefont {Reis}\ \emph {et~al.}(2017)\citenamefont {Reis},
  \citenamefont {Li}, \citenamefont {Dudy}, \citenamefont {Bauernfeind},
  \citenamefont {Glass}, \citenamefont {Hanke}, \citenamefont {Thomale},
  \citenamefont {Sch{\"a}fer},\ and\ \citenamefont {Claessen}}]{Reis17}%
  \BibitemOpen
  \bibfield  {author} {\bibinfo {author} {\bibfnamefont {F.}~\bibnamefont
  {Reis}}, \bibinfo {author} {\bibfnamefont {G.}~\bibnamefont {Li}}, \bibinfo
  {author} {\bibfnamefont {L.}~\bibnamefont {Dudy}}, \bibinfo {author}
  {\bibfnamefont {M.}~\bibnamefont {Bauernfeind}}, \bibinfo {author}
  {\bibfnamefont {S.}~\bibnamefont {Glass}}, \bibinfo {author} {\bibfnamefont
  {W.}~\bibnamefont {Hanke}}, \bibinfo {author} {\bibfnamefont
  {R.}~\bibnamefont {Thomale}}, \bibinfo {author} {\bibfnamefont
  {J.}~\bibnamefont {Sch{\"a}fer}}, \ and\ \bibinfo {author} {\bibfnamefont
  {R.}~\bibnamefont {Claessen}},\ }\href {\doibase 10.1126/science.aai8142}
  {\bibfield  {journal} {\bibinfo  {journal} {Science}\ } (\bibinfo {year}
  {2017}),\ 10.1126/science.aai8142}\BibitemShut {NoStop}%
\bibitem [{\citenamefont {Shi}\ \emph {et~al.}(2019)\citenamefont {Shi},
  \citenamefont {Kahn}, \citenamefont {Niu}, \citenamefont {Fei}, \citenamefont
  {Sun}, \citenamefont {Cai}, \citenamefont {Francisco}, \citenamefont {Wu},
  \citenamefont {Shen}, \citenamefont {Xu}, \citenamefont {Cobden},\ and\
  \citenamefont {Cui}}]{Shi2019}%
  \BibitemOpen
  \bibfield  {author} {\bibinfo {author} {\bibfnamefont {Y.}~\bibnamefont
  {Shi}}, \bibinfo {author} {\bibfnamefont {J.}~\bibnamefont {Kahn}}, \bibinfo
  {author} {\bibfnamefont {B.}~\bibnamefont {Niu}}, \bibinfo {author}
  {\bibfnamefont {Z.}~\bibnamefont {Fei}}, \bibinfo {author} {\bibfnamefont
  {B.}~\bibnamefont {Sun}}, \bibinfo {author} {\bibfnamefont {X.}~\bibnamefont
  {Cai}}, \bibinfo {author} {\bibfnamefont {B.~A.}\ \bibnamefont {Francisco}},
  \bibinfo {author} {\bibfnamefont {D.}~\bibnamefont {Wu}}, \bibinfo {author}
  {\bibfnamefont {Z.-X.}\ \bibnamefont {Shen}}, \bibinfo {author}
  {\bibfnamefont {X.}~\bibnamefont {Xu}}, \bibinfo {author} {\bibfnamefont
  {D.~H.}\ \bibnamefont {Cobden}}, \ and\ \bibinfo {author} {\bibfnamefont
  {Y.-T.}\ \bibnamefont {Cui}},\ }\href {\doibase 10.1126/sciadv.aat8799}
  {\bibfield  {journal} {\bibinfo  {journal} {Science Advances}\ }\textbf
  {\bibinfo {volume} {5}} (\bibinfo {year} {2019}),\
  10.1126/sciadv.aat8799}\BibitemShut {NoStop}%
\bibitem [{\citenamefont {Wu}\ \emph {et~al.}(2018)\citenamefont {Wu},
  \citenamefont {Fatemi}, \citenamefont {Gibson}, \citenamefont {Watanabe},
  \citenamefont {Taniguchi}, \citenamefont {Cava},\ and\ \citenamefont
  {Jarillo-Herrero}}]{Wu2018}%
  \BibitemOpen
  \bibfield  {author} {\bibinfo {author} {\bibfnamefont {S.}~\bibnamefont
  {Wu}}, \bibinfo {author} {\bibfnamefont {V.}~\bibnamefont {Fatemi}}, \bibinfo
  {author} {\bibfnamefont {Q.~D.}\ \bibnamefont {Gibson}}, \bibinfo {author}
  {\bibfnamefont {K.}~\bibnamefont {Watanabe}}, \bibinfo {author}
  {\bibfnamefont {T.}~\bibnamefont {Taniguchi}}, \bibinfo {author}
  {\bibfnamefont {R.~J.}\ \bibnamefont {Cava}}, \ and\ \bibinfo {author}
  {\bibfnamefont {P.}~\bibnamefont {Jarillo-Herrero}},\ }\href {\doibase
  10.1126/science.aan6003} {\bibfield  {journal} {\bibinfo  {journal}
  {Science}\ }\textbf {\bibinfo {volume} {359}},\ \bibinfo {pages} {76}
  (\bibinfo {year} {2018})}\BibitemShut {NoStop}%
\bibitem [{\citenamefont {Tang}\ \emph {et~al.}(2017)\citenamefont {Tang},
  \citenamefont {Zhang}, \citenamefont {Wong}, \citenamefont {Pedramrazi},
  \citenamefont {Tsai}, \citenamefont {Jia}, \citenamefont {Moritz},
  \citenamefont {Claassen}, \citenamefont {Ryu}, \citenamefont {Kahn},
  \citenamefont {Jiang}, \citenamefont {Yan}, \citenamefont {Hashimoto},
  \citenamefont {Lu}, \citenamefont {Moore}, \citenamefont {Hwang},
  \citenamefont {Hwang}, \citenamefont {Hussain}, \citenamefont {Chen},
  \citenamefont {Ugeda}, \citenamefont {Liu}, \citenamefont {Xie},
  \citenamefont {Devereaux}, \citenamefont {Crommie}, \citenamefont {Mo},\ and\
  \citenamefont {Shen}}]{Tang2017}%
  \BibitemOpen
  \bibfield  {author} {\bibinfo {author} {\bibfnamefont {S.}~\bibnamefont
  {Tang}}, \bibinfo {author} {\bibfnamefont {C.}~\bibnamefont {Zhang}},
  \bibinfo {author} {\bibfnamefont {D.}~\bibnamefont {Wong}}, \bibinfo {author}
  {\bibfnamefont {Z.}~\bibnamefont {Pedramrazi}}, \bibinfo {author}
  {\bibfnamefont {H.-Z.}\ \bibnamefont {Tsai}}, \bibinfo {author}
  {\bibfnamefont {C.}~\bibnamefont {Jia}}, \bibinfo {author} {\bibfnamefont
  {B.}~\bibnamefont {Moritz}}, \bibinfo {author} {\bibfnamefont
  {M.}~\bibnamefont {Claassen}}, \bibinfo {author} {\bibfnamefont
  {H.}~\bibnamefont {Ryu}}, \bibinfo {author} {\bibfnamefont {S.}~\bibnamefont
  {Kahn}}, \bibinfo {author} {\bibfnamefont {J.}~\bibnamefont {Jiang}},
  \bibinfo {author} {\bibfnamefont {H.}~\bibnamefont {Yan}}, \bibinfo {author}
  {\bibfnamefont {M.}~\bibnamefont {Hashimoto}}, \bibinfo {author}
  {\bibfnamefont {D.}~\bibnamefont {Lu}}, \bibinfo {author} {\bibfnamefont
  {R.~G.}\ \bibnamefont {Moore}}, \bibinfo {author} {\bibfnamefont {C.-C.}\
  \bibnamefont {Hwang}}, \bibinfo {author} {\bibfnamefont {C.}~\bibnamefont
  {Hwang}}, \bibinfo {author} {\bibfnamefont {Z.}~\bibnamefont {Hussain}},
  \bibinfo {author} {\bibfnamefont {Y.}~\bibnamefont {Chen}}, \bibinfo {author}
  {\bibfnamefont {M.~M.}\ \bibnamefont {Ugeda}}, \bibinfo {author}
  {\bibfnamefont {Z.}~\bibnamefont {Liu}}, \bibinfo {author} {\bibfnamefont
  {X.}~\bibnamefont {Xie}}, \bibinfo {author} {\bibfnamefont {T.~P.}\
  \bibnamefont {Devereaux}}, \bibinfo {author} {\bibfnamefont {M.~F.}\
  \bibnamefont {Crommie}}, \bibinfo {author} {\bibfnamefont {S.-K.}\
  \bibnamefont {Mo}}, \ and\ \bibinfo {author} {\bibfnamefont {Z.-X.}\
  \bibnamefont {Shen}},\ }\href {\doibase 10.1038/nphys4174} {\bibfield
  {journal} {\bibinfo  {journal} {Nature Physics}\ }\textbf {\bibinfo {volume}
  {13}},\ \bibinfo {pages} {683} (\bibinfo {year} {2017})}\BibitemShut
  {NoStop}%
\bibitem [{\citenamefont {Li}\ \emph {et~al.}(2018{\natexlab{a}})\citenamefont
  {Li}, \citenamefont {Hanke}, \citenamefont {Hankiewicz}, \citenamefont
  {Reis}, \citenamefont {Sch\"afer}, \citenamefont {Claessen}, \citenamefont
  {Wu},\ and\ \citenamefont {Thomale}}]{Li2018}%
  \BibitemOpen
  \bibfield  {author} {\bibinfo {author} {\bibfnamefont {G.}~\bibnamefont
  {Li}}, \bibinfo {author} {\bibfnamefont {W.}~\bibnamefont {Hanke}}, \bibinfo
  {author} {\bibfnamefont {E.~M.}\ \bibnamefont {Hankiewicz}}, \bibinfo
  {author} {\bibfnamefont {F.}~\bibnamefont {Reis}}, \bibinfo {author}
  {\bibfnamefont {J.}~\bibnamefont {Sch\"afer}}, \bibinfo {author}
  {\bibfnamefont {R.}~\bibnamefont {Claessen}}, \bibinfo {author}
  {\bibfnamefont {C.}~\bibnamefont {Wu}}, \ and\ \bibinfo {author}
  {\bibfnamefont {R.}~\bibnamefont {Thomale}},\ }\href {\doibase
  10.1103/PhysRevB.98.165146} {\bibfield  {journal} {\bibinfo  {journal} {Phys.
  Rev. B}\ }\textbf {\bibinfo {volume} {98}},\ \bibinfo {pages} {165146}
  (\bibinfo {year} {2018}{\natexlab{a}})}\BibitemShut {NoStop}%
\bibitem [{\citenamefont {Liu}\ \emph {et~al.}(2020)\citenamefont {Liu},
  \citenamefont {Culcer}, \citenamefont {Wang}, \citenamefont {Edmonds},\ and\
  \citenamefont {Fuhrer}}]{Liu2020}%
  \BibitemOpen
  \bibfield  {author} {\bibinfo {author} {\bibfnamefont {C.}~\bibnamefont
  {Liu}}, \bibinfo {author} {\bibfnamefont {D.}~\bibnamefont {Culcer}},
  \bibinfo {author} {\bibfnamefont {Z.}~\bibnamefont {Wang}}, \bibinfo {author}
  {\bibfnamefont {M.~T.}\ \bibnamefont {Edmonds}}, \ and\ \bibinfo {author}
  {\bibfnamefont {M.~S.}\ \bibnamefont {Fuhrer}},\ }\href {\doibase
  10.1021/acs.nanolett.0c01649} {\bibfield  {journal} {\bibinfo  {journal}
  {Nano Letters}\ }\textbf {\bibinfo {volume} {20}},\ \bibinfo {pages} {6306}
  (\bibinfo {year} {2020})}\BibitemShut {NoStop}%
\bibitem [{\citenamefont {Knez}\ \emph {et~al.}(2011)\citenamefont {Knez},
  \citenamefont {Du},\ and\ \citenamefont {Sullivan}}]{Knez11}%
  \BibitemOpen
  \bibfield  {author} {\bibinfo {author} {\bibfnamefont {I.}~\bibnamefont
  {Knez}}, \bibinfo {author} {\bibfnamefont {R.-R.}\ \bibnamefont {Du}}, \ and\
  \bibinfo {author} {\bibfnamefont {G.}~\bibnamefont {Sullivan}},\ }\href
  {\doibase 10.1103/PhysRevLett.107.136603} {\bibfield  {journal} {\bibinfo
  {journal} {Phys. Rev. Lett.}\ }\textbf {\bibinfo {volume} {107}},\ \bibinfo
  {pages} {136603} (\bibinfo {year} {2011})}\BibitemShut {NoStop}%
\bibitem [{\citenamefont {Liu}\ \emph {et~al.}(2008)\citenamefont {Liu},
  \citenamefont {Hughes}, \citenamefont {Qi}, \citenamefont {Wang},\ and\
  \citenamefont {Zhang}}]{Liu08}%
  \BibitemOpen
  \bibfield  {author} {\bibinfo {author} {\bibfnamefont {C.}~\bibnamefont
  {Liu}}, \bibinfo {author} {\bibfnamefont {T.~L.}\ \bibnamefont {Hughes}},
  \bibinfo {author} {\bibfnamefont {X.-L.}\ \bibnamefont {Qi}}, \bibinfo
  {author} {\bibfnamefont {K.}~\bibnamefont {Wang}}, \ and\ \bibinfo {author}
  {\bibfnamefont {S.-C.}\ \bibnamefont {Zhang}},\ }\href {\doibase
  10.1103/PhysRevLett.100.236601} {\bibfield  {journal} {\bibinfo  {journal}
  {Phys. Rev. Lett.}\ }\textbf {\bibinfo {volume} {100}},\ \bibinfo {pages}
  {236601} (\bibinfo {year} {2008})}\BibitemShut {NoStop}%
\bibitem [{\citenamefont {Pribiag}\ \emph {et~al.}(2015)\citenamefont
  {Pribiag}, \citenamefont {Beukman}, \citenamefont {Qu}, \citenamefont
  {Cassidy}, \citenamefont {Charpentier}, \citenamefont {Wegscheider},\ and\
  \citenamefont {Kouwenhoven}}]{Pribiag2015}%
  \BibitemOpen
  \bibfield  {author} {\bibinfo {author} {\bibfnamefont {V.~S.}\ \bibnamefont
  {Pribiag}}, \bibinfo {author} {\bibfnamefont {A.~J.~A.}\ \bibnamefont
  {Beukman}}, \bibinfo {author} {\bibfnamefont {F.}~\bibnamefont {Qu}},
  \bibinfo {author} {\bibfnamefont {M.~C.}\ \bibnamefont {Cassidy}}, \bibinfo
  {author} {\bibfnamefont {C.}~\bibnamefont {Charpentier}}, \bibinfo {author}
  {\bibfnamefont {W.}~\bibnamefont {Wegscheider}}, \ and\ \bibinfo {author}
  {\bibfnamefont {L.~P.}\ \bibnamefont {Kouwenhoven}},\ }\href {\doibase
  10.1038/nnano.2015.86} {\bibfield  {journal} {\bibinfo  {journal} {Nature
  Nanotechnology}\ }\textbf {\bibinfo {volume} {10}},\ \bibinfo {pages} {593}
  (\bibinfo {year} {2015})}\BibitemShut {NoStop}%
\bibitem [{\citenamefont {Roth}\ \emph {et~al.}(2009)\citenamefont {Roth},
  \citenamefont {Brune}, \citenamefont {Buhmann}, \citenamefont {Molenkamp},
  \citenamefont {Maciejko}, \citenamefont {Qi},\ and\ \citenamefont
  {Zhang}}]{Roth2009}%
  \BibitemOpen
  \bibfield  {author} {\bibinfo {author} {\bibfnamefont {A.}~\bibnamefont
  {Roth}}, \bibinfo {author} {\bibfnamefont {C.}~\bibnamefont {Brune}},
  \bibinfo {author} {\bibfnamefont {H.}~\bibnamefont {Buhmann}}, \bibinfo
  {author} {\bibfnamefont {L.~W.}\ \bibnamefont {Molenkamp}}, \bibinfo {author}
  {\bibfnamefont {J.}~\bibnamefont {Maciejko}}, \bibinfo {author}
  {\bibfnamefont {X.-L.}\ \bibnamefont {Qi}}, \ and\ \bibinfo {author}
  {\bibfnamefont {S.-C.}\ \bibnamefont {Zhang}},\ }\href {\doibase
  10.1126/science.1174736} {\bibfield  {journal} {\bibinfo  {journal}
  {Science}\ }\textbf {\bibinfo {volume} {325}},\ \bibinfo {pages} {294}
  (\bibinfo {year} {2009})}\BibitemShut {NoStop}%
\bibitem [{\citenamefont {Shamim}\ \emph {et~al.}(2021)\citenamefont {Shamim},
  \citenamefont {Beugeling}, \citenamefont {Shekhar}, \citenamefont {Bendias},
  \citenamefont {Lunczer}, \citenamefont {Kleinlein}, \citenamefont {Buhmann},\
  and\ \citenamefont {Molenkamp}}]{Shamim2021}%
  \BibitemOpen
  \bibfield  {author} {\bibinfo {author} {\bibfnamefont {S.}~\bibnamefont
  {Shamim}}, \bibinfo {author} {\bibfnamefont {W.}~\bibnamefont {Beugeling}},
  \bibinfo {author} {\bibfnamefont {P.}~\bibnamefont {Shekhar}}, \bibinfo
  {author} {\bibfnamefont {K.}~\bibnamefont {Bendias}}, \bibinfo {author}
  {\bibfnamefont {L.}~\bibnamefont {Lunczer}}, \bibinfo {author} {\bibfnamefont
  {J.}~\bibnamefont {Kleinlein}}, \bibinfo {author} {\bibfnamefont
  {H.}~\bibnamefont {Buhmann}}, \ and\ \bibinfo {author} {\bibfnamefont
  {L.~W.}\ \bibnamefont {Molenkamp}},\ }\href {\doibase
  10.1038/s41467-021-23262-1} {\bibfield  {journal} {\bibinfo  {journal}
  {Nature Communications}\ }\textbf {\bibinfo {volume} {12}} (\bibinfo {year}
  {2021}),\ 10.1038/s41467-021-23262-1}\BibitemShut {NoStop}%
\bibitem [{\citenamefont {Br\"{u}ne}\ \emph {et~al.}(2012)\citenamefont
  {Br\"{u}ne}, \citenamefont {Roth}, \citenamefont {Buhmann}, \citenamefont
  {Hankiewicz}, \citenamefont {Molenkamp}, \citenamefont {Maciejko},
  \citenamefont {Qi},\ and\ \citenamefont {Zhang}}]{Bruene2012}%
  \BibitemOpen
  \bibfield  {author} {\bibinfo {author} {\bibfnamefont {C.}~\bibnamefont
  {Br\"{u}ne}}, \bibinfo {author} {\bibfnamefont {A.}~\bibnamefont {Roth}},
  \bibinfo {author} {\bibfnamefont {H.}~\bibnamefont {Buhmann}}, \bibinfo
  {author} {\bibfnamefont {E.~M.}\ \bibnamefont {Hankiewicz}}, \bibinfo
  {author} {\bibfnamefont {L.~W.}\ \bibnamefont {Molenkamp}}, \bibinfo {author}
  {\bibfnamefont {J.}~\bibnamefont {Maciejko}}, \bibinfo {author}
  {\bibfnamefont {X.-L.}\ \bibnamefont {Qi}}, \ and\ \bibinfo {author}
  {\bibfnamefont {S.-C.}\ \bibnamefont {Zhang}},\ }\href {\doibase
  10.1038/nphys2322} {\bibfield  {journal} {\bibinfo  {journal} {Nature
  Physics}\ }\textbf {\bibinfo {volume} {8}},\ \bibinfo {pages} {485} (\bibinfo
  {year} {2012})}\BibitemShut {NoStop}%
\bibitem [{\citenamefont {Hart}\ \emph {et~al.}(2014)\citenamefont {Hart},
  \citenamefont {Ren}, \citenamefont {Wagner}, \citenamefont {Leubner},
  \citenamefont {Mühlbauer}, \citenamefont {Brüne}, \citenamefont {Buhmann},
  \citenamefont {Molenkamp},\ and\ \citenamefont {Yacoby}}]{Hart2014}%
  \BibitemOpen
  \bibfield  {author} {\bibinfo {author} {\bibfnamefont {S.}~\bibnamefont
  {Hart}}, \bibinfo {author} {\bibfnamefont {H.}~\bibnamefont {Ren}}, \bibinfo
  {author} {\bibfnamefont {T.}~\bibnamefont {Wagner}}, \bibinfo {author}
  {\bibfnamefont {P.}~\bibnamefont {Leubner}}, \bibinfo {author} {\bibfnamefont
  {M.}~\bibnamefont {Mühlbauer}}, \bibinfo {author} {\bibfnamefont
  {C.}~\bibnamefont {Brüne}}, \bibinfo {author} {\bibfnamefont
  {H.}~\bibnamefont {Buhmann}}, \bibinfo {author} {\bibfnamefont {L.~W.}\
  \bibnamefont {Molenkamp}}, \ and\ \bibinfo {author} {\bibfnamefont
  {A.}~\bibnamefont {Yacoby}},\ }\href {\doibase 10.1038/nphys3036} {\bibfield
  {journal} {\bibinfo  {journal} {Nature Physics}\ }\textbf {\bibinfo {volume}
  {10}},\ \bibinfo {pages} {638} (\bibinfo {year} {2014})}\BibitemShut
  {NoStop}%
\bibitem [{\citenamefont {Wiedenmann}\ \emph {et~al.}(2016)\citenamefont
  {Wiedenmann}, \citenamefont {Bocquillon}, \citenamefont {Deacon},
  \citenamefont {Hartinger}, \citenamefont {Herrmann}, \citenamefont
  {Klapwijk}, \citenamefont {Maier}, \citenamefont {Ames}, \citenamefont
  {Br\"{u}ne}, \citenamefont {Gould}, \citenamefont {Oiwa}, \citenamefont
  {Ishibashi}, \citenamefont {Tarucha}, \citenamefont {Buhmann},\ and\
  \citenamefont {Molenkamp}}]{Wiedenmann2016}%
  \BibitemOpen
  \bibfield  {author} {\bibinfo {author} {\bibfnamefont {J.}~\bibnamefont
  {Wiedenmann}}, \bibinfo {author} {\bibfnamefont {E.}~\bibnamefont
  {Bocquillon}}, \bibinfo {author} {\bibfnamefont {R.~S.}\ \bibnamefont
  {Deacon}}, \bibinfo {author} {\bibfnamefont {S.}~\bibnamefont {Hartinger}},
  \bibinfo {author} {\bibfnamefont {O.}~\bibnamefont {Herrmann}}, \bibinfo
  {author} {\bibfnamefont {T.~M.}\ \bibnamefont {Klapwijk}}, \bibinfo {author}
  {\bibfnamefont {L.}~\bibnamefont {Maier}}, \bibinfo {author} {\bibfnamefont
  {C.}~\bibnamefont {Ames}}, \bibinfo {author} {\bibfnamefont {C.}~\bibnamefont
  {Br\"{u}ne}}, \bibinfo {author} {\bibfnamefont {C.}~\bibnamefont {Gould}},
  \bibinfo {author} {\bibfnamefont {A.}~\bibnamefont {Oiwa}}, \bibinfo {author}
  {\bibfnamefont {K.}~\bibnamefont {Ishibashi}}, \bibinfo {author}
  {\bibfnamefont {S.}~\bibnamefont {Tarucha}}, \bibinfo {author} {\bibfnamefont
  {H.}~\bibnamefont {Buhmann}}, \ and\ \bibinfo {author} {\bibfnamefont
  {L.~W.}\ \bibnamefont {Molenkamp}},\ }\href {\doibase 10.1038/ncomms10303}
  {\bibfield  {journal} {\bibinfo  {journal} {Nature Communications}\ }\textbf
  {\bibinfo {volume} {7}} (\bibinfo {year} {2016}),\
  10.1038/ncomms10303}\BibitemShut {NoStop}%
\bibitem [{\citenamefont {Bocquillon}\ \emph {et~al.}(2016)\citenamefont
  {Bocquillon}, \citenamefont {Deacon}, \citenamefont {Wiedenmann},
  \citenamefont {Leubner}, \citenamefont {Klapwijk}, \citenamefont {Br\"{u}ne},
  \citenamefont {Ishibashi}, \citenamefont {Buhmann},\ and\ \citenamefont
  {Molenkamp}}]{Bocquillon2016}%
  \BibitemOpen
  \bibfield  {author} {\bibinfo {author} {\bibfnamefont {E.}~\bibnamefont
  {Bocquillon}}, \bibinfo {author} {\bibfnamefont {R.~S.}\ \bibnamefont
  {Deacon}}, \bibinfo {author} {\bibfnamefont {J.}~\bibnamefont {Wiedenmann}},
  \bibinfo {author} {\bibfnamefont {P.}~\bibnamefont {Leubner}}, \bibinfo
  {author} {\bibfnamefont {T.~M.}\ \bibnamefont {Klapwijk}}, \bibinfo {author}
  {\bibfnamefont {C.}~\bibnamefont {Br\"{u}ne}}, \bibinfo {author}
  {\bibfnamefont {K.}~\bibnamefont {Ishibashi}}, \bibinfo {author}
  {\bibfnamefont {H.}~\bibnamefont {Buhmann}}, \ and\ \bibinfo {author}
  {\bibfnamefont {L.~W.}\ \bibnamefont {Molenkamp}},\ }\href {\doibase
  10.1038/nnano.2016.159} {\bibfield  {journal} {\bibinfo  {journal} {Nature
  Nanotechnology}\ }\textbf {\bibinfo {volume} {12}},\ \bibinfo {pages} {137}
  (\bibinfo {year} {2016})}\BibitemShut {NoStop}%
\bibitem [{\citenamefont {Bocquillon}\ \emph {et~al.}(2018)\citenamefont
  {Bocquillon}, \citenamefont {Wiedenmann}, \citenamefont {Deacon},
  \citenamefont {Klapwijk}, \citenamefont {Buhmann},\ and\ \citenamefont
  {Molenkamp}}]{Bocquillon2018}%
  \BibitemOpen
  \bibfield  {author} {\bibinfo {author} {\bibfnamefont {E.}~\bibnamefont
  {Bocquillon}}, \bibinfo {author} {\bibfnamefont {J.}~\bibnamefont
  {Wiedenmann}}, \bibinfo {author} {\bibfnamefont {R.~S.}\ \bibnamefont
  {Deacon}}, \bibinfo {author} {\bibfnamefont {T.~M.}\ \bibnamefont
  {Klapwijk}}, \bibinfo {author} {\bibfnamefont {H.}~\bibnamefont {Buhmann}}, \
  and\ \bibinfo {author} {\bibfnamefont {L.~W.}\ \bibnamefont {Molenkamp}},\
  }in\ \href {\doibase 10.1007/978-3-319-76388-0_5} {\emph {\bibinfo
  {booktitle} {Topological Matter}}}\ (\bibinfo  {publisher} {Springer
  International Publishing},\ \bibinfo {year} {2018})\ pp.\ \bibinfo {pages}
  {115--148}\BibitemShut {NoStop}%
\bibitem [{\citenamefont {Tkachov}\ and\ \citenamefont
  {Hankiewicz}(2013)}]{Tkachov2013}%
  \BibitemOpen
  \bibfield  {author} {\bibinfo {author} {\bibfnamefont {G.}~\bibnamefont
  {Tkachov}}\ and\ \bibinfo {author} {\bibfnamefont {E.~M.}\ \bibnamefont
  {Hankiewicz}},\ }\href {\doibase 10.1103/physrevb.88.075401} {\bibfield
  {journal} {\bibinfo  {journal} {Phys. Rev. B}\ }\textbf {\bibinfo {volume}
  {88}},\ \bibinfo {pages} {075401} (\bibinfo {year} {2013})}\BibitemShut
  {NoStop}%
\bibitem [{\citenamefont {Beenakker}\ \emph {et~al.}(2013)\citenamefont
  {Beenakker}, \citenamefont {Pikulin}, \citenamefont {Hyart}, \citenamefont
  {Schomerus},\ and\ \citenamefont {Dahlhaus}}]{Beenakker2013}%
  \BibitemOpen
  \bibfield  {author} {\bibinfo {author} {\bibfnamefont {C.~W.~J.}\
  \bibnamefont {Beenakker}}, \bibinfo {author} {\bibfnamefont {D.~I.}\
  \bibnamefont {Pikulin}}, \bibinfo {author} {\bibfnamefont {T.}~\bibnamefont
  {Hyart}}, \bibinfo {author} {\bibfnamefont {H.}~\bibnamefont {Schomerus}}, \
  and\ \bibinfo {author} {\bibfnamefont {J.~P.}\ \bibnamefont {Dahlhaus}},\
  }\href {\doibase 10.1103/physrevlett.110.017003} {\bibfield  {journal}
  {\bibinfo  {journal} {Phys. Rev. Lett.}\ }\textbf {\bibinfo {volume} {110}},\
  \bibinfo {pages} {017003} (\bibinfo {year} {2013})}\BibitemShut {NoStop}%
\bibitem [{\citenamefont {Sothmann}\ and\ \citenamefont
  {Hankiewicz}(2016)}]{Sothmann2016}%
  \BibitemOpen
  \bibfield  {author} {\bibinfo {author} {\bibfnamefont {B.}~\bibnamefont
  {Sothmann}}\ and\ \bibinfo {author} {\bibfnamefont {E.~M.}\ \bibnamefont
  {Hankiewicz}},\ }\href {\doibase 10.1103/PhysRevB.94.081407} {\bibfield
  {journal} {\bibinfo  {journal} {Phys. Rev. B}\ }\textbf {\bibinfo {volume}
  {94}},\ \bibinfo {pages} {081407(R)} (\bibinfo {year} {2016})}\BibitemShut
  {NoStop}%
\bibitem [{\citenamefont {Tanaka}\ \emph {et~al.}(2009)\citenamefont {Tanaka},
  \citenamefont {Yokoyama}, \citenamefont {Balatsky},\ and\ \citenamefont
  {Nagaosa}}]{Tanaka2009}%
  \BibitemOpen
  \bibfield  {author} {\bibinfo {author} {\bibfnamefont {Y.}~\bibnamefont
  {Tanaka}}, \bibinfo {author} {\bibfnamefont {T.}~\bibnamefont {Yokoyama}},
  \bibinfo {author} {\bibfnamefont {A.~V.}\ \bibnamefont {Balatsky}}, \ and\
  \bibinfo {author} {\bibfnamefont {N.}~\bibnamefont {Nagaosa}},\ }\href
  {\doibase 10.1103/PhysRevB.79.060505} {\bibfield  {journal} {\bibinfo
  {journal} {Phys. Rev. B}\ }\textbf {\bibinfo {volume} {79}},\ \bibinfo
  {pages} {060505} (\bibinfo {year} {2009})}\BibitemShut {NoStop}%
\bibitem [{\citenamefont {Li}\ \emph {et~al.}(2016)\citenamefont {Li},
  \citenamefont {Pan}, \citenamefont {Bernevig},\ and\ \citenamefont
  {Lutchyn}}]{Li2016}%
  \BibitemOpen
  \bibfield  {author} {\bibinfo {author} {\bibfnamefont {J.}~\bibnamefont
  {Li}}, \bibinfo {author} {\bibfnamefont {W.}~\bibnamefont {Pan}}, \bibinfo
  {author} {\bibfnamefont {B.~A.}\ \bibnamefont {Bernevig}}, \ and\ \bibinfo
  {author} {\bibfnamefont {R.~M.}\ \bibnamefont {Lutchyn}},\ }\href {\doibase
  10.1103/PhysRevLett.117.046804} {\bibfield  {journal} {\bibinfo  {journal}
  {Phys. Rev. Lett.}\ }\textbf {\bibinfo {volume} {117}},\ \bibinfo {pages}
  {046804} (\bibinfo {year} {2016})}\BibitemShut {NoStop}%
\bibitem [{\citenamefont {Pikulin}\ \emph {et~al.}(2016)\citenamefont
  {Pikulin}, \citenamefont {Komijani},\ and\ \citenamefont
  {Affleck}}]{Pikulin2016}%
  \BibitemOpen
  \bibfield  {author} {\bibinfo {author} {\bibfnamefont {D.~I.}\ \bibnamefont
  {Pikulin}}, \bibinfo {author} {\bibfnamefont {Y.}~\bibnamefont {Komijani}}, \
  and\ \bibinfo {author} {\bibfnamefont {I.}~\bibnamefont {Affleck}},\ }\href
  {\doibase 10.1103/PhysRevB.93.205430} {\bibfield  {journal} {\bibinfo
  {journal} {Phys. Rev. B}\ }\textbf {\bibinfo {volume} {93}},\ \bibinfo
  {pages} {205430} (\bibinfo {year} {2016})},\ \Eprint
  {http://arxiv.org/abs/1511.06319v4} {1511.06319v4} \BibitemShut {NoStop}%
\bibitem [{\citenamefont {Laroche}\ \emph {et~al.}(2019)\citenamefont
  {Laroche}, \citenamefont {Bouman}, \citenamefont {van Woerkom}, \citenamefont
  {Proutski}, \citenamefont {Murthy}, \citenamefont {Pikulin}, \citenamefont
  {Nayak}, \citenamefont {van Gulik}, \citenamefont {Nyg{\aa}rd}, \citenamefont
  {Krogstrup}, \citenamefont {Kouwenhoven},\ and\ \citenamefont
  {Geresdi}}]{Laroche2019}%
  \BibitemOpen
  \bibfield  {author} {\bibinfo {author} {\bibfnamefont {D.}~\bibnamefont
  {Laroche}}, \bibinfo {author} {\bibfnamefont {D.}~\bibnamefont {Bouman}},
  \bibinfo {author} {\bibfnamefont {D.~J.}\ \bibnamefont {van Woerkom}},
  \bibinfo {author} {\bibfnamefont {A.}~\bibnamefont {Proutski}}, \bibinfo
  {author} {\bibfnamefont {C.}~\bibnamefont {Murthy}}, \bibinfo {author}
  {\bibfnamefont {D.~I.}\ \bibnamefont {Pikulin}}, \bibinfo {author}
  {\bibfnamefont {C.}~\bibnamefont {Nayak}}, \bibinfo {author} {\bibfnamefont
  {R.~J.~J.}\ \bibnamefont {van Gulik}}, \bibinfo {author} {\bibfnamefont
  {J.}~\bibnamefont {Nyg{\aa}rd}}, \bibinfo {author} {\bibfnamefont
  {P.}~\bibnamefont {Krogstrup}}, \bibinfo {author} {\bibfnamefont {L.~P.}\
  \bibnamefont {Kouwenhoven}}, \ and\ \bibinfo {author} {\bibfnamefont
  {A.}~\bibnamefont {Geresdi}},\ }\href {\doibase 10.1038/s41467-018-08161-2}
  {\bibfield  {journal} {\bibinfo  {journal} {Nat. Commun.}\ }\textbf {\bibinfo
  {volume} {10}},\ \bibinfo {pages} {245} (\bibinfo {year} {2019})}\BibitemShut
  {NoStop}%
\bibitem [{\citenamefont {Tkachov}\ \emph {et~al.}(2015)\citenamefont
  {Tkachov}, \citenamefont {Burset}, \citenamefont {Trauzettel},\ and\
  \citenamefont {Hankiewicz}}]{Tkachov2015}%
  \BibitemOpen
  \bibfield  {author} {\bibinfo {author} {\bibfnamefont {G.}~\bibnamefont
  {Tkachov}}, \bibinfo {author} {\bibfnamefont {P.}~\bibnamefont {Burset}},
  \bibinfo {author} {\bibfnamefont {B.}~\bibnamefont {Trauzettel}}, \ and\
  \bibinfo {author} {\bibfnamefont {E.~M.}\ \bibnamefont {Hankiewicz}},\ }\href
  {\doibase 10.1103/physrevb.92.045408} {\bibfield  {journal} {\bibinfo
  {journal} {Physical Review B}\ }\textbf {\bibinfo {volume} {92}},\ \bibinfo
  {pages} {045408} (\bibinfo {year} {2015})}\BibitemShut {NoStop}%
\bibitem [{\citenamefont {Baxevanis}\ \emph {et~al.}(2015)\citenamefont
  {Baxevanis}, \citenamefont {Ostroukh},\ and\ \citenamefont
  {Beenakker}}]{Baxevanis2015}%
  \BibitemOpen
  \bibfield  {author} {\bibinfo {author} {\bibfnamefont {B.}~\bibnamefont
  {Baxevanis}}, \bibinfo {author} {\bibfnamefont {V.~P.}\ \bibnamefont
  {Ostroukh}}, \ and\ \bibinfo {author} {\bibfnamefont {C.~W.~J.}\ \bibnamefont
  {Beenakker}},\ }\href {\doibase 10.1103/PhysRevB.91.041409} {\bibfield
  {journal} {\bibinfo  {journal} {Phys. Rev. B}\ }\textbf {\bibinfo {volume}
  {91}},\ \bibinfo {pages} {041409(R)} (\bibinfo {year} {2015})}\BibitemShut
  {NoStop}%
\bibitem [{\citenamefont {Dartiailh}\ \emph {et~al.}(2021)\citenamefont
  {Dartiailh}, \citenamefont {Cuozzo}, \citenamefont {Elfeky}, \citenamefont
  {Mayer}, \citenamefont {Yuan}, \citenamefont {Wickramasinghe}, \citenamefont
  {Rossi},\ and\ \citenamefont {Shabani}}]{Dartiailh2021}%
  \BibitemOpen
  \bibfield  {author} {\bibinfo {author} {\bibfnamefont {M.~C.}\ \bibnamefont
  {Dartiailh}}, \bibinfo {author} {\bibfnamefont {J.~J.}\ \bibnamefont
  {Cuozzo}}, \bibinfo {author} {\bibfnamefont {B.~H.}\ \bibnamefont {Elfeky}},
  \bibinfo {author} {\bibfnamefont {W.}~\bibnamefont {Mayer}}, \bibinfo
  {author} {\bibfnamefont {J.}~\bibnamefont {Yuan}}, \bibinfo {author}
  {\bibfnamefont {K.~S.}\ \bibnamefont {Wickramasinghe}}, \bibinfo {author}
  {\bibfnamefont {E.}~\bibnamefont {Rossi}}, \ and\ \bibinfo {author}
  {\bibfnamefont {J.}~\bibnamefont {Shabani}},\ }\href {\doibase
  10.1038/s41467-020-20382-y} {\bibfield  {journal} {\bibinfo  {journal} {Nat
  Commun}\ }\textbf {\bibinfo {volume} {12}},\ \bibinfo {pages} {78} (\bibinfo
  {year} {2021})}\BibitemShut {NoStop}%
\bibitem [{\citenamefont {de~Vries}\ \emph {et~al.}(2018)\citenamefont
  {de~Vries}, \citenamefont {Timmerman}, \citenamefont {Ostroukh},
  \citenamefont {van Veen}, \citenamefont {Beukman}, \citenamefont {Qu},
  \citenamefont {Wimmer}, \citenamefont {Nguyen}, \citenamefont {Kiselev},
  \citenamefont {Yi}, \citenamefont {Sokolich}, \citenamefont {Manfra},
  \citenamefont {Marcus},\ and\ \citenamefont {Kouwenhoven}}]{deVries2018}%
  \BibitemOpen
  \bibfield  {author} {\bibinfo {author} {\bibfnamefont {F.~K.}\ \bibnamefont
  {de~Vries}}, \bibinfo {author} {\bibfnamefont {T.}~\bibnamefont {Timmerman}},
  \bibinfo {author} {\bibfnamefont {V.~P.}\ \bibnamefont {Ostroukh}}, \bibinfo
  {author} {\bibfnamefont {J.}~\bibnamefont {van Veen}}, \bibinfo {author}
  {\bibfnamefont {A.~J.~A.}\ \bibnamefont {Beukman}}, \bibinfo {author}
  {\bibfnamefont {F.}~\bibnamefont {Qu}}, \bibinfo {author} {\bibfnamefont
  {M.}~\bibnamefont {Wimmer}}, \bibinfo {author} {\bibfnamefont {B.-M.}\
  \bibnamefont {Nguyen}}, \bibinfo {author} {\bibfnamefont {A.~A.}\
  \bibnamefont {Kiselev}}, \bibinfo {author} {\bibfnamefont {W.}~\bibnamefont
  {Yi}}, \bibinfo {author} {\bibfnamefont {M.}~\bibnamefont {Sokolich}},
  \bibinfo {author} {\bibfnamefont {M.~J.}\ \bibnamefont {Manfra}}, \bibinfo
  {author} {\bibfnamefont {C.~M.}\ \bibnamefont {Marcus}}, \ and\ \bibinfo
  {author} {\bibfnamefont {L.~P.}\ \bibnamefont {Kouwenhoven}},\ }\href
  {\doibase 10.1103/PhysRevLett.120.047702} {\bibfield  {journal} {\bibinfo
  {journal} {Phys. Rev. Lett.}\ }\textbf {\bibinfo {volume} {120}},\ \bibinfo
  {pages} {047702} (\bibinfo {year} {2018})}\BibitemShut {NoStop}%
\bibitem [{\citenamefont {de~Vries}\ \emph {et~al.}(2019)\citenamefont
  {de~Vries}, \citenamefont {Sol}, \citenamefont {Gazibegovic}, \citenamefont
  {Veld}, \citenamefont {Balk}, \citenamefont {Car}, \citenamefont {Bakkers},
  \citenamefont {Kouwenhoven},\ and\ \citenamefont {Shen}}]{deVries2019}%
  \BibitemOpen
  \bibfield  {author} {\bibinfo {author} {\bibfnamefont {F.~K.}\ \bibnamefont
  {de~Vries}}, \bibinfo {author} {\bibfnamefont {M.~L.}\ \bibnamefont {Sol}},
  \bibinfo {author} {\bibfnamefont {S.}~\bibnamefont {Gazibegovic}}, \bibinfo
  {author} {\bibfnamefont {R.~L. M.~o.}\ \bibnamefont {Veld}}, \bibinfo
  {author} {\bibfnamefont {S.~C.}\ \bibnamefont {Balk}}, \bibinfo {author}
  {\bibfnamefont {D.}~\bibnamefont {Car}}, \bibinfo {author} {\bibfnamefont
  {E.~P. A.~M.}\ \bibnamefont {Bakkers}}, \bibinfo {author} {\bibfnamefont
  {L.~P.}\ \bibnamefont {Kouwenhoven}}, \ and\ \bibinfo {author} {\bibfnamefont
  {J.}~\bibnamefont {Shen}},\ }\href {\doibase
  10.1103/PhysRevResearch.1.032031} {\bibfield  {journal} {\bibinfo  {journal}
  {Phys. Rev. Research}\ }\textbf {\bibinfo {volume} {1}},\ \bibinfo {pages}
  {032031(R)} (\bibinfo {year} {2019})}\BibitemShut {NoStop}%
\bibitem [{\citenamefont {Blasi}\ \emph {et~al.}(2020)\citenamefont {Blasi},
  \citenamefont {Taddei}, \citenamefont {Arrachea}, \citenamefont {Carrega},\
  and\ \citenamefont {Braggio}}]{Blasi2020}%
  \BibitemOpen
  \bibfield  {author} {\bibinfo {author} {\bibfnamefont {G.}~\bibnamefont
  {Blasi}}, \bibinfo {author} {\bibfnamefont {F.}~\bibnamefont {Taddei}},
  \bibinfo {author} {\bibfnamefont {L.}~\bibnamefont {Arrachea}}, \bibinfo
  {author} {\bibfnamefont {M.}~\bibnamefont {Carrega}}, \ and\ \bibinfo
  {author} {\bibfnamefont {A.}~\bibnamefont {Braggio}},\ }\href {\doibase
  10.1103/PhysRevLett.124.227701} {\bibfield  {journal} {\bibinfo  {journal}
  {Phys. Rev. Lett.}\ }\textbf {\bibinfo {volume} {124}},\ \bibinfo {pages}
  {227701} (\bibinfo {year} {2020})}\BibitemShut {NoStop}%
\bibitem [{\citenamefont {Haidekker~Galambos}\ \emph
  {et~al.}(2020)\citenamefont {Haidekker~Galambos}, \citenamefont {Hoffman},
  \citenamefont {Recher}, \citenamefont {Klinovaja},\ and\ \citenamefont
  {Loss}}]{Galambos2020}%
  \BibitemOpen
  \bibfield  {author} {\bibinfo {author} {\bibfnamefont {T.}~\bibnamefont
  {Haidekker~Galambos}}, \bibinfo {author} {\bibfnamefont {S.}~\bibnamefont
  {Hoffman}}, \bibinfo {author} {\bibfnamefont {P.}~\bibnamefont {Recher}},
  \bibinfo {author} {\bibfnamefont {J.}~\bibnamefont {Klinovaja}}, \ and\
  \bibinfo {author} {\bibfnamefont {D.}~\bibnamefont {Loss}},\ }\href {\doibase
  10.1103/PhysRevLett.125.157701} {\bibfield  {journal} {\bibinfo  {journal}
  {Phys. Rev. Lett.}\ }\textbf {\bibinfo {volume} {125}},\ \bibinfo {pages}
  {157701} (\bibinfo {year} {2020})}\BibitemShut {NoStop}%
\bibitem [{\citenamefont {Vigliotti}\ \emph {et~al.}(2022)\citenamefont
  {Vigliotti}, \citenamefont {Calzona}, \citenamefont {Trauzettel},
  \citenamefont {Sassetti},\ and\ \citenamefont {Ziani}}]{Vigliotti2022}%
  \BibitemOpen
  \bibfield  {author} {\bibinfo {author} {\bibfnamefont {L.}~\bibnamefont
  {Vigliotti}}, \bibinfo {author} {\bibfnamefont {A.}~\bibnamefont {Calzona}},
  \bibinfo {author} {\bibfnamefont {B.}~\bibnamefont {Trauzettel}}, \bibinfo
  {author} {\bibfnamefont {M.}~\bibnamefont {Sassetti}}, \ and\ \bibinfo
  {author} {\bibfnamefont {N.~T.}\ \bibnamefont {Ziani}},\ }\href@noop {}
  {\enquote {\bibinfo {title} {Anomalous flux periodicity in proximitised
  quantum spin hall constrictions},}\ } (\bibinfo {year} {2022}),\ \Eprint
  {http://arxiv.org/abs/2201.03259} {arXiv:2201.03259 [cond-mat.mes-hall]}
  \BibitemShut {NoStop}%
\bibitem [{\citenamefont {Danon}\ \emph {et~al.}(2020)\citenamefont {Danon},
  \citenamefont {Hellenes}, \citenamefont {Hansen}, \citenamefont {Casparis},
  \citenamefont {Higginbotham},\ and\ \citenamefont {Flensberg}}]{Danon2020}%
  \BibitemOpen
  \bibfield  {author} {\bibinfo {author} {\bibfnamefont {J.}~\bibnamefont
  {Danon}}, \bibinfo {author} {\bibfnamefont {A.~B.}\ \bibnamefont {Hellenes}},
  \bibinfo {author} {\bibfnamefont {E.~B.}\ \bibnamefont {Hansen}}, \bibinfo
  {author} {\bibfnamefont {L.}~\bibnamefont {Casparis}}, \bibinfo {author}
  {\bibfnamefont {A.~P.}\ \bibnamefont {Higginbotham}}, \ and\ \bibinfo
  {author} {\bibfnamefont {K.}~\bibnamefont {Flensberg}},\ }\href {\doibase
  10.1103/PhysRevLett.124.036801} {\bibfield  {journal} {\bibinfo  {journal}
  {Phys. Rev. Lett.}\ }\textbf {\bibinfo {volume} {124}},\ \bibinfo {pages}
  {036801} (\bibinfo {year} {2020})}\BibitemShut {NoStop}%
\bibitem [{\citenamefont {Nichele}\ \emph {et~al.}(2017)\citenamefont
  {Nichele}, \citenamefont {Drachmann}, \citenamefont {Whiticar}, \citenamefont
  {O'Farrell}, \citenamefont {Suominen}, \citenamefont {Fornieri},
  \citenamefont {Wang}, \citenamefont {Gardner}, \citenamefont {Thomas},
  \citenamefont {Hatke}, \citenamefont {Krogstrup}, \citenamefont {Manfra},
  \citenamefont {Flensberg},\ and\ \citenamefont {Marcus}}]{Nichele2017}%
  \BibitemOpen
  \bibfield  {author} {\bibinfo {author} {\bibfnamefont {F.}~\bibnamefont
  {Nichele}}, \bibinfo {author} {\bibfnamefont {A.~C.~C.}\ \bibnamefont
  {Drachmann}}, \bibinfo {author} {\bibfnamefont {A.~M.}\ \bibnamefont
  {Whiticar}}, \bibinfo {author} {\bibfnamefont {E.~C.~T.}\ \bibnamefont
  {O'Farrell}}, \bibinfo {author} {\bibfnamefont {H.~J.}\ \bibnamefont
  {Suominen}}, \bibinfo {author} {\bibfnamefont {A.}~\bibnamefont {Fornieri}},
  \bibinfo {author} {\bibfnamefont {T.}~\bibnamefont {Wang}}, \bibinfo {author}
  {\bibfnamefont {G.~C.}\ \bibnamefont {Gardner}}, \bibinfo {author}
  {\bibfnamefont {C.}~\bibnamefont {Thomas}}, \bibinfo {author} {\bibfnamefont
  {A.~T.}\ \bibnamefont {Hatke}}, \bibinfo {author} {\bibfnamefont
  {P.}~\bibnamefont {Krogstrup}}, \bibinfo {author} {\bibfnamefont {M.~J.}\
  \bibnamefont {Manfra}}, \bibinfo {author} {\bibfnamefont {K.}~\bibnamefont
  {Flensberg}}, \ and\ \bibinfo {author} {\bibfnamefont {C.~M.}\ \bibnamefont
  {Marcus}},\ }\href {\doibase 10.1103/PhysRevLett.119.136803} {\bibfield
  {journal} {\bibinfo  {journal} {Phys. Rev. Lett.}\ }\textbf {\bibinfo
  {volume} {119}},\ \bibinfo {pages} {136803} (\bibinfo {year}
  {2017})}\BibitemShut {NoStop}%
\bibitem [{\citenamefont {Schindele}\ \emph {et~al.}(2014)\citenamefont
  {Schindele}, \citenamefont {Baumgartner}, \citenamefont {Maurand},
  \citenamefont {Weiss},\ and\ \citenamefont
  {Sch\"onenberger}}]{Schindele2014}%
  \BibitemOpen
  \bibfield  {author} {\bibinfo {author} {\bibfnamefont {J.}~\bibnamefont
  {Schindele}}, \bibinfo {author} {\bibfnamefont {A.}~\bibnamefont
  {Baumgartner}}, \bibinfo {author} {\bibfnamefont {R.}~\bibnamefont
  {Maurand}}, \bibinfo {author} {\bibfnamefont {M.}~\bibnamefont {Weiss}}, \
  and\ \bibinfo {author} {\bibfnamefont {C.}~\bibnamefont {Sch\"onenberger}},\
  }\href {\doibase 10.1103/PhysRevB.89.045422} {\bibfield  {journal} {\bibinfo
  {journal} {Phys. Rev. B}\ }\textbf {\bibinfo {volume} {89}},\ \bibinfo
  {pages} {045422} (\bibinfo {year} {2014})}\BibitemShut {NoStop}%
\bibitem [{\citenamefont {Strunz}\ \emph {et~al.}(2019)\citenamefont {Strunz},
  \citenamefont {Wiedenmann}, \citenamefont {Fleckenstein}, \citenamefont
  {Lunczer}, \citenamefont {Beugeling}, \citenamefont {M\"{u}ller},
  \citenamefont {Shekhar}, \citenamefont {Ziani}, \citenamefont {Shamim},
  \citenamefont {Kleinlein}, \citenamefont {Buhmann}, \citenamefont
  {Trauzettel},\ and\ \citenamefont {Molenkamp}}]{Strunz2019}%
  \BibitemOpen
  \bibfield  {author} {\bibinfo {author} {\bibfnamefont {J.}~\bibnamefont
  {Strunz}}, \bibinfo {author} {\bibfnamefont {J.}~\bibnamefont {Wiedenmann}},
  \bibinfo {author} {\bibfnamefont {C.}~\bibnamefont {Fleckenstein}}, \bibinfo
  {author} {\bibfnamefont {L.}~\bibnamefont {Lunczer}}, \bibinfo {author}
  {\bibfnamefont {W.}~\bibnamefont {Beugeling}}, \bibinfo {author}
  {\bibfnamefont {V.~L.}\ \bibnamefont {M\"{u}ller}}, \bibinfo {author}
  {\bibfnamefont {P.}~\bibnamefont {Shekhar}}, \bibinfo {author} {\bibfnamefont
  {N.~T.}\ \bibnamefont {Ziani}}, \bibinfo {author} {\bibfnamefont
  {S.}~\bibnamefont {Shamim}}, \bibinfo {author} {\bibfnamefont
  {J.}~\bibnamefont {Kleinlein}}, \bibinfo {author} {\bibfnamefont
  {H.}~\bibnamefont {Buhmann}}, \bibinfo {author} {\bibfnamefont
  {B.}~\bibnamefont {Trauzettel}}, \ and\ \bibinfo {author} {\bibfnamefont
  {L.~W.}\ \bibnamefont {Molenkamp}},\ }\href {\doibase
  10.1038/s41567-019-0692-4} {\bibfield  {journal} {\bibinfo  {journal} {Nature
  Physics}\ }\textbf {\bibinfo {volume} {16}},\ \bibinfo {pages} {83} (\bibinfo
  {year} {2019})}\BibitemShut {NoStop}%
\bibitem [{1()}]{1}%
  \BibitemOpen
  \href@noop {} {}\bibinfo {note} {We assume the superconducting regions to be
  much larger than the superconducting coherence lengths $\xi _r = v/(\pi
  \Delta _r)$ ($v$ is the Fermi velocity) so that the upper helical edge is
  completed decoupled from other gapless parts of the 2DTI on the outer side of
  each superconducting electrode.}\BibitemShut {Stop}%
\bibitem [{\citenamefont {Delagrange}\ \emph {et~al.}(2015)\citenamefont
  {Delagrange}, \citenamefont {Luitz}, \citenamefont {Weil}, \citenamefont
  {Kasumov}, \citenamefont {Meden}, \citenamefont {Bouchiat},\ and\
  \citenamefont {Deblock}}]{Delagrange2015}%
  \BibitemOpen
  \bibfield  {author} {\bibinfo {author} {\bibfnamefont {R.}~\bibnamefont
  {Delagrange}}, \bibinfo {author} {\bibfnamefont {D.~J.}\ \bibnamefont
  {Luitz}}, \bibinfo {author} {\bibfnamefont {R.}~\bibnamefont {Weil}},
  \bibinfo {author} {\bibfnamefont {A.}~\bibnamefont {Kasumov}}, \bibinfo
  {author} {\bibfnamefont {V.}~\bibnamefont {Meden}}, \bibinfo {author}
  {\bibfnamefont {H.}~\bibnamefont {Bouchiat}}, \ and\ \bibinfo {author}
  {\bibfnamefont {R.}~\bibnamefont {Deblock}},\ }\href {\doibase
  10.1103/physrevb.91.241401} {\bibfield  {journal} {\bibinfo  {journal}
  {Physical Review B}\ }\textbf {\bibinfo {volume} {91}},\ \bibinfo {pages}
  {241401(R)} (\bibinfo {year} {2015})}\BibitemShut {NoStop}%
\bibitem [{\citenamefont {Li}\ \emph {et~al.}(2018{\natexlab{b}})\citenamefont
  {Li}, \citenamefont {de~Boer}, \citenamefont {de~Ronde}, \citenamefont
  {Ramankutty}, \citenamefont {van Heumen}, \citenamefont {Huang},
  \citenamefont {de~Visser}, \citenamefont {Golubov}, \citenamefont {Golden},\
  and\ \citenamefont {Brinkman}}]{Li2018_4pi}%
  \BibitemOpen
  \bibfield  {author} {\bibinfo {author} {\bibfnamefont {C.}~\bibnamefont
  {Li}}, \bibinfo {author} {\bibfnamefont {J.~C.}\ \bibnamefont {de~Boer}},
  \bibinfo {author} {\bibfnamefont {B.}~\bibnamefont {de~Ronde}}, \bibinfo
  {author} {\bibfnamefont {S.~V.}\ \bibnamefont {Ramankutty}}, \bibinfo
  {author} {\bibfnamefont {E.}~\bibnamefont {van Heumen}}, \bibinfo {author}
  {\bibfnamefont {Y.}~\bibnamefont {Huang}}, \bibinfo {author} {\bibfnamefont
  {A.}~\bibnamefont {de~Visser}}, \bibinfo {author} {\bibfnamefont {A.~A.}\
  \bibnamefont {Golubov}}, \bibinfo {author} {\bibfnamefont {M.~S.}\
  \bibnamefont {Golden}}, \ and\ \bibinfo {author} {\bibfnamefont
  {A.}~\bibnamefont {Brinkman}},\ }\href {\doibase 10.1038/s41563-018-0158-6}
  {\bibfield  {journal} {\bibinfo  {journal} {Nature Materials}\ }\textbf
  {\bibinfo {volume} {17}},\ \bibinfo {pages} {875} (\bibinfo {year}
  {2018}{\natexlab{b}})}\BibitemShut {NoStop}%
\bibitem [{\citenamefont {Murani}\ \emph {et~al.}(2017)\citenamefont {Murani},
  \citenamefont {Kasumov}, \citenamefont {Sengupta}, \citenamefont {Kasumov},
  \citenamefont {Volkov}, \citenamefont {Khodos}, \citenamefont {Brisset},
  \citenamefont {Delagrange}, \citenamefont {Chepelianskii}, \citenamefont
  {Deblock}, \citenamefont {Bouchiat},\ and\ \citenamefont
  {Gu{\'{e}}ron}}]{Murani2017}%
  \BibitemOpen
  \bibfield  {author} {\bibinfo {author} {\bibfnamefont {A.}~\bibnamefont
  {Murani}}, \bibinfo {author} {\bibfnamefont {A.}~\bibnamefont {Kasumov}},
  \bibinfo {author} {\bibfnamefont {S.}~\bibnamefont {Sengupta}}, \bibinfo
  {author} {\bibfnamefont {Y.~A.}\ \bibnamefont {Kasumov}}, \bibinfo {author}
  {\bibfnamefont {V.~T.}\ \bibnamefont {Volkov}}, \bibinfo {author}
  {\bibfnamefont {I.~I.}\ \bibnamefont {Khodos}}, \bibinfo {author}
  {\bibfnamefont {F.}~\bibnamefont {Brisset}}, \bibinfo {author} {\bibfnamefont
  {R.}~\bibnamefont {Delagrange}}, \bibinfo {author} {\bibfnamefont
  {A.}~\bibnamefont {Chepelianskii}}, \bibinfo {author} {\bibfnamefont
  {R.}~\bibnamefont {Deblock}}, \bibinfo {author} {\bibfnamefont
  {H.}~\bibnamefont {Bouchiat}}, \ and\ \bibinfo {author} {\bibfnamefont
  {S.}~\bibnamefont {Gu{\'{e}}ron}},\ }\href {\doibase 10.1038/ncomms15941}
  {\bibfield  {journal} {\bibinfo  {journal} {Nature Communications}\ }\textbf
  {\bibinfo {volume} {8}} (\bibinfo {year} {2017}),\
  10.1038/ncomms15941}\BibitemShut {NoStop}%
\bibitem [{\citenamefont {Cr\'epin}\ \emph {et~al.}(2014)\citenamefont
  {Cr\'epin}, \citenamefont {Trauzettel},\ and\ \citenamefont
  {Dolcini}}]{Crepin2014}%
  \BibitemOpen
  \bibfield  {author} {\bibinfo {author} {\bibfnamefont {F.}~\bibnamefont
  {Cr\'epin}}, \bibinfo {author} {\bibfnamefont {B.}~\bibnamefont
  {Trauzettel}}, \ and\ \bibinfo {author} {\bibfnamefont {F.}~\bibnamefont
  {Dolcini}},\ }\href {\doibase 10.1103/PhysRevB.89.205115} {\bibfield
  {journal} {\bibinfo  {journal} {Phys. Rev. B}\ }\textbf {\bibinfo {volume}
  {89}},\ \bibinfo {pages} {205115} (\bibinfo {year} {2014})}\BibitemShut
  {NoStop}%
\bibitem [{Sup()}]{SuppMat}%
  \BibitemOpen
  \href@noop {} {}\bibinfo {note} {See Supplementary Material.}\BibitemShut
  {Stop}%
\bibitem [{2()}]{2}%
  \BibitemOpen
  \href@noop {} {}\bibinfo {note} {The labeling for the hole amplitudes is
  completely analogous.}\BibitemShut {Stop}%
\bibitem [{\citenamefont {Beenakker}(1991)}]{Beenakker1991}%
  \BibitemOpen
  \bibfield  {author} {\bibinfo {author} {\bibfnamefont {C.~W.~J.}\
  \bibnamefont {Beenakker}},\ }\href {\doibase 10.1103/physrevlett.67.3836}
  {\bibfield  {journal} {\bibinfo  {journal} {Physical Review Letters}\
  }\textbf {\bibinfo {volume} {67}},\ \bibinfo {pages} {3836} (\bibinfo {year}
  {1991})}\BibitemShut {NoStop}%
\bibitem [{\citenamefont {Ferraro}\ \emph {et~al.}(2014)\citenamefont
  {Ferraro}, \citenamefont {Wahl}, \citenamefont {Rech}, \citenamefont
  {Jonckheere},\ and\ \citenamefont {Martin}}]{Ferraro2014}%
  \BibitemOpen
  \bibfield  {author} {\bibinfo {author} {\bibfnamefont {D.}~\bibnamefont
  {Ferraro}}, \bibinfo {author} {\bibfnamefont {C.}~\bibnamefont {Wahl}},
  \bibinfo {author} {\bibfnamefont {J.}~\bibnamefont {Rech}}, \bibinfo {author}
  {\bibfnamefont {T.}~\bibnamefont {Jonckheere}}, \ and\ \bibinfo {author}
  {\bibfnamefont {T.}~\bibnamefont {Martin}},\ }\href {\doibase
  10.1103/PhysRevB.89.075407} {\bibfield  {journal} {\bibinfo  {journal} {Phys.
  Rev. B}\ }\textbf {\bibinfo {volume} {89}},\ \bibinfo {pages} {075407}
  (\bibinfo {year} {2014})}\BibitemShut {NoStop}%
\bibitem [{\citenamefont {Calzona}\ and\ \citenamefont
  {Trauzettel}(2019)}]{Calzona2019}%
  \BibitemOpen
  \bibfield  {author} {\bibinfo {author} {\bibfnamefont {A.}~\bibnamefont
  {Calzona}}\ and\ \bibinfo {author} {\bibfnamefont {B.}~\bibnamefont
  {Trauzettel}},\ }\href {\doibase 10.1103/PhysRevResearch.1.033212} {\bibfield
   {journal} {\bibinfo  {journal} {Phys. Rev. Research}\ }\textbf {\bibinfo
  {volume} {1}},\ \bibinfo {pages} {033212} (\bibinfo {year}
  {2019})}\BibitemShut {NoStop}%
\bibitem [{\citenamefont {Str\"om}\ and\ \citenamefont
  {Johannesson}(2009)}]{Stroem2009}%
  \BibitemOpen
  \bibfield  {author} {\bibinfo {author} {\bibfnamefont {A.}~\bibnamefont
  {Str\"om}}\ and\ \bibinfo {author} {\bibfnamefont {H.}~\bibnamefont
  {Johannesson}},\ }\href {\doibase 10.1103/PhysRevLett.102.096806} {\bibfield
  {journal} {\bibinfo  {journal} {Phys. Rev. Lett.}\ }\textbf {\bibinfo
  {volume} {102}},\ \bibinfo {pages} {096806} (\bibinfo {year}
  {2009})}\BibitemShut {NoStop}%
\bibitem [{\citenamefont {Inhofer}\ and\ \citenamefont
  {Bercioux}(2013)}]{Inhofer2013}%
  \BibitemOpen
  \bibfield  {author} {\bibinfo {author} {\bibfnamefont {A.}~\bibnamefont
  {Inhofer}}\ and\ \bibinfo {author} {\bibfnamefont {D.}~\bibnamefont
  {Bercioux}},\ }\href {\doibase 10.1103/physrevb.88.235412} {\bibfield
  {journal} {\bibinfo  {journal} {Physical Review B}\ }\textbf {\bibinfo
  {volume} {88}},\ \bibinfo {pages} {235412} (\bibinfo {year}
  {2013})}\BibitemShut {NoStop}%
\bibitem [{\citenamefont {Dolcini}(2011)}]{Dolcini2010}%
  \BibitemOpen
  \bibfield  {author} {\bibinfo {author} {\bibfnamefont {F.}~\bibnamefont
  {Dolcini}},\ }\href {\doibase 10.1103/PhysRevB.83.165304} {\bibfield
  {journal} {\bibinfo  {journal} {Phys. Rev. B}\ }\textbf {\bibinfo {volume}
  {83}},\ \bibinfo {pages} {165304} (\bibinfo {year} {2011})}\BibitemShut
  {NoStop}%
\bibitem [{3()}]{3}%
  \BibitemOpen
  \href@noop {} {}\bibinfo {note} {Note that our setup is effectively a
  multi-terminal device, featuring two metallic and two superconducting
  electrodes. The latter ones can act as sinks/sources of pairs of electrons
  via Andreev reflection.}\BibitemShut {Stop}%
\bibitem [{4()}]{4}%
  \BibitemOpen
  \href@noop {} {}\bibinfo {note} {Note that Andreev transmission processes
  necessarily involve both spin-preserving and spin-flipping tunnelings. Hence,
  for $\lambda _f=0$, we have $c^{he}_\leftrightarrow =0$ and $0\leq G_{12/21}
  \leq e^2/h$.}\BibitemShut {Stop}%
\bibitem [{\citenamefont {Chang}\ and\ \citenamefont
  {Bagwell}(1994)}]{Chang1994}%
  \BibitemOpen
  \bibfield  {author} {\bibinfo {author} {\bibfnamefont {L.-F.}\ \bibnamefont
  {Chang}}\ and\ \bibinfo {author} {\bibfnamefont {P.~F.}\ \bibnamefont
  {Bagwell}},\ }\href {\doibase 10.1103/PhysRevB.49.15853} {\bibfield
  {journal} {\bibinfo  {journal} {Phys. Rev. B}\ }\textbf {\bibinfo {volume}
  {49}},\ \bibinfo {pages} {15853} (\bibinfo {year} {1994})}\BibitemShut
  {NoStop}%
\bibitem [{5()}]{5}%
  \BibitemOpen
  \href@noop {} {}\bibinfo {note} {Since we consider $\Delta _R \sim \Delta _L$
  and finite junction lengths in order to accomodate the QPC (e.g. $D=6\xi _L$
  in Figs. 2 and 3), we expect to find ABS for all the values of the phase
  difference $\chi $ \cite {Chang1994}.}\BibitemShut {Stop}%
\bibitem [{\citenamefont {Fleckenstein}\ \emph {et~al.}(2021)\citenamefont
  {Fleckenstein}, \citenamefont {Ziani}, \citenamefont {Calzona}, \citenamefont
  {Sassetti},\ and\ \citenamefont {Trauzettel}}]{Fleckenstein2021}%
  \BibitemOpen
  \bibfield  {author} {\bibinfo {author} {\bibfnamefont {C.}~\bibnamefont
  {Fleckenstein}}, \bibinfo {author} {\bibfnamefont {N.~T.}\ \bibnamefont
  {Ziani}}, \bibinfo {author} {\bibfnamefont {A.}~\bibnamefont {Calzona}},
  \bibinfo {author} {\bibfnamefont {M.}~\bibnamefont {Sassetti}}, \ and\
  \bibinfo {author} {\bibfnamefont {B.}~\bibnamefont {Trauzettel}},\ }\href
  {\doibase 10.1103/PhysRevB.103.125303} {\bibfield  {journal} {\bibinfo
  {journal} {Phys. Rev. B}\ }\textbf {\bibinfo {volume} {103}},\ \bibinfo
  {pages} {125303} (\bibinfo {year} {2021})}\BibitemShut {NoStop}%
\bibitem [{sup()}]{supp}%
  \BibitemOpen
  \href@noop {} {\enquote {\bibinfo {title} {See supp. material.}}\
  }\BibitemShut {NoStop}%
\bibitem [{6()}]{6}%
  \BibitemOpen
  \href@noop {} {}\bibinfo {note} {Strictly speaking, the observation of
  $\Sigma =0$ implies either $\theta _R=\theta _L=0$ (i.e. pure electronic
  backscattering) or $\theta _R=\theta _L=\pi /2$ (i.e. pure AR). It is however
  straightforward to distinguish between these to limits. For example, in
  presence of pure electronic backscattering, no Andreev processes are present
  and $G_{12/21}$ cannot be negative. Moreover, in this case, the
  superconductors should not play any role and the conductances are thus
  expected not to depend on $\chi $.}\BibitemShut {Stop}%
\end{thebibliography}%

\onecolumngrid
\appendix
%\counterwithin{figure}{section}
%\renewcommand\appendixname{Section}

\section{Perfect Andreev reflections}
\label{app:AR}
The aim of this section is to derive Eq.~(2) %\eqref{eq:ar}#
of the main text, which describes the phases associated with AR within a JJ. In the setup considered in the main text, the JJ is defined on the upper edge of a 2DTI which has negative helicity, that is $r\sigma = \zeta= -1$. However, for the sake of generality, in the following, we investigate a single helical edge with a generic helicity $\zeta$ (for this reason we can drop the index $r$ of the fermionic fields). Its Hamiltonian reads {(in this Section we assume $v=1$)}
\begin{equation}
	H_\zeta = \int dx \sum_{\sigma} \zeta \sigma \psi_{\sigma}^\dagger(x) (-i\partial_x) \psi_\sigma(x).
\end{equation}
In the portions of the helical system which are proximitized by a standard superconductor, an additional proximity-induced pairing is present
\begin{equation}
	H_\Delta = \int dx\,  \Delta(x) \left[ e^{-i\chi} \psi^\dagger_{\uparrow}(x)  \psi^\dagger_{\downarrow}(x) +e^{i\chi} \psi_{\downarrow}(x)  \psi_{\uparrow}(x) \right].
\end{equation}
The resulting Hamiltonian density in th Bogoliubov-de Gennes (BdG) form reads
\begin{equation}
	\mathcal{H}_{\rm BdG}^\zeta = \begin{pmatrix}
		\zeta p & 0 & 0 & \Delta(x) e^{-i\chi}\\
		0&-\zeta p &-\Delta(x) e^{-i\chi} & 0\\
		0& -\Delta(x) e^{i\chi} &	\zeta p & 0 \\
		\Delta(x) e^{i\chi} & 0&0&-\zeta p \\
	\end{pmatrix}
	%= \zeta p \sigma_z \tau_0 - \Delta \cos(\chi) \sigma_y \tau_y - \Delta \sin(\chi) i \sigma_y \tau_x 
\end{equation}
with $p=-i\partial_x$ and written in the basis $\Psi = (\psi_\uparrow, \psi_\downarrow, \psi_\uparrow^\dagger, \psi_\downarrow^\dagger)^T$.

Focusing on a given proximitized region (and assuming the superconducting pairing to be homogeneous), the BdG equation of the system is given by 
\begin{equation}
	E \begin{pmatrix}
		u_\sigma\\v_{\tilde \sigma}
	\end{pmatrix} =  \sigma  \begin{pmatrix}
		\zeta p&\Delta e^{-i\chi}\\
		\Delta e^{i\chi}&-\zeta p
	\end{pmatrix}
	\begin{pmatrix}
		u_\sigma\\v_{\tilde \sigma}
	\end{pmatrix},
\end{equation}
or equivalently
\begin{equation}
	-i\partial_x \begin{pmatrix}
		u_\sigma\\-v_{\tilde \sigma}
	\end{pmatrix} = 
	\zeta 
	\begin{pmatrix}
		\sigma E&\Delta e^{-i\chi}\\
		-\Delta e^{i\chi}&-\sigma E
	\end{pmatrix}
	\begin{pmatrix}
		u_\sigma\\-v_{\tilde \sigma}
	\end{pmatrix},
\end{equation}
where we denote the eigenvectors of $\mathcal{H}_{\rm BdG}^\zeta$, with eigenvalue $E$, as $\phi=(u_\uparrow, u_\downarrow, v_\uparrow, v_\downarrow)^T$. Its solutions read
\begin{equation}
	\label{eq:sol_SC}
	\begin{pmatrix}
		u_\sigma\\v_{\tilde \sigma}
	\end{pmatrix}
	=
	A_+ e^{\kappa x}\begin{pmatrix}
		e^{-i\chi} (\zeta\sigma E-i\kappa)\\\zeta\Delta
	\end{pmatrix}
	+
	A_- e^{-\kappa x}\begin{pmatrix}
		e^{-i\chi} (\zeta\sigma E+i\kappa)\\\zeta\Delta
	\end{pmatrix},
\end{equation}
where we introduce $\kappa = \sqrt{\Delta^2-E^2}$. By contrast, in the non-proximitized helical region, we have $\Delta = 0$ and the solutions of the BdG  equations are simply plane waves
\begin{equation}
	\begin{pmatrix}
		u_\sigma\\v_{\tilde \sigma}
	\end{pmatrix} = \begin{pmatrix}
		A e^{i \zeta\sigma E x}\\B e^{-i \zeta\sigma E x}
	\end{pmatrix} = \begin{pmatrix}
		A e^{-i \zeta\tilde \sigma E x}\\B e^{i \zeta\tilde \sigma E x}
	\end{pmatrix}.
\end{equation}

Let us now consider the reflections from two semi-infinite superconductors located at $x=x_r$, with $r=R,L$. At first, we focus on an incident electronic plane wave with energy $E$ and amplitude $\beta_{r \sigma}^e$ [see Fig.~1%\ref{fig:setup}
(c) of the main text for a sketch of our notation]. The Andreev reflected hole has an amplitude given by the solution of 
\begin{equation}
	\begin{cases}
		\beta_{r \sigma}^e e^{irEx_r} = C e^{-i\chi_r} (rE+ri\kappa_r)\\
		\alpha_{\tilde r \tilde \sigma}^h e^{-irEx_r} = C r \sigma \Delta_r
	\end{cases}
\end{equation} 
which gives us 
\begin{equation}
	\alpha_{\tilde r \tilde \sigma}^h = e^{i2rEx_r}  e^{i\chi_r} \frac{\sigma \Delta_r}{E+i\kappa_r} \beta_{r \sigma}^e =  \sigma e^{i2rEx_r}  e^{i\chi_r} e^{-i \arccos(E/\Delta_r)} \beta_{r \sigma}^e.
\end{equation}
As for the scattering of holes with energy $E$ and amplitude $\beta_{\tilde r \tilde \sigma}^h$, we get 
\begin{equation}
	\begin{cases}
		\alpha_{\tilde r \tilde \sigma}^e e^{-irEx_r} = C e^{-i\chi_r} (-rE+ri\kappa_r)\\
		\beta_{r \sigma}^h e^{irEx_r} = - C r \tilde \sigma \Delta_r
	\end{cases}
\end{equation} 
which results in 
\begin{equation}
	\alpha_{\tilde r \tilde \sigma}^e = e^{i2rEx_r} e^{-i\chi_r} \frac{E-i\kappa_r}{\tilde \sigma \Delta_r} \beta_{r \sigma}^h = \tilde \sigma e^{i2rEx_r} e^{-i\chi_r} e^{-i\arccos(E/\Delta_r)} \beta_{r \sigma}^h. 
	%= 
	%i \sigma  \; e^{-i \, \chi_{r}} e^{i\Phi_{r}(E)} \beta_{r \sigma}^{h}
\end{equation}
Note that we use the identity ($|E|<\Delta$)  
\begin{equation}	
	\frac{\Delta}{E+i\sqrt{\Delta^2-E^2}} = \exp\left[ -i \arccos \frac{\Delta}{E}\right].
\end{equation}
In summary, we obtain ($c=e/h=\pm 1$, $\tilde e = h$)
\begin{align}
	\label{eq:And_Ref_OLD}
	{
		\alpha_{r \sigma}^{c} = c\sigma \exp \left[
		- i c \chi_{\tilde r} + 2 \tilde r i E  x_{\tilde{r}} -i \arccos(E/\Delta_{\tilde r})\right]\beta_{\tilde r \tilde \sigma}^{\tilde c}}.
\end{align}
The parameters $\chi_r$ and $\Delta_r$ are the superconducting phase and pairing potential of the right ($r=R$) and left ($r=L$) superconducting lead. Note that, if the two superconductors are located at $x_r = r D/2$, one simply gets $
2 \tilde r i E  x_{\tilde{r}} = i E D $
regardless of the value of $r$. Eq.~\eqref{eq:And_Ref_OLD} demonstrates the validity of Eq.~(2) %\eqref{eq:ar} 
of the main text. 

{Incidentally, we observe that Eq.\ \eqref{eq:And_Ref_OLD} is compatible with particle-hole symmetry. The latter exchanges particles and holes $\alpha^c_{r\sigma}\to \alpha^{\tilde c}_{r\sigma}$ and $\beta^c_{r\sigma}\to \beta^{\tilde c}_{r\sigma}$, adds a complex conjugation $i\to -i$, and flips the sign of the energy $E\to-E$. It is easy to verify that \eqref{eq:And_Ref_OLD} is invariant under particle-hole symmetry
	\begin{equation}
		\begin{split}
			&\alpha_{r \sigma}^{c} = c\sigma \exp \left[
			- i c \chi_{\tilde r} + 2 \tilde r i E  x_{\tilde{r}}\right]  \frac{\Delta_{\tilde r}}{E+i \kappa_{\tilde r}} \beta_{\tilde r \tilde \sigma}^{\tilde c} \quad
			\overset{\tiny{P.H.}}{\Rightarrow} \quad \alpha_{r \sigma}^{\tilde c} = -c\sigma \exp \left[
			+ i c \chi_{\tilde r} + 2 \tilde r i E  x_{\tilde{r}}\right] \frac{\Delta_{\tilde r}}{E+i \kappa_{\tilde r}} \beta_{\tilde r \tilde \sigma}^{c}%\\
			%\Rightarrow \quad & \alpha_{r \sigma}^{c} = -c\sigma \exp \left[
			%- i c \chi_{\tilde r} + 2 \tilde r i E  x_{\tilde{r}}\right] \frac{\Delta_{\tilde r}}{-E-i %\kappa_{\tilde r}} \beta_{\tilde r \tilde \sigma}^{\tilde c}. 
		\end{split}
	\end{equation}
	As for time-reversal symmetry, it add a complex conjugation $i\to- i$, exchanges incoming and outgoing states, flips the propagation direction and the spin, and add a spin-dependent sign to the amplitudes, i.e. $\alpha^c_{r\sigma}\to \sigma \beta^{ c}_{\tilde r \tilde \sigma}$ and  $\beta^c_{r\sigma}\to \sigma \alpha^{ c}_{\tilde r \tilde \sigma}$. Eq.~\eqref{eq:And_Ref_OLD} is not invariant under time-reversal symmetry, since the latter flips the sign of the superconducting phases. In particular, one obtains 
	\begin{equation}
		\begin{split}
			&\alpha_{r \sigma}^{c} = c\sigma \exp \left[
			- i c \chi_{\tilde r} + 2 \tilde r i E  x_{\tilde{r}}\right]  \frac{\Delta_{\tilde r}}{E+i \kappa_{\tilde r}} \beta_{\tilde r \tilde \sigma}^{\tilde c}%\\
			\quad \overset{\tiny{T.R.}}{\Rightarrow} \quad
			%&\sigma \beta_{\tilde r \tilde \sigma}^{c} = c\sigma \exp \left[
			%+ i c \chi_{\tilde r} - 2 \tilde r i E  x_{\tilde{r}}\right]  \frac{\Delta_{\tilde r}}{E-i %\kappa_{\tilde r}} \tilde \sigma \alpha_{r \sigma}^{\tilde c}\\
			%\Rightarrow \quad &
			%\alpha_{r \sigma}^{\tilde c} = - \beta_{\tilde r \tilde \sigma}^{c} 
			%c\sigma \exp \left[
			%- i c \chi_{\tilde r} + 2 \tilde r i E  x_{\tilde{r}}\right]  \frac{E-i \kappa_{\tilde %r}}{\Delta_{\tilde r}} \\
			%\Rightarrow \quad &
			\alpha_{r \sigma}^{c} = 
			c\sigma \exp \left[
			+ i c \chi_{\tilde r} + 2 \tilde r i E  x_{\tilde{r}}\right]  \frac{\Delta_{\tilde r}}{E+i \kappa_{\tilde r}} \beta_{\tilde r \tilde \sigma}^{\tilde c} .
		\end{split}
\end{equation}}

%If we assume, for simplicity, $\Delta_R=\Delta_L=\Delta$ as well as $x_r = rL/2$, the matrix takes the simple form
%\begin{equation}
%	S_A = e^{i EL} \frac{\Delta}{E+i \sqrt{\Delta^2-E^2}} \begin{pmatrix}
%	0 & Y_A\\-Y_A^* & 0
%	\end{pmatrix}
%\end{equation}
%with 
%\begin{equation}
%	Y_A = \begin{pmatrix}
%	0&0&0&e^{-i\chi_R} \\
%	0&0&-e^{-i\chi_R}&0 \\
%	0&e^{-i\chi_L}&0&0 \\
%	-e^{-i\chi_L}&0&0&0 
%	\end{pmatrix}.
%\end{equation}

\section{Scattering matrix of the QPC}
\label{app:QPC}
The goal of this section is to derive the scattering matrix of the QPC, both for electron ($S^e_{\rm QPC}$) and hole ($S^h_{\rm QPC}$) states, starting from the Hamiltonian $H_0$ of the helical edges [see Eq.~(1) %\eqref{eq:h0}
of the main text] and from the one describing tunneling at the QPC $ H_{\rm QPC}$ [see Eq.~(4) %\eqref{eq:h_qpc}
of the main text] located at $x=\bar x$. We remind the reader that, in order for the QPC to be time-reversal invariant, we consider real amplitudes for both spin-preserving ($\lambda_{p}$) and spin-flipping ($\lambda_{f}$) tunneling amplitudes. It is straightforward to compute the equations of motion of the field operators, i.e. $i \partial_t \psi_{r\sigma} = - [H_0+H_{\rm QPC},\psi_{r\sigma}]$, that give
\begin{align}
	i \partial_t \psi_{r \sigma} = -i \vartheta_r v \partial_x \psi_{r\sigma}+ 2 v \delta(x-\bar x) \left(\lambda_p \psi_{\tilde r \sigma} + \vartheta_r \lambda_f \psi_{r \tilde \sigma} \right)
\end{align}
\timerev{\textcolor{red}{
		\begin{align}
			i \partial_t \psi_{L\uparrow} &= +i  v \partial_x \psi_{L\uparrow}+ 2 v \delta(x-\bar x) \left(\lambda_p^* \psi_{R\uparrow} -\lambda_f \psi_{L\downarrow} \right)\\
			i \partial_t \psi_{L\downarrow} &= +i v \partial_x \psi_{L\downarrow}+ 2 v \delta(x-\bar x) \left(\lambda_p^* \psi_{R\downarrow} - \lambda_f^* \psi_{L\uparrow} \right)\\
			i \partial_t \psi_{R\uparrow} &= -i v \partial_x \psi_{R\uparrow}+ 2 v \delta(x-\bar x) \left(\lambda_p \psi_{L\uparrow} +\lambda_f \psi_{R\downarrow} \right)\\
			i \partial_t \psi_{R\downarrow} &= -i v \partial_x \psi_{R\downarrow}+ 2 v \delta(x-\bar x) \left(\lambda_p \psi_{L\downarrow} +  \lambda_f^* \psi_{R\uparrow} \right)
		\end{align}
}}
where $\tilde \uparrow = \downarrow$, $\tilde R = L$ and vice-versa. By using the plane-wave ansatz 
\begin{align}
	\psi_{r \sigma}(x) = \frac{e^{-i Et}}{\sqrt{\hbar v}} \begin{cases}
		\alpha_{r \sigma} e^{i r E x/v} \qquad & r (x-\bar x) <0\\
		\beta_{r \sigma} e^{i r E x/v} \qquad & r (x-\bar x) >0
	\end{cases},
\end{align}
we can relate incoming and outgoing amplitudes as
\begin{align}
	i (\beta_{r \sigma} - \alpha_{r \sigma})  = \lambda_p e^{-2\vartheta_r iE \bar x/v} (\beta_{\tilde r \sigma} + \alpha_{\tilde r \sigma}) + \vartheta_r \lambda_f (\beta_{r \tilde \sigma} +\alpha_{r \tilde \sigma}).
\end{align}
\timerev{\textcolor{red}{
		\begin{align}
			i (\beta_{L\uparrow} - \alpha_{L\uparrow})  = \lambda_p^* e^{+2 iE \bar x/v} (\beta_{R\uparrow} + \alpha_{R\uparrow}) - \lambda_f (\beta_{L\downarrow} +\alpha_{L\downarrow})\\
			i (\beta_{L\downarrow} - \alpha_{L\downarrow})  = \lambda_p^* e^{+2 iE \bar x/v} (\beta_{R\downarrow} + \alpha_{R\downarrow}) - \lambda_f^* (\beta_{L\uparrow} +\alpha_{L\uparrow})\\
			i (\beta_{R\uparrow} - \alpha_{R\uparrow})  = \lambda_p e^{-2 iE \bar x/v} (\beta_{L\uparrow} + \alpha_{L\uparrow}) + \lambda_f (\beta_{R\downarrow} +\alpha_{R\downarrow})\\
			i (\beta_{R\downarrow} - \alpha_{R\downarrow})  = \lambda_p e^{-2 iE \bar x/v} (\beta_{L\downarrow} + \alpha_{L\downarrow}) + \lambda_f^* (\beta_{R\uparrow} +\alpha_{R\uparrow})
		\end{align}
}}
By rearranging terms, we get 
\begin{equation}
	\begin{split}
		&\begin{pmatrix}
			i& \lambda_f&-\lambda_p e^{2iE\bar x/v}&\\
			\lambda_f&i&&-\lambda_p e^{2iE\bar x/v}\\
			-\lambda_p e^{-2iE\bar x/v}&&i&-\lambda_f\\
			&-\lambda_p e^{-2iE\bar x/v}&- \lambda_f&i\\
		\end{pmatrix}
		\begin{pmatrix}
			\beta_{L\uparrow}\\
			\beta_{L\downarrow}\\
			\beta_{R\uparrow}\\
			\beta_{R\downarrow}
		\end{pmatrix} = \\ & \qquad 
		\begin{pmatrix}
			i&- \lambda_f&\lambda_p e^{2iE\bar x/v}&\\
			- \lambda_f&i&&\lambda_p e^{2iE\bar x/v}\\
			\lambda_p e^{-2iE\bar x/v}&&i& \lambda_f\\
			&\lambda_p e^{-2iE\bar x/v}& \lambda_f&i\\
		\end{pmatrix}
		\begin{pmatrix}
			\alpha_{L\uparrow}\\
			\alpha_{L\downarrow}\\
			\alpha_{R\uparrow}\\
			\alpha_{R\downarrow}
		\end{pmatrix},
	\end{split}
\end{equation}
\timerev{\textcolor{red}{
		\begin{equation}
			\begin{split}
				&\begin{pmatrix}
					i& \lambda_f&-\lambda_p^* e^{2iE\bar x/v}&0\\
					\lambda_f^*&i&0&-\lambda_p^* e^{2iE\bar x/v}\\
					-\lambda_p e^{-2iE\bar x/v}&0&i&-\lambda_f\\
					0&-\lambda_p e^{-2iE\bar x/v}&- \lambda_f^*&i\\
				\end{pmatrix}
				\begin{pmatrix}
					\beta_{L\uparrow}\\
					\beta_{L\downarrow}\\
					\beta_{R\uparrow}\\
					\beta_{R\downarrow}
				\end{pmatrix} = \\ & \qquad 
				\begin{pmatrix}
					i&- \lambda_f&\lambda_p^* e^{2iE\bar x/v}&0\\
					- \lambda_f^*&i&0&\lambda_p^* e^{2iE\bar x/v}\\
					\lambda_p e^{-2iE\bar x/v}&0&i& \lambda_f\\
					0&\lambda_p e^{-2iE\bar x/v}& \lambda_f^*&i\\
				\end{pmatrix}
				\begin{pmatrix}
					\alpha_{L\uparrow}\\
					\alpha_{L\downarrow}\\
					\alpha_{R\uparrow}\\
					\alpha_{R\downarrow}
				\end{pmatrix}
			\end{split}
\end{equation}}}
which allows us to readily derive the electronic scattering matrix $S^e_{\rm QPC}$ as
\begin{equation} 
	\label{eq:app:SQPCe}
	\begin{pmatrix}
		\beta^e_{L\uparrow}\\
		\beta^e_{L\downarrow}\\
		\beta^e_{R\uparrow}\\
		\beta^e_{R\downarrow}
	\end{pmatrix} = 
	\begin{pmatrix}
		\Lambda_{pf} & \Lambda_{ff} & \Lambda_{pb}& 0\\
		\Lambda_{ff}&\Lambda_{pf} &0&\Lambda_{pb}\\
		-\Lambda_{pb}^*&0&\Lambda_{pf} &-\Lambda_{ff}\\
		0&-\Lambda_{pb}^*&-\Lambda_{ff}&\Lambda_{pf} \\
	\end{pmatrix}
	\begin{pmatrix}
		\alpha^e_{L\uparrow}\\
		\alpha^e_{L\downarrow}\\
		\alpha^e_{R\uparrow}\\
		\alpha^e_{R\downarrow}
	\end{pmatrix}
\end{equation}
\timerev{\textcolor{red}{
		\begin{equation} 
			\label{eq:app:SQPCe_NoTR}
			\begin{pmatrix}
				\beta^e_{L\uparrow}\\
				\beta^e_{L\downarrow}\\
				\beta^e_{R\uparrow}\\
				\beta^e_{R\downarrow}
			\end{pmatrix} = 
			\begin{pmatrix}
				\Lambda_{pf} & \Lambda_{ff} & \Lambda_{pb}& 0\\
				-\Lambda_{ff}^*&\Lambda_{pf} &0&\Lambda_{pb}\\
				-\Lambda_{pb}^*&0&\Lambda_{pf} &-\Lambda_{ff}\\
				0&-\Lambda_{pb}^*&\Lambda_{ff}^*&\Lambda_{pf} \\
			\end{pmatrix}
			\begin{pmatrix}
				\alpha^e_{L\uparrow}\\
				\alpha^e_{L\downarrow}\\
				\alpha^e_{R\uparrow}\\
				\alpha^e_{R\downarrow}
			\end{pmatrix}
\end{equation}}}
with
\begin{align}
	\label{eq:pf}
	\Lambda_{pf} & = \frac{1-\lambda_{f}^2-\lambda_{p}^2}{1+\lambda_{f}^2+\lambda_{p}^2},\\
	\Lambda_{ff} & = \frac{2 i \lambda_f}{1+\lambda_{f}^2+\lambda_{p}^2},\\
	\label{eq:pb}
	\Lambda_{pb} & = -\frac{2 i \lambda_p e^{2iE\bar x/v}}{1+\lambda_{f}^2+\lambda_{p}^2}.
\end{align}
\timerev{\textcolor{red}{
		\begin{align}
			\label{eq:pf_NoTR}
			\Lambda_{pf} & = \frac{1-|\lambda_{f}|^2-|\lambda_{p}|^2}{1+|\lambda_{f}|^2+|\lambda_{p}|^2}\\
			\Lambda_{ff} & = \frac{2 i \lambda_f}{1+|\lambda_{f}|^2+|\lambda_{p}|^2}\\
			\label{eq:pb_NoTR}
			\Lambda_{pb} & = -\frac{2 i \lambda_p^* e^{2iE\bar x/v}}{1+|\lambda_{f}|^2+|\lambda_{p}|^2}.
\end{align}}}
In presence of superconductivity, it is convenient to compute the scattering matrix also for incoming and outgoing holes. In this case, the amplitudes $\alpha^h_{r \sigma}$ and $\beta^h_{r \sigma}$ 
%\begin{align}
%\psi_{r \sigma}^\dagger(x) = \frac{e^{i Et/\hbar}}{\sqrt{\hbar v}} \begin{cases}
%\alpha^h_{r \sigma} e^{-i \vartheta_r k x} \qquad & \vartheta_r (x-\bar x) <0\\
%\beta^h_{r \sigma} e^{-i \vartheta_r k x} \qquad & \vartheta_r (x-\bar x) >0
%\end{cases}
%\end{align}
%which, for a given energy $E$, 
are related by the scattering matrix $S^h(\lambda_p,\lambda_f)= S^e(-\lambda_p,-\lambda_f)$, i.e.
\begin{equation} 
	\label{eq:app:SQPCh}
	\begin{pmatrix}
		\beta^h_{L\uparrow}\\
		\beta^h_{L\downarrow}\\
		\beta^h_{R\uparrow}\\
		\beta^h_{R\downarrow}
	\end{pmatrix} = 
	\begin{pmatrix}
		\Lambda_{pf} & -\Lambda_{ff} &- \Lambda_{pb}& 0\\
		-\Lambda_{ff}&\Lambda_{pf} &0&-\Lambda_{pb}\\
		\Lambda_{pb}^*&0&\Lambda_{pf} &\Lambda_{ff}\\
		0&\Lambda_{pb}^*&\Lambda_{ff}&\Lambda_{pf} \\
	\end{pmatrix}
	\begin{pmatrix}
		\alpha^h_{L\uparrow}\\
		\alpha^h_{L\downarrow}\\
		\alpha^h_{R\uparrow}\\
		\alpha^h_{R\downarrow}
	\end{pmatrix}.
\end{equation}
\timerev{\textcolor{red}{
		\begin{equation} 
			\label{eq:app:SQPCh_NTR}
			\begin{pmatrix}
				\beta^h_{L\uparrow}\\
				\beta^h_{L\downarrow}\\
				\beta^h_{R\uparrow}\\
				\beta^h_{R\downarrow}
			\end{pmatrix} = 
			\begin{pmatrix}
				\Lambda_{pf} & -\Lambda_{ff} &- \Lambda_{pb}& 0\\
				\Lambda_{ff}^*&\Lambda_{pf} &0&-\Lambda_{pb}\\
				\Lambda_{pb}^*&0&\Lambda_{pf} &\Lambda_{ff}\\
				0&\Lambda_{pb}^*&-\Lambda_{ff}^*&\Lambda_{pf} \\
			\end{pmatrix}
			\begin{pmatrix}
				\alpha^h_{L\uparrow}\\
				\alpha^h_{L\downarrow}\\
				\alpha^h_{R\uparrow}\\
				\alpha^h_{R\downarrow}
			\end{pmatrix}.
\end{equation}}}

\section{Scattering matrix on the lower edge}
In this section, we combine the perfect AR on the upper edge with the scattering at the QPC, aiming at deriving the scattering matrix for the amplitudes on the lower helical edge. By solving the resulting linear system for $\beta^e_{R\uparrow}, \beta^e_{L\downarrow}, \beta^h_{R\uparrow}$, and $\beta^h_{L\downarrow}$ we readily obtain Eq.~(5) %\eqref{eq:SG}
of the main text. The latter reads
\begin{equation}
	\label{eq:app:trca}
	\begin{pmatrix}
		\beta_R^e\\
		\beta_L^e\\
		\beta_R^h\\
		\beta_L^h
	\end{pmatrix}
	= \begin{pmatrix}
		t^{ee}_{\rightarrow} &   r^{ee}_{\hookrightarrow}&  c^{eh}_{\rightarrow} & a^{eh}_{\hookrightarrow} \\
		r^{ee}_{\hookleftarrow}	& t^{ee}_{\leftarrow}& a^{eh}_{\hookleftarrow}& c^{eh}_{\leftarrow}\\
		c^{he}_{\rightarrow}&  a^{he}_{\hookrightarrow}& t^{hh}_{\rightarrow}& r^{hh}_{\hookrightarrow}\\
		a^{he}_{\hookleftarrow}& c^{he}_{\leftarrow}& r^{hh}_{\hookleftarrow}& t^{hh}_{\leftarrow}	
	\end{pmatrix}
	\begin{pmatrix}
		\alpha_R^e\\
		\alpha_L^e\\
		\alpha_R^h\\
		\alpha_L^h
	\end{pmatrix}
\end{equation}
where we suppressed the redundant spin index, which is fixed by the (positive) helicity of the lower helical edge. 

It is interesting to show and comment on the analytical expression of the transmission and reflections coefficients. To this end, it is particularly convenient to use the functions $\Phi_r(E)$ functions defined in the main text. The coefficients associated with an incoming electron with spin-up (i.e. from the $C_1$ lead) reads
\begin{align}
	\label{eq:app:tee}
	t^{ee}_\rightarrow &=  \frac{(\lambda_p^2+\lambda_f^2)^2-1}{\Omega} \Big[e^{i(\chi+2\Phi_R+2\Phi_L)}(\lambda_p^2+\lambda_p^2-1)^2 + e^{i\chi} (\lambda_p^2+\lambda_p^2+1)^2 \\ \notag &\qquad + e^{i(\Phi_L+\Phi_R+2\chi)} (\lambda_p^4 + 2 \lambda_p^2 (\lambda_f^2-1) + (1+\lambda_f^2)^2) +  e^{i(\Phi_L+\Phi_R)} (\lambda_f^4 + 2 \lambda_f^2 (\lambda_p^2-1) + (1+\lambda_p^2)^2) \Big]\\
	\label{app:eq:r_ee}
	r^{ee}_{\hookleftarrow} &= \frac{4 e^{2i E \bar x /v}}{\Omega}  e^{i(\Phi_L+\Phi_R)} (e^{2i \chi}-1) \lambda_f\lambda_p ((\lambda_p^2+\lambda_f^2)^2-1)\\
	c^{he}_\rightarrow &= \frac{8 \lambda_f\lambda_p e^{i \chi}}{\Omega}  \Big[(\lambda_p^2+\lambda_f^2-1)^2 e^{i 3(\Phi_L+\Phi_R)/2} \sin \tfrac{4 E \bar x/v +\chi + \Phi_L-\Phi_R }{2}  +(\lambda_p^2+\lambda_f^2+1)^2 e^{i(\Phi_L+\Phi_R)/2} \sin \tfrac{4 E \bar x/v+ \Phi_L-\Phi_R-\chi}{2}\Big]\\
	a^{he}_{\hookleftarrow} &= -\frac{4 i e^{i \chi/2}}{\Omega} \Big[
	\lambda_f^2 (\lambda_p^2+\lambda_f^2-1)^2 e^{i(\Phi_L+2\Phi_R)}+
	\lambda_p^2 (\lambda_p^2+\lambda_f^2-1)^2 e^{i(4 E \bar x /v + 2\Phi_L+\Phi_R+\chi)}\\ \notag & \qquad +
	\lambda_f^2 (\lambda_p^2+\lambda_f^2+1)^2 e^{i(\Phi_R+\chi)}+
	\lambda_p^2 (\lambda_p^2+\lambda_f^2+1)^2 e^{i(\Phi_L+4 E \bar x/v)}
	\Big]
\end{align}
with the denominator 
\begin{equation}
	\Omega = \Big[(\lambda_p^2+\lambda_f^2+1)^2+(\lambda_p^2+\lambda_f^2-1)^2e^{i(\Phi_L+\Phi_R+\chi)}\Big]
	\Big[(\lambda_p^2+\lambda_f^2+1)^2 e^{i\chi}+(\lambda_p^2+\lambda_f^2-1)^2e^{i(\Phi_L+\Phi_R)}\Big].
\end{equation}
The other coefficients of the scattering matrix are given by
\begin{align}
	t^{hh/ee}_\leftarrow &= t^{ee/hh}_\rightarrow \\
	t^{ee}_\leftarrow(\lambda_p,\lambda_f) &= t^{ee}_\rightarrow(\lambda_f,\lambda_p)\\
	r^{hh}_{\hookleftarrow} &= - r^{ee}_{\hookleftarrow} \\
	\label{eq:app:r}
	r^{hh}_{\hookrightarrow} &= - r^{ee}_{\hookrightarrow} = - e^{-4 i  E \bar x/v} r^{ee}_{\hookleftarrow} \\
	\label{eq:app:c}
	c^{he/eh}_\leftarrow &= -c^{he/eh}_\rightarrow\\
	c^{eh}_\rightarrow(-\chi) &= - e^{-2i \chi} c^{he}_\rightarrow(\chi)\\
	a^{eh}_{\hookleftarrow}(-\chi) &= -e^{-2i\chi} a^{he}_{\hookleftarrow}(\chi)\\
	a^{eh}_{\hookrightarrow}(-\chi) &= -e^{-2i\chi} a^{he}_{\hookrightarrow}(\chi)\\
	\label{eq:app:ahe}
	a^{he}_{\hookrightarrow}(\lambda_p,\lambda_f) &= - e^{-4iE \bar x/v} a^{he}_{\hookleftarrow}(\lambda_f,\lambda_p).
\end{align}

Note that the modulus squared of the standard transmission and reflection coefficients does not depend on the position $\bar x$ of the QPC. By contrast, the Andreev transmission and reflection coefficients do depend on $\bar x$ at finite energy. %The reason for this difference is that standard processes always involve an even number of reflections at the two topological superconductors while the Andreev processes involve an odd number of these reflections: In the latter case, the distance between the QPC and one specific topological superconductors matters. 
We also point out that Andreev transmission $c$ and standard reflection $r$ are both proportional to $\lambda_p\lambda_f$: Those processes are therefore present only when both spin-preserving and spin-flipping tunneling events are allowed. As expected, the standard reflections vanish for $\chi = 0,\pi$, i.e. when the whole system is time-reversal symmetric and backscattering within a single helical edge is therefore forbidden. 

\subsection{Particular values of phase and energy}
We observe that the standard transmission coefficient vanishes for $\chi=\pi$ and $E=0$ (which implies $\Phi_L=\Phi_R=0$) that is, when the JJ host a Kramers pair of Majorana zero modes and only Andreev processes are allowed. In this case, the other coefficients reads
\begin{align}
	r^{ee}_{\hookleftarrow} &= 0\\
	c^{he}_\rightarrow &= \frac{-2\lambda_f\lambda_p}{\lambda_f^2+\lambda_p^2}\\
	a^{he}_{\hookleftarrow} &= \frac{\lambda_f^2-\lambda_p^2}{\lambda_f^2+\lambda_p^2},
\end{align}
which lead to $G_{12}(0,\pi)=G_{21}(0,\pi)=-(e^2/h)\,{4\lambda_f^2\lambda_p^2}/{(\lambda_f^2+\lambda_p^2)^2}$. As stated in the main text, we thus have $-e^2/h\leq G_{12}(0,\pi)=G_{21}(0,\pi)\leq 0$ and the value of $-e^2/h$ is reached for $\lambda_p=\lambda_f$.  

It is illustrative to consider the case $\Delta_R=\Delta_L$, which leads to $\Phi_L=\Phi_R=\Phi$. In this case, given the general relation in Eq.~(3) %\eqref{eq:Epr}
of the main text, the energy-phase relation for the $\nu$ class of ABSs is simply given by 
\begin{equation}
	\Phi^\nu(\chi)=(\pi -\nu \chi)/2. 
\end{equation}
By plugging this relation into the expressions of the coefficients, the latter greatly simplify. This allows us to obtain more concise expressions for the conductances. In particular, if we consider the difference between the two non-local conductances $G_{12}$ and $G_{21}$, computed for the same $\chi$ and for the same energy $E$ which satisfy $\Phi(E) = \Phi^\nu(\chi)$, we get 
\begin{equation}
	\delta G (\chi,\nu) = \nu \frac{e^2}{h} \frac{(\lambda_p^2-\lambda_f^2)((\lambda_p^2+\lambda_f^2)^2-1)^2 \sin(\chi)^2}{4(\lambda_f^2+\lambda_p^2)^3 \cos(\chi)^2+(\lambda_p^2+\lambda_f^2)((\lambda_p^2+\lambda_f^2)^2-1)^2 \sin(\chi)^2}.
\end{equation}  
Such an expression vanishes for $\chi=0,\pi$, i.e. when ABSs belonging to different classes are present at the same time, and reaches its largest absolute value for $\chi=\pm \pi/2$. For $\chi=\pi/2$ and $\nu=+1$, we obtain the expression reported in Eq.~(8) %\ref{eq:deltaG}
of the main text. 

\section{Spurious backscattering on the upper edge}
By combining the scattering at the QPC, described in Eqs.~\eqref{eq:app:SQPCe} and \eqref{eq:app:SQPCh}, with the generic reflection matrix in Eq.~(11) %\eqref{eq:ref_gen} 
of the main text, it is straightforward to compute the coefficients of the resulting scattering matrix on the lower edge (whose structure is shown in Eq.~(5) %\eqref{eq:SG}
of the main text). The analytical expressions of those coefficients, which allow us to compute the conductances discussed in the main text, are extremely lengthy and it is not convenient to explicitly write them here. It is, however, extremely useful to consider particular combinations of those coefficients, such as the quantity $\Sigma$ defined in Eq.~(10) %\eqref{eq:Xi}
of the main text. Since the scattering matrix is unitary, the modulus squared of the coefficients of Andreev reflections ($|a^{he}|$) can be expressed in terms of the others. This allows us to write  
\begin{equation}
	\begin{split}
		\Sigma &= G_{12}+G_{11} -G_{21}-G_{22} \\
		& = (|t^{ee}_{\rightarrow}|^2 - |c^{he}_{\rightarrow}|^2) - (|t^{ee}_{\leftarrow}|^2 - |c^{he}_{\leftarrow}|^2) + (1 - |r^{ee}_{\hookleftarrow}|^2 + |a^{he}_{\hookleftarrow}|^2) - (1 - |r^{ee}_{\hookrightarrow}|^2 + |a^{he}_{\hookrightarrow}|^2) \\
		& = (|t^{ee}_{\rightarrow}|^2 - |c^{he}_{\rightarrow}|^2) - (|t^{ee}_{\leftarrow}|^2 - |c^{he}_{\leftarrow}|^2) + (2 - 2 |r^{ee}_{\hookleftarrow}|^2 - |t^{ee}_{\rightarrow}|^2 - |c^{he}_{\rightarrow}|^2) - (2 - 2 |r^{ee}_{\hookrightarrow}|^2 - |t^{ee}_{\leftarrow}|^2 - |c^{he}_{\leftarrow}|^2) \\
		& = 2\left[
		|r^{ee}_{\hookrightarrow}|^2 + |c^{he}_{\leftarrow}|^2 - |r^{ee}_{\hookleftarrow}|^2 - |c^{he}_{\rightarrow}|^2 
		\right].
	\end{split}
\end{equation}
Given Eqs. \eqref{eq:app:r} and \eqref{eq:app:c}, it is straightforward to show that the presence of hABSs necessarily leads to $|r^{ee}_{\hookrightarrow}|^2 = |r^{ee}_{\hookleftarrow}|^2$ and $ |c^{he}_{\leftarrow}|^2 = |c^{he}_{\rightarrow}|^2$, and thus to $\Sigma  =0$. Importantly, in the main text we claim that $\Sigma=0$ also implies the existence of hABSs. Before giving a more mathematical proof of this statement, we discuss the physical picture behind it. 

\subsection{Physical intuition}
\begin{figure}
	\centering
	\includegraphics[width=0.8\linewidth]{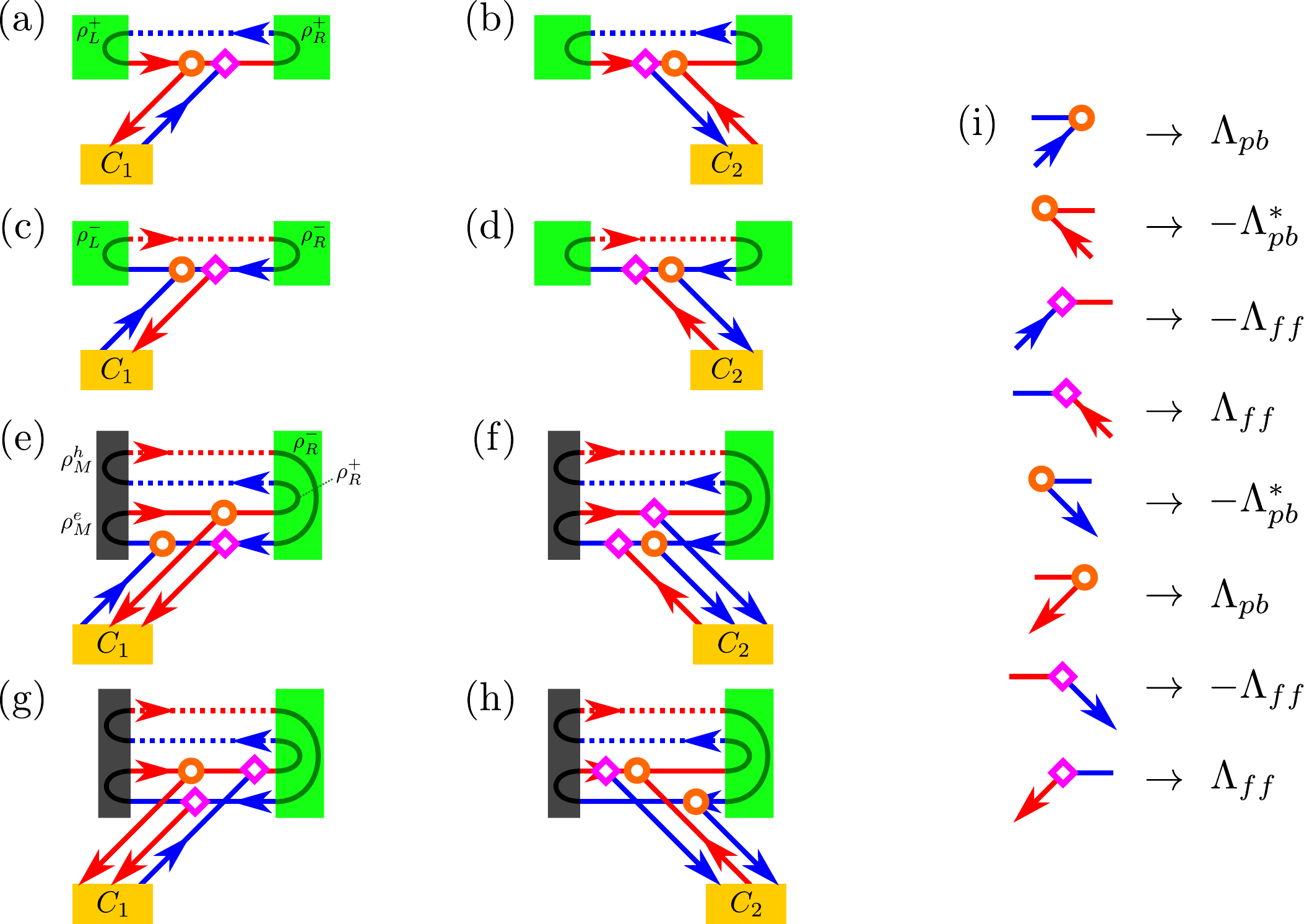}
	\caption{Sketch of the processes that contribute to the reflection at the QPC of electrons on the lower edge. As in Fig.~1 %\ref{fig:setup}
		of the main text, blue (red) lines indicate spin-up (spin-down) channels, while solid (dotted) line indicate electron (hole) channels. Moreover, green (yellow) rectangles indicate superconducting (normal) electrodes. Spin-preserving (spin-flipping) tunneling events at the QPC are denoted by orange circles (pink diamonds). In (a-d) we consider the presence of a single hABS with $\nu=-1$, i.e. consisting of left-moving spin-up electrons and right-moving spin-down holes. In (e-h) we consider a bound state resulting from the presence of a strong magnetic region (dark gray) to the left of the QPC, which induces perfect electronic backscattering. In (i), according to Eq.\ \eqref{eq:app:SQPCe}, we show the amplitudes associated with each tunneling event. }
	\label{fig:reflectionapp}
\end{figure}
{
	Let us carefully analyze the reflection coefficients for electrons emerging from the contacts $C_1$ and $C_2$, i.e.  $r_{\hookleftarrow}^{ee}$ and $r_{\hookrightarrow}^{ee}$. %The Andreev transmission coefficients $c^{he}$ behave in an analogous way. 
	In particular, we want to highlight the differences between two scenarios. In case (I), we consider the presence of hABSs on the upper edge. In case (II), we consider the existence of a completely different mid-gap state on the upper edge, resulting from the presence of perfect electronic backscattering to the left of the QPC. Case (II) could stem from the presence of a magnetic barrier placed to the left of the QPC or, equivalently, to a point-like magnetic impurity, described by the Hamiltonian $H_M$ in the main text, in the limit $m=1$ (see Sec. \ref{sec:app:mag} for more details).}

{
	These two cases allow us to show how the nature of the bound state located on the upper edge has a direct effect on the possible paths that an incoming electron, from the lower edge, can follow before being eventually reflected back. Those paths are sketched in Fig.\ \ref{fig:reflectionapp}, where spin-preserving and spin-flipping tunnelings at the QPC are highlighted with orange circles and pink diamonds, respectively. The corresponding amplitudes, according to Eq.\ \eqref{eq:app:SQPCe}, are summarized for clarity in Fig.~\ref{fig:reflectionapp}(i). Unitary reflections happening at the interfaces with superconductors (in green) are depicted with curved lines and associated with the complex phases $\rho^\pm_{L/R}$. Analogously, unitary reflections at the interface with the magnetic barrier (in gray) are associated with the complex phases $\rho_M^{e/h}$. Note that spin-preserving forward-scattering events at the QPC (associated with the amplitude $\Lambda_{pf}$ in Eq.\ \eqref{eq:app:SQPCe}) are not explicitly shown. The reflection coefficients $r_{\hookleftarrow}^{ee}$ and $r_{\hookrightarrow}^{ee}$ can be calculated by summing all the amplitudes associated with the allowed paths.}

{
	Let us focus on case (I), sketched in Fig.~\ref{fig:reflectionapp}(a-d). % We set our device so that the energy $E$ and the superconducting phase difference $\chi$ match the energy-phase relation of a specific hABS, say with $\nu=-1$. That is the case depicted in Fig. \ref{fig:reflectionapp}(a,b). The existence of the hABS assures that the compatibility relation $
	%	\rho_L \rho_R = 1 $
	%is verified.
	A spin-up electron impinging on the QPC from contact $C_1$ can tunnel to the upper edge either via a spin-flipping [panel (a)] or spin-preserving [panel (c)] tunneling event, coupling to one of the two different classes of hABSs with $\nu=+1$ and $\nu=-1$. However, the only way for this electron to be reflected to contact $C_1$ is to tunnel back into the lower edge with via a tunneling event of opposite nature. The whole reflection process, therefore, necessarily consists of both a spin-preserving and a spin-flipping tunneling event at the QPC. By summing over all the possible paths, we can express the reflection amplitude as
	\begin{equation}
		\begin{split}
			r_{\hookleftarrow}^{ee} & = - \sum_{\nu=\pm} \nu \Lambda_{pb} \Lambda_{pf} \Lambda_{ff}\,\rho_L^\nu \rho_R^\nu \, \sum_{k=0}^{\infty} (\rho_L^\nu \rho_R^\nu \Lambda_{pf}^{2})^k \\&
			=  - \Lambda_{pb} \Lambda_{pf} \Lambda_{ff} \sum_{\nu=\pm} \frac{ \nu \rho_L^\nu \rho_R^\nu}{1-\rho_L^\nu \rho_R^\nu \Lambda_{pf}^{2}},
		\end{split}
	\end{equation}
	where we take into account the possibility of $k$ additional reflections between the two superconductors on the upper edge. As for spin-down electrons coming from contact $C_2$, the case depicted in Fig.\ \ref{fig:reflectionapp}(b,d), we obtain
	\begin{equation}
		\begin{split}
			r_{\hookrightarrow}^{ee} & =  + \Lambda_{pb}^* \Lambda_{pf} \Lambda_{ff} \sum_{\nu=\pm} \frac{ \nu \rho_L^\nu \rho_R^\nu}{1-\rho_L^\nu \rho_R^\nu \Lambda_{pf}^{2}}
		\end{split}
	\end{equation}
	These expressions, which are compatible with Eqs.~\eqref{app:eq:r_ee} and \eqref{eq:app:r}, merely differ by a global phase and they clearly satisfy $|r_{\hookleftarrow}^{ee}|^2 =| r_{\hookrightarrow}^{ee}|^2$.}

{
	The scattering processes are distinctively different in case (II), displayed in Fig.~\ref{fig:reflectionapp}(e-h). Again, a spin-up electron coming from the contact $C_1$ can tunnel to the upper edge both via a spin-preserving [panel (e)] and via a spin-flipping tunneling event [panel (g)]. However, regardless of the nature of this first tunneling event, the electron can tunnel back to the lower edge and reach contact $C_1$ in two different ways, i.e. by preserving of flipping its spin. As a result, there are four different kinds of paths that contribute to $r_{\hookleftarrow}^{ee}$. Their sum reads
	\begin{equation}
		\begin{split}
			r_{\hookleftarrow}^{ee} &= \frac{1}{1-\rho_M^e\rho_R^+\rho_M^h\rho_R^-\Lambda_{pf}^4} \Big[
			\Lambda_{pb} \rho_M^e \Lambda_{pb} + \Lambda_{pb} \rho_M^e\rho_R^+\rho_M^h\rho_R^- \Lambda_{pf}^3 \Lambda_{ff} - \Lambda_{ff} \rho_R^+\rho_M^h\rho_R^- \Lambda_{pf}^2 \Lambda_{ff} - \Lambda_{ff} \rho_M^e\rho_R^+\rho_M^h\rho_R^- \Lambda_{pf}^3 \Lambda_{pb} 
			\Big]\\
			&= \frac{1}{1-\rho_M^e\rho_R^+\rho_M^h\rho_R^-\Lambda_{pf}^4} \Big[
			\Lambda_{pb}^2 \rho_M^e - \Lambda_{ff}^2 \Lambda_{pf}^2 \rho_R^+\rho_M^h\rho_R^- \Big]. 
		\end{split}
	\end{equation}
	As for spin-down electrons coming from contact $C_2$, by looking at Fig.\ \ref{fig:reflectionapp}(f,h), we get
	\begin{equation}
		\begin{split}
			r_{\hookrightarrow}^{ee} &= \frac{1}{1-\rho_M^e\rho_R^+\rho_M^h\rho_R^-\Lambda_{pf}^4}  \Big[
			(\Lambda_{pb}^*)^2  \Lambda_{pf}^2 \rho_R^+\rho_M^h\rho_R^- - \Lambda_{ff}^2 \rho_M^e  \Big]. 
		\end{split}
	\end{equation}
	Those two terms differ more than just for a global phase factor, leading to $|r_{\hookleftarrow}^{ee}|^2 \neq | r_{\hookrightarrow}^{ee}|^2$. The same argument applies to the Andreev transmission coefficients. Hence, a mid-gap state consisting of electronic channels with both spin orientations (i.e. not an hABS), which we mimic in our model by the presence of a magnetic scatterer, results in $\Sigma \neq 0$.}

\subsection{General proof}

In the following, under general assumptions, we demonstrate that $\Sigma = 0$ implies $\theta_r = 0, \pi/2$ (mod $\pi$), thus proving the existence of hABSs. To this end, we analytically compute $\Sigma$ as a function of all the parameters of the systems with the generic reflection matrices [see Eq.~(11) %\eqref{eq:ref_gen}
of the main text], i.e.
\begin{equation}
	\Sigma(\lambda_p,\lambda_f,E,\bar x, \theta_L, \xi^{ee}_L,\xi^{eh}_L,\xi^{he}_L,\xi^{hh}_L, \theta_R, \xi^{ee}_R,\xi^{eh}_R,\xi^{he}_R,\xi^{hh}_R) = \frac{\mathcal{N}}{\mathcal{D}}.
\end{equation}
The denominator $\mathcal{D}$ is a bounded function and the numerator $\mathcal{N}$ is a polynomial in $\lambda_p$ and $\lambda_f$. Requiring that $\Sigma=0$ regardless of the specific value of the tunneling amplitudes at the QPC is equivalent to require that each coefficient $\mathcal{C}_{n,m}$ that multiplies $\lambda_p^n \lambda_f^m$ in $\mathcal{N}$ vanishes. Among the several resulting conditions that have to be met, we focus on one of them, $\mathcal{C}_{2,6} + 2 \mathcal{C}_{4,6} = 0$, which can be conveniently expressed as 
\begin{equation}
	\label{eq:app:cond}
	f(\bar x,E,\eta_L,\eta_R) + f(-\bar x,E,\eta_R,\eta_L) = 0,
\end{equation}
with
\begin{equation}
	\begin{split}
		f(\bar x,E,\eta_L,\eta_R) &= \cos ( \theta_L) \sin ( \theta_R) \Big[2 \sin ( \theta_L) (\cos (2 E \bar x+\xi^{ee}_L-\xi
		^{eh}_L-\xi^{he}_R) -\cos (2 E \bar x+\xi^{ee}_L-\xi^{eh}_R-\xi^{he}_L)\\&\qquad+\cos (2 E \bar x-\xi^{ee}_L+\xi^{he}_L-\xi^{he}_R)+\cos (2 E x+\xi^{he}_L-\xi^{he}_R-\xi^{hh}_L))\\&\qquad+2 \sin (\theta
		_R) (\cos (2 E \bar x+\xi^{ee}_L)+\cos (2 E \bar x+\xi^{hh}_L))\Big],
	\end{split}
\end{equation}
and where we introduce the variable $\eta_r$ that stands for $\theta_r, \xi^{ee}_r,\xi^{eh}_r,\xi^{he}_r,\xi^{hh}_r$. Eq.~\eqref{eq:app:cond} is clearly verified for $\theta_L=\theta_R = 0$ or $\theta_L=\theta_R = \pi/2$. However, for generic values of $\theta_L$ and $\theta_R$, because of the intricate dependence on $\eta_r$ and the position of the QPC $\bar x$, we expect that Eq.~\eqref{eq:app:cond} can only be valid for specific fine-tuned points in the parameter space. {In order to rule out the possibility that the observation of $\Sigma =0$ stems from the fact of having hit one these fine-tuned points, we recommend to measure $\Sigma$ for several different parameter choices. In particular, the $(E-\chi)$ diagram could be sampled [as in Fig. 3 (d)] and several samples could be inspected, featuring e.g. different QPC transparencies (i.e. different $\lambda_p$ and $\lambda_f$), QPC positions (i.e. different $\bar x$) and JJ length $D$. In this sense,} the consistent observation of $\Sigma=0$ over a wide range of parameters represents a proof of $\theta_r = 0, \pi/2$. Note that it is straightforward to distinguish between these two limiting cases and to rule out the $\theta_r = 0$ scenario. For example, since the latter does not allow for any Andreev process, negative values of $G_{12/21}$ would not be possible. Their observation, together with $\Sigma =0$, represents therefore a proof of the existence of hABSs. \\
\begin{figure}
	\centering
	\includegraphics[height=0.2\linewidth]{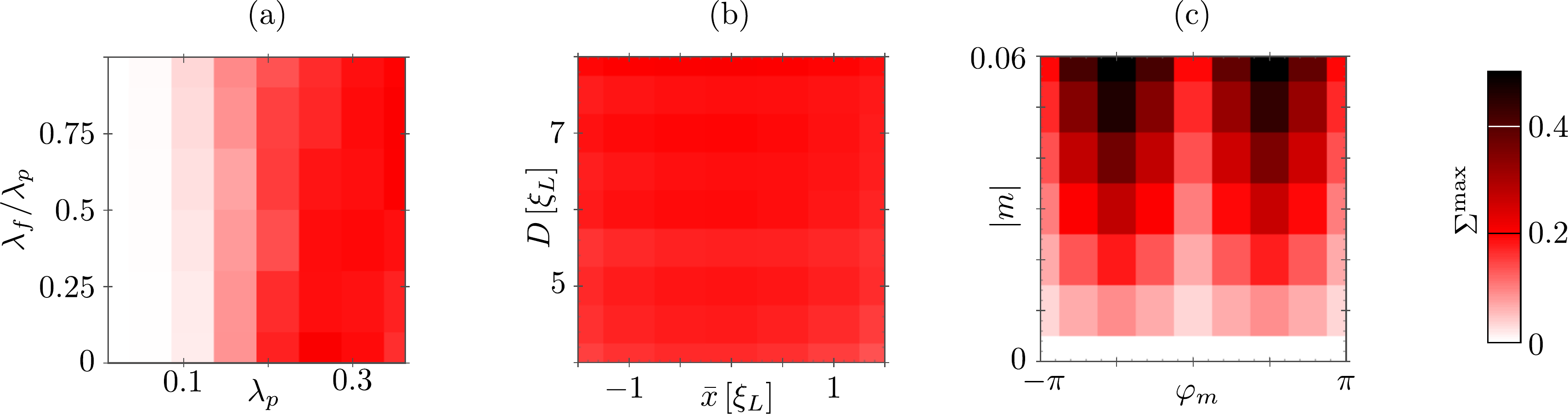}
	\caption{{Values of $\Sigma^{\rm max}$ (in units of $e^2/h$), as a function of different parameters, in presence of a single magnetic impurity with strength $m$ and located at $x_M = 2\xi_L$ (with $\xi_L=v/(\pi \Delta_L)$). All the panels share the same color bar. In panel (a), we study the dependence of $\Sigma^{\rm max}$ on the tunneling amplitude $\lambda_p$ and on the ratio $\lambda_f/\lambda_p$ between spin-flipping and spin-preserving tunneling amplitudes at the QPC. Panel (b) highlights the very weak dependence of $\Sigma^{\rm max}$ on $D$ and $\bar x$ (both in units of $\xi_L$). In Panel (c), we consider different impurity strengths, by varying both the magnitude $|m|$ and the phase $\varphi_m = \arg(m)$. Each panel shares its \textit{fixed} parameters with Fig.~3 of the main text, that is $D=6\xi_L$, $\Delta_L=\Delta, \Delta_R=1.2\Delta, \bar x=\xi_L$, $m=0.05$, $x_{\rm M}=2\xi_L$, $\lambda_p=0.25$, $\lambda_f=0.15$.}
	}
	\label{fig:plotsm1}
\end{figure}

{To better discuss the robustness of our analysis and stress the importance of sampling multiple points in parameter space, we numerically compute the maximum value of $|\Sigma|$ over the whole $(E-\chi)$ diagram, which we denote with $\Sigma^{\rm max}$, for several different scenarios. In Fig.~\ref{fig:plotsm1}, we consider the presence of a single magnetic impurity, as in Fig. 3 of the main text. We plot $\Sigma^{\rm max}$ for different combinations of parameters: $\lambda_p$ and $\lambda_p/\lambda_f$ [Fig.~\ref{fig:plotsm1} (a)], $\bar x$ and $D$ [Fig.~\ref{fig:plotsm1}(b)], and $|m|$ and $\varphi_{m}=\arg(m)$ [Fig.~\ref{fig:plotsm1}(c)]. Note that $\varphi_m$ controls the direction of the impurity magnetization, which lies on the plane perpendicular to the spin quantization axis. These plots show the robustness of the results displayed in Fig.~3 of the main text, where we observe $\Sigma^{\rm max}\sim 0.2\, e^2/h$. The quantity $\Sigma^{\rm max}$ remains indeed considerably and consistently different from zero, with the exception of the trivial limits $|m|\to0$ (i.e. without magnetic impurity) and $\lambda_f,\lambda_p \to 0$ (i.e. without QPC).}

{	
	In Fig.~\ref{fig:plotsm2}, we perform a similar analysis considering the presence of \textit{two} magnetic impurities with strengths $m_i$ ($i=1,2$), each one described by the Hamiltonian $H_{\rm M}= 2vm_i\, \psi_{\uparrow}^\dagger(x_{M_i}) \psi_{\downarrow}(x_{M_i}) + h.c.$ [see Eq. \eqref{eq:HM}], and located on the two sides of the QPC (i.e. $-D/2<x_{M_1}<\bar x<x_{M_2}<D/2$). In this case, we identify one specific scenario that results in $\Sigma=0$, even in presence of magnetic scatterers. It corresponds to the the fully symmetric configuration with \textit{real} $m_1=m_2$, $\Delta_L=\Delta_R$, $\bar x=0$, and $x_{M_2}=-x_{M_1}$ (see the white spots in both panels of Fig.~\ref{fig:plotsm2}). Importantly, however, small deviations from this fine-tuned scenario lead to a rapid increase of $\Sigma^{\rm max}$ to detectable finite values. In particular, in Fig.~\ref{fig:plotsm2}(a), we show how variations in phase and magnitude of $m_1$, while keeping $m_2=0.5$ fixed, result in a finite $\Sigma^{\rm max}$. In Fig.~\ref{fig:plotsm2}(b), we show how, even for symmetric and real strengths $m_1=m_2=0.05$, it is still possible to get finite $\Sigma^{\rm max}$ just by changing the QPC position $\bar x$ and/or the position of one impurity ($x_{\rm M_2}$), while keeping the other one fixed at $x_{\rm M_1}=-2\xi_L$.   
}

\begin{figure}
	\centering
	\includegraphics[height=0.2\linewidth]{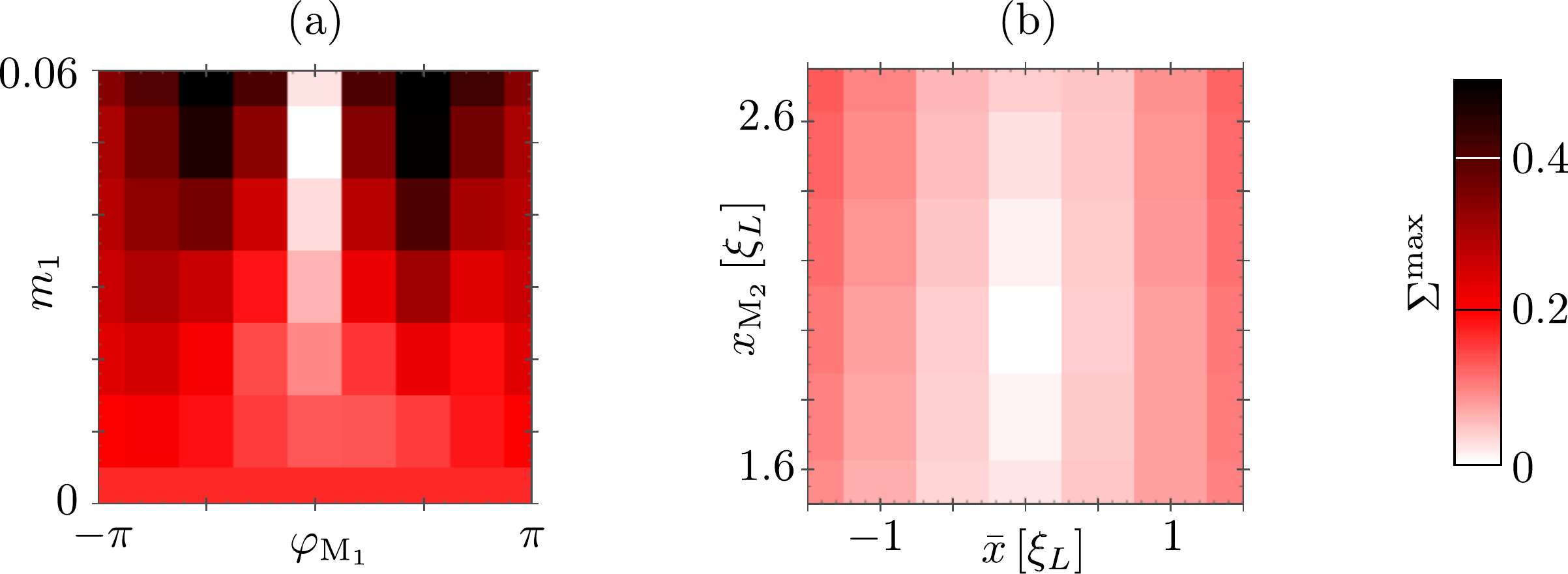}
	\caption{{Values of $\Sigma^{\rm max}$ (in units of $e^2/h$), as a function of different parameters, in presence of two magnetic impurities, with strengths $m_1$ and $m_2$, located on both sides of the QPC (i.e $x_{M_1}<\bar x<x_{M_2}$). The two panels share the same color bar. 
			In panel (a), we study the dependence of $\Sigma^{\rm max}$ on the strength of impurity $1$, by varying both its magnitude $|m_1|$ and phase $\varphi_{m_1} = \arg(m_1)$, while keeping $m_2=0.05$ fixed. The QPC is at $\bar x=0$ and the impurities are located at $x_{\rm M_2}=-x_{\rm M_1}=2\xi_L$. 
			In panel (b), we plot $\Sigma^{\rm max}$ as a function of the positions of the QPC ($\bar x$) and the impurity $2$ ($x_{\rm M_2}$), both in units of $\xi_L$, for $m_1=m_2=0.05$ and $x_{\rm M_1}=-2\xi_L$.  
			Both panels share the remaining parameters, which read $D=6\xi_L$, $\Delta_L=\Delta_R=\Delta$,
			$\lambda_p=0.25$, $\lambda_f=0.15$.}}
	\label{fig:plotsm2}
\end{figure}

%We can therefore conclude that the consistent observation of $\Xi = 0$ for different values of the parameters (e.g. different values of energy and superconducting phase difference, with $\bar x \neq 0$) represents a proof of $\theta_r = 0, \pi/2$. These two extreme cases, only electronic backscattering and only superconducting PAR, should be easy to distinguish between each other, for example because pure electronic backscattering should be independent on the superconducting phase difference of the Josephson junction. 

\section{Magnetic impurity}% \textcolor{red}{TO BE CHECKED!}}
\label{sec:app:mag}
Here, we consider the effect of a delta-like magnetic impurity along the upper helical edge, described by the Hamiltonian (we suppress the redundant index $r$)
\begin{equation}
	\label{eq:HM}
	H_{\rm M}= 2vm\, \psi_{\uparrow}^\dagger(x_M) \psi_{\downarrow}(x_M) + h.c. .
\end{equation}
The equations of motion for the field operators, i.e. $i \partial_t \psi_{\sigma} = - [H_{\zeta=-1}+H_{\rm M},\psi_{\sigma}]$, become
\begin{align}
	i\partial_t \psi_\uparrow &= +iv\partial_x \psi_{\uparrow} + 2mv \delta(x-x_M) \psi_\downarrow\\ 
	i\partial_t \psi_\downarrow &= -iv\partial_x \psi_{\downarrow} +2 m^*v \delta(x-x_M) \psi_\uparrow.
\end{align}
Using again the plane wave ansatz
\begin{align}
	\psi_{\sigma}(x) = \frac{e^{-i Et}}{\sqrt{\hbar v}} \begin{cases}
		\tau^e_{\sigma} e^{-i \sigma E x/v} \qquad & \sigma(x- x_M) >0\\
		\omega^e_{\sigma} e^{-i \sigma E x/v} \qquad & \sigma (x- x_M) <0
	\end{cases},
\end{align}
we can relate the incoming ($\tau$) and outgoing amplitudes ($\omega$) as
\begin{align}
	i(\omega^e_{\uparrow}-\tau^e_{\uparrow}) &= m e^{2i Ex_M/v} (\omega^e_{\downarrow}+\tau^e_{\downarrow})\\
	i(\omega^e_{\downarrow}-\tau^e_{\downarrow}) &= {m^*} e^{-2i Ex_M/v}(\omega^e_{\uparrow}+\tau^e_{\uparrow}).
\end{align}
The resulting (electronic) scattering matrix reads
\begin{equation}
	\begin{pmatrix}
		\omega^e_{\uparrow}\\
		\omega^e_{\downarrow}
	\end{pmatrix} = 
	\frac{1}{1+|m|^2} \begin{pmatrix}
		1-|m|^2  & -2i e^{2i Ex_M/v} m\\
		-2ie^{-2i Ex_M/v} m^*&1-|m|^2
	\end{pmatrix}
	\begin{pmatrix}
		\tau^e_{\uparrow}\\
		\tau^e_{\downarrow}
	\end{pmatrix}.
\end{equation}
We observe that, for $m=1$, the transmission coefficients vanish. In this limit, therefore, the magnetic impurity described by $H_{\rm M}$ induces perfect electronic backscattering (with spin-flip) at $x=x_{M}$. {Introducing the hole amplitudes, we get
	\begin{equation}
		\begin{pmatrix}
			\omega^e_{\uparrow}\\
			\omega^e_{\downarrow}\\
			\omega^h_{\uparrow}\\
			\omega^h_{\downarrow}
		\end{pmatrix} = S_{\rm M}(m,x_M,E)
		\begin{pmatrix}
			\tau^e_{\uparrow}\\
			\tau^e_{\downarrow}\\
			\tau^h_{\uparrow}\\
			\tau^h_{\downarrow}
		\end{pmatrix} 
	\end{equation}
	with 
	\begin{equation}
		\label{eq:app:SM}
		S_{\rm M}(m,x_M,E) = 
		\frac{1}{1+|m|^2} \begin{pmatrix}
			1-|m|^2  & -2i e^{2i Ex_M/v} m&0&0\\
			-2i e^{-2i Ex_M/v} m^*&1-|m|^2&0&0\\
			0&0&1-|m|^2  & 2i e^{2i Ex_M/v} m^*&\\
			0&0&2ie^{-2i Ex_M/v}  m&1-|m|^2\\
		\end{pmatrix}
\end{equation}}

\section{{Properites of the non-local differential conductances}}
{The aim of this section is to address the properties of the non-local differential conductances $G_{12/21}(E,\chi)$, with respect to the inversion of $E$ and/or $\chi$.  }

{
	\subsection{Energy inversion}
	From the observation of Fig.~2(a-c) in the main text, one can immediately notice the asymmetry $G_{12/21} (E,\chi) \neq G_{12/21} (-E,\chi)$. The latter stems precisely from the capability of our system to selectively detect only one class of hABS (and not its particle-hole symmetric partner, with opposite spin structure). This feature is strictly present for $\lambda_f=0$ [Fig. 2(a) of the main text]. However, the asymmetry survives also in presence of a weak to moderate $\lambda_f <\lambda_p$ [see Fig. 2(b,c) of the main text]. It disappears only for the special case $\lambda_f=\lambda_p$, i.e. when the tunneling at the QPC is completely spin insensitive and both classes of hABS give the same signal in the conductances.}

{
	\subsection{Energy and phase inversion}
	Interestingly, in presence of perfect hABSs, the non-local differential conductance satisfy 
	\begin{equation}
		\label{eq:app:symm}
		G_{12/21} (E,\chi) = G_{12/21}(-E,-\chi).
	\end{equation}
	This is a direct consequence of the energy-phase relation of each class of hABS [see Eq.(3) in the main text], which is indeed invariant under the transformation (note that the functions $\Phi_r$ are odd with respect to E)
	\begin{equation}
		\label{eq:tr}
		\begin{cases}
			E\to -E\\
			\chi \to -\chi.
		\end{cases}
	\end{equation} 
	At the mathematical level, we observe that the transformation \eqref{eq:tr} modifies the reflection coefficients at the QPC and at the superconducting interfaces as [see Eqs. (2) and Eqs. (\ref	{eq:app:SQPCe}-\ref{eq:app:SQPCh})]
	\begin{align}
		S_{\rm QPC}(E,\lambda_p,\lambda_f) &= S_{\rm QPC}(-E,-\lambda_p,-\lambda_f)^* \\
		\frac{\alpha^c_{r\sigma}}{\beta^{\tilde c}_{{\tilde r}{\tilde \sigma}}}\Big|_{(E,\chi)} &=
		-\left[
		\frac{\alpha^c_{r\sigma}}{\beta^{\tilde c}_{{\tilde r}{\tilde \sigma}}}\Big|_{(-E,-\chi)}
		\right]^*.
	\end{align}
	Those changes are irrelevant for the computation of the absolute values of the transmission and reflection coefficients on the lower edge [i.e. $t,r,c,a$ in Eq. (5) in the main text and in Eq.~\eqref{eq:app:trca}]. As a consequence, they have no effect on the differential conductances either, as confirmed by the analytical expressions of the coefficients in Eqs. (\ref{eq:app:tee}-\ref{eq:app:ahe}).}

{
	The situation is different, however, in presence of additional scattering mechanisms on the upper edge. For the sake of concreteness, let us focus on the presence of magnetic impurities. In this case, the transformation \eqref{eq:tr} modifies the corresponding scattering matrix as $S_{\rm M}(E,m) = S_{\rm M}(-E,-m^*)^*$ [see Eq.~\eqref{eq:app:SM}]. Every amplitude appearing in Eq.~\eqref{eq:app:trca} results from the interference of paths featuring different numbers of reflections on the magnetic impurity. As long as $m$ is not imaginary, therefore, the transformation $m\to -m^*$ modifies the interference and thus the differential conductances. This explains why, the relation $G_{12/21} (E,\chi) = G_{12/21}(-E,-\chi)$ does not hold in Fig. 3 of the main text. However, it is possible to construct scattering mechanisms that preserve Eq.~\eqref{eq:app:symm} while still hybridizing and destroying the hABSs. It is the case, for example, for magnetic impurities with imaginary $m$, i.e. impurities with magnetization along the $y$ direction (assuming the spin quantization axis to be along $z$). Such a scenario is analyzed in Fig.~\ref{fig:plotsmF}, where we plot the differential conductances $G_{12/21/11}$ and $\Sigma$ using the same parameters considered in Fig.~3 of the main text, but with an imaginary $m=0.05 \, i$. The two non-local conductances $G_{12/21}$ clearly satisfy Eq.~\eqref{eq:app:symm}, even in presence of magnetic scatterers. Importantly, the absence of hABSs is correctly signaled by the quantity $\Sigma$ [Fig.~\ref{fig:plotsmF}(d)], which features large deviations from zero. 
}

\begin{figure}
	\centering
	\includegraphics[height=0.2\linewidth]{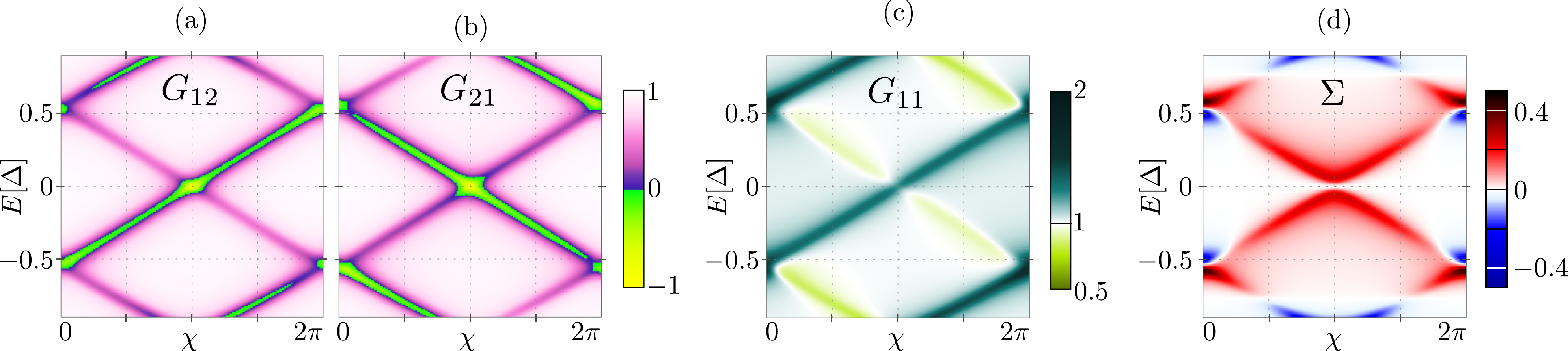}
	\caption{{Differential conductances (in units of $e^2/h$) in presence of electronic backscattering on the upper edge, induced by a magnetic impurity with $m=0.05\, i$ and located at $x_{\rm M} = 2\xi_L$ (with $\xi_L=v/(\pi \Delta_L)$). The remaining parameters are the same as in Figs. 2(b,c) and 3 of the main text. They read $D=6\xi_L$, $\lambda_p=0.25$, $\lambda_f=0.15$, $\bar x = \xi_L$, $\Delta_L=\Delta$, $\Delta_R = 1.2 \Delta$.}}
	\label{fig:plotsmF}
\end{figure}

\end{document}